\documentclass[preprint,1p]{elsarticle}

\usepackage{hyperref}

\journal{Astroparticle Physics}
\usepackage{bm}       
\usepackage{graphicx}
\usepackage{hyperref}      
\usepackage{color}
\usepackage{amssymb,amsmath}

\newcommand*{\mysub}[2]{\ensuremath{#1_{\mathrm{#2}}}}

\newcommand*{\fgas}{\mysub{f}{gas}\,}
\newcommand*{\Omegam}{\mysub{\Omega}{m}}
\newcommand*{\Omegab}{\mysub{\Omega}{b}}
\newcommand*{\Omegal}{\ensuremath{\Omega_{\Lambda}}}
\newcommand*{\LCDM}{\ensuremath{\Lambda}CDM}

\begin{document}

\begin{frontmatter}

\title{Distance Probes of Dark Energy}

\author[a]{A.~G.~Kim}
\author[b]{N.~Padmanabhan}
\author[a]{G.~Aldering}
\author[c,d]{S.~W.~Allen}
\author[b]{C.~Baltay}
\author[a]{R.~N.~Cahn}
\author[e]{C.~B.~D'Andrea}
\author[f]{N.~Dalal}
\author[g]{K.~S.~Dawson}
\author[h]{K.~D.~Denney}
\author[i]{D.~J.~Eisenstein}
\author[j]{D.~A.~Finley}
\author[k]{W.~L.~Freedman}
\author[l]{S.~Ho}
\author[m]{D.~E.~Holz}
\author[n,o]{D.~Kasen}
\author[p]{S.~M.~Kent}
\author[q]{R.~Kessler}
\author[r]{S.~Kuhlmann}
\author[a,s]{E.~V.~Linder}
\author[h]{P.~Martini}
\author[t]{P.~E.~Nugent}
\author[a,o]{S.~Perlmutter}
\author[h]{B.~M.~Peterson}
\author[u]{A.~G.~Riess}
\author[a]{D.~Rubin}
\author[v]{M.~Sako}
\author[w]{N.~V.~Suntzeff}
\author[a,g]{N.~Suzuki}
\author[t]{R.~C.~Thomas}
\author[x]{W.~M.~Wood-Vasey}
\author[y]{S.~E.~Woosley}

\address[a]{Physics Division, Lawrence Berkeley National Laboratory, 1 Cyclotron Road, Berkeley, CA 94720, USA}
\address[b]{Physics Department, Yale University, PO Box 208121, New Haven CT 06520, USA}
\address[c]{Kavli Institute for Particle Astrophysics and Cosmology, SLAC National Accelerator Laboratory, 2575 Sand Hill Road, Menlo Park, CA 94025, USA}
\address[d]{Department of Physics, Stanford University, 382 Via Pueblo Mall, Stanford, CA 94305, USA}
\address[e]{Institute of Cosmology and Gravitation, University of Portsmouth, Dennis Sciama Building, Burnaby Road, Portsmouth PO1 3FX, UK}
\address[f]{Astronomy Department, 
University of Illinois at Urbana-Champaign,
1002 W. Green Street, Urbana, IL 61801,
USA}
\address[g]{Department of Physics and Astronomy,
University of Utah, Salt Lake City, UT 84112, USA}
\address[h]{Department of Astronomy and Center for Cosmology and Astroparticle Physics,
The Ohio State University
140 West 18th Avenue
Columbus, OH 43210, USA}
\address[i]{Harvard University.
Harvard-Smithsonian Center for Astrophysics, 60 Garden St.,
Cambridge, MA 02138, USA}
\address[j]{Particle Physics  Division, Fermi National Accelerator Laboratory, P.O. Box 500, Batavia, IL 60510, USA}
\address[k]{Carnegie Observatories, 813 Santa Barbara Street, Pasadena, CA 91101, USA}
\address[l]{McWilliams Center for Cosmology, Department of Physics, Carnegie Mellon University, 5000 Forbes Ave, Pittsburgh, PA 15213, USA}
\address[m]{Enrico Fermi Institute, Department of Physics, and Kavli Institute for Cosmological Physics University of Chicago, Chicago, IL 60637, USA}
\address[n]{Physics Department,
University of California, Berkeley,
366 LeConte Hall,
Berkeley, CA, 94720, USA}
\address[o]{Nuclear Science Division, Lawrence Berkeley National Laboratory, 1 Cyclotron Road, Berkeley, CA 94720, USA}
\address[p]{Scientific Computing Division, Fermi National Accelerator Laboratory, P.O. Box 500, Batavia, IL 60510, USA}
\address[q]{Department of Astronomy and Astrophysics, University of Chicago, 5640 South Ellis Avenue, Chicago, IL 60637, USA}
\address[r]{Argonne National Laboratory,
9700 S. Cass Avenue,
Lemont, IL 60439, USA}
\address[s]{
Space Sciences Laboratory, University of California, Berkeley, 94720 USA}
\address[t]{Computational Research Division, Lawrence Berkeley National Laboratory, 1 Cyclotron Road, Berkeley, CA 94720, USA}
\address[u]{Department of Physics and Astronomy, Johns Hopkins University, 3400 North Charles Street, Baltimore, Maryland 21218, USA}
\address[v]{Department of Physics and Astronomy, University of Pennsylvania, 209 South 33rd Street, Philadelphia, PA 19104, USA}
\address[w]{George P. and Cynthia Woods Mitchell Institute for Fundamental Physics and Astronomy, Department of Physics and Astronomy, Texas A\&M University, College Station, TX 77843, USA}
\address[x]{Pittsburgh Particle Physics, Astrophysics, and Cosmology Center (Pitt-PACC), University of Pittsburgh, Pittsburgh, PA 15260, USA}
\address[y]{Department of Astronomy and Astrophysics, University of California, Santa Cruz, CA 95064, USA}

\begin{abstract}
This document presents the results from the Distances subgroup of the Cosmic Frontier Community Planning Study (Snowmass 2013). We summarize the current state of the field as well as future prospects and challenges. In addition to the established probes using Type~Ia supernovae and baryon acoustic oscillations, we also consider prospective methods based on clusters, active galactic nuclei, gravitational wave sirens and strong lensing time delays.
\end{abstract}


\end{frontmatter}















\section{Executive Summary}
A basic paradigm of physics is that if one
measures the distance to an object as a function of time, one can
determine its velocity and acceleration.  Add dynamics and one can find
the forces, either from $F=ma$ or $G_{\mu\nu} = 8\pi T_{\mu\nu}$, and from
these one can determine the nature of the underlying matter. Already
distance measurements have shown that the
cosmological constant, long disowned as being no more than theoretically
allowable, is in fact a
necessity. 
What remains to be seen is whether the Universe is pervaded by a uniform and never-changing energy density,
an energy density that varies in time and perhaps position, or whether describing the Universe as a whole by General Relativity fails. Whatever future experiments reveal, the simple plot of the distance scale of the Universe as a function of time will be one of the primary icons of physical science.  As the Copernican picture of the solar system removed our privileged perspective at the center of the Universe, the cosmological distance plot shows that the baryons, the matter that composes our physical
essence,
represent a minority fraction of the energy content of the Universe.

The data for this plot so far come primarily from measurements of Type Ia supernovae (SNe~Ia) and baryon acoustic oscillations (BAO) and these will be the sources of streams of data in coming years.  The Dark Energy Survey (DES) and Large Synoptic Survey Telescope (LSST) will provide an essentially limitless supply of supernova, thousands, then hundreds of thousands.  The challenge is to make measurements thoroughly enough to mitigate systematic uncertainties, especially those that are functions of redshift.  Detailed studies of nearby supernovae are beginning to provide clues for how to do this.  Much would be gained if observations could be made from space, but some of the gain could be achieved if we could make ground-based observations that avoid the atmospheric lines in the near infrared.

The subtle pattern of anisotropy in the cosmic microwave background, just one part in $10^5$, is just the two dimensional boundary of a three-dimensional feature, the fluctuations in matter density throughout space.  The counterpart of the oscillations in the CMB power spectrum is a peak in the correlation between the densities at points separated by 150 Mpc, left behind by baryon acoustic oscillations in the early Universe.  This very large meter stick can be observed at redshifts out as far as $z=1.6$ using galaxies as traces of matter density, and even out to $z=3$ using light from quasars.  The current experiment, the Baryon Oscillation Spectroscopic Survey (BOSS) \cite{2013AJ....145...10D}, is likely to report a distance measurement soon at 1\% accuracy and ultimately will provide two or perhaps three.  The successor BAO experiment, Dark Energy Spectroscopic Instrument (DESI), should provide more than a dozen  independent distance measurements.

If our basic understanding is correct, the supernova and BAO measurements should be in absolute agreement.  The distance-versus-time curve of the Universe is so fundamental that exploring it with completely different techniques is essential.  A basic disagreement would challenge our current picture, just as the discovery of the accelerating Universe upset the earlier picture.  Provided the measurements agree, we can go on to see that they are consistent with expectations for a Universe
containing 30\% matter and 70\% nearly-constant energy density.  Finally, we will ask whether the nearly constant part is really constant. 
How well can data exclude a cosmological constant?

To measure progress in determining the expansion history of the Universe a simple quantitative characterization was proposed by the 2006 Dark Energy Task Force.  The  equation of state of the dark energy, $w=p/\rho$, which is $-1$ for the cosmological constant can be expanded  as $w=w_0+(1-a)w_a$, where $a=1/(1+z)$ is the size-scale of the Universe relative to its size today, when $a=1$ and $z=0$.  The DETF figure-of-merit is simply the reciprocal of the area of the error ellipse in the $w_0 - w_a$ plane, suitably normalized.  This figure-of-merit is calculated using input from projections from the full Planck survey and from existing or projected results from the various dark energy experiments.  

The cosmic expansion history is a fundamental element of the physics of our 
Universe. Ideally we would map it accurately at all redshifts. Within a 
cosmological model such as cold dark matter plus dynamical dark energy, 
the precision on the dark energy equation of state $w(z)=w_0+w_a z/(1+z)$ 
starts to plateau for measurements beyond $z\approx1.5-2$. However, even 
within a cosmological constant model the dark energy contributes nearly 10\% 
of the cosmic energy density at $z=2$ and alters the deceleration parameter 
by 25\%. Surprises could certainly await as we probe to these redshifts and 
beyond. Thus next generation experiments aim to map cosmic distances to 
$z\approx2$, as outlined in the Rocky III report, while keeping in mind 
potential techniques to improve our understanding further. 

\begin{figure}
\centering
\includegraphics[width=0.485\textwidth]{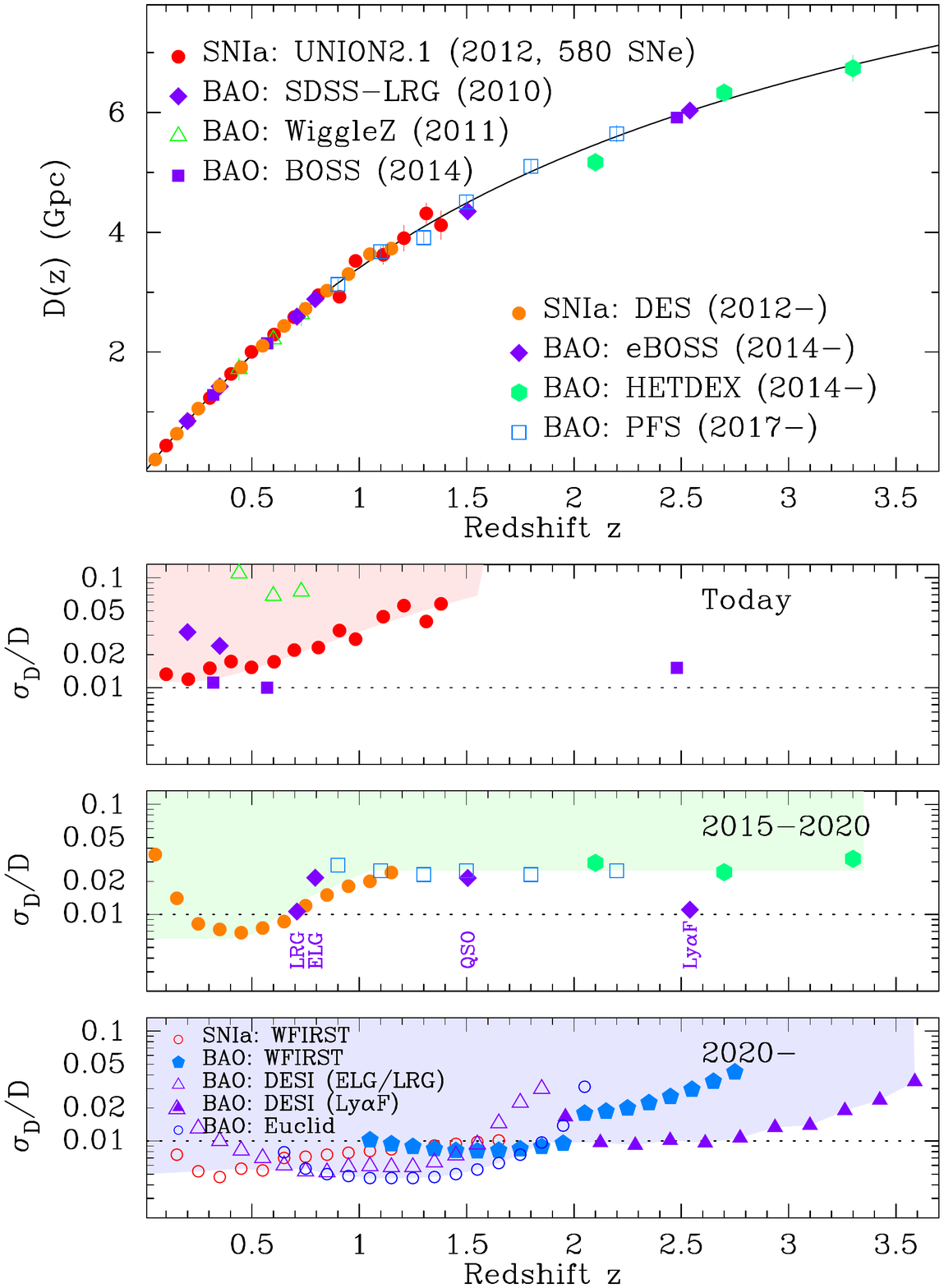} 
\includegraphics[width=0.485\textwidth]{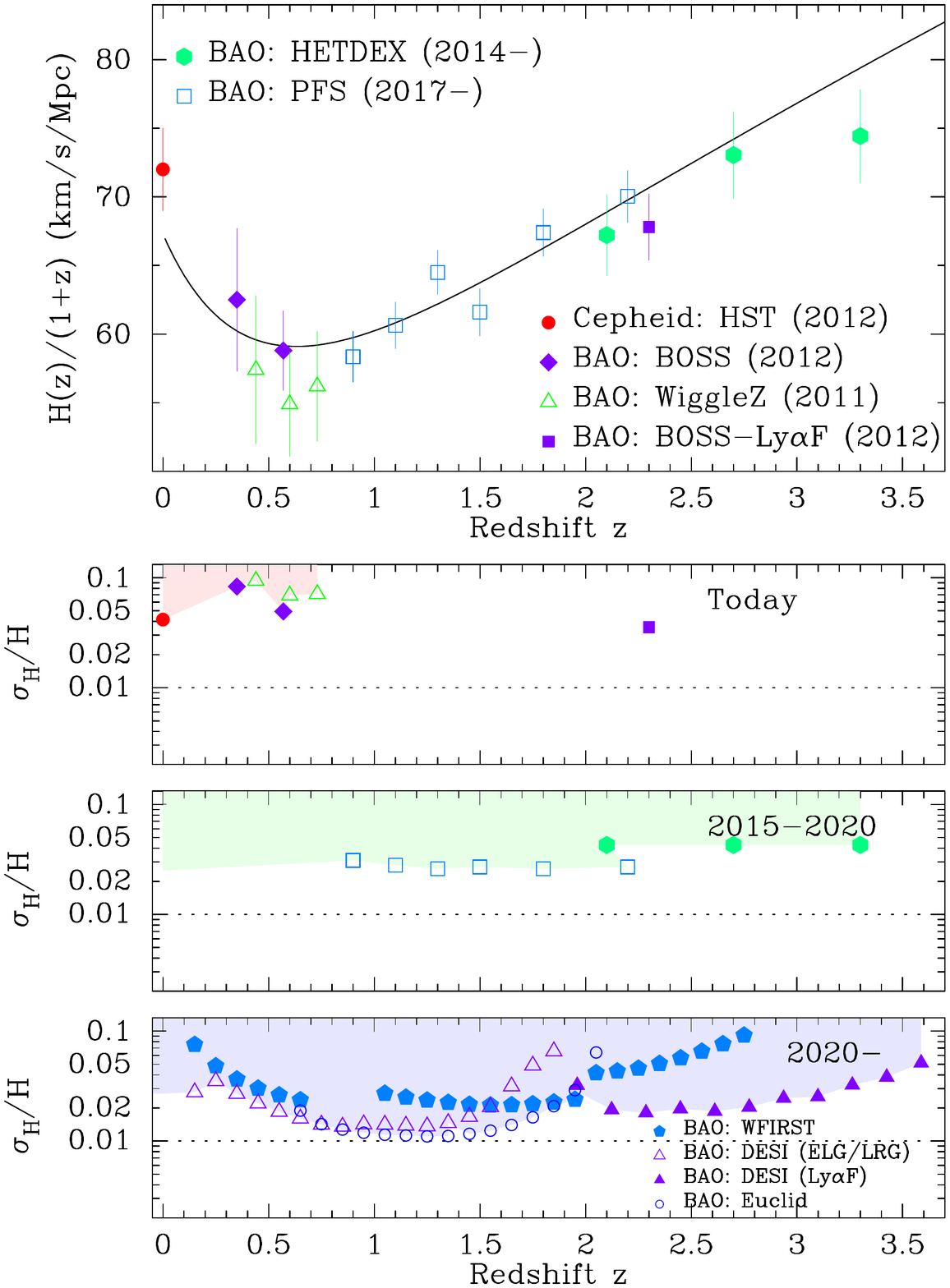} 
\caption{Measurement uncertainties of the distance scale from current and near-term SN~Ia and BAO projects.
The top panels represent the projected distance measurements as a function of redshift.  The second, third, and fourth rows of panels represent the current uncertainties, projected uncertainties in 2015--2020, and projected uncertainties after 2020, respectively. Uncertainty envelopes are shaded to guide the eye. Projections are based on a $\Lambda$CDM Cosmology ($h = 0.673$, $\Omega_M= 0.315$, $\Omega_\Lambda = 0.685$).
Left:  Compilation of current and future measurements of conformal distance $D(z)$.  Right:  Compilation of current and future measurements of expansion rate history $H(z)/(1 + z)$.  \label{future_SN_BAO}}
\end{figure}

Anticipated progress in direct distance measurements is shown in Fig. \ref{future_SN_BAO}.
Today, 580 SNe Ia lead to 1\% precision measurements at the lowest redshifts, with uncertainties climbing to roughly 5\% over the redshift interval $1<z<1.5$.  DES will lower uncertainties in the 2015-2020 timeframe, while LSST and WFIRST will have a larger impact in the longer-term.  Measurements of the BAO feature in the Lyman-$\alpha$  forest with BOSS confirm deceleration at $z=2.4$. In the next 5 years, eBOSS will provide three new 1-2\% precision measurements over the interval $0.6<z<2$, while the combination of Prime Focus Spectrograph (PFS) and Hobby-Eberly Telescope Dark Energy Experiment (HETDEX) will offer nine measurements at $\sim 2$\% precision at fairly uniform spacing over the interval $0.8<z<3.5$.  More generally, the future experiments DESI, WFIRST, and Euclid
are expected to fill in the entire expansion history of the Universe from deceleration to acceleration.

While supernovae and BAO are established techniques,
other distance probes could provide independent reduction of statistical uncertainty, check 
of systematic bias, and different sensitivity to dark energy parameters.
Galaxy clusters, gravitational lensing time delays, reverberation mapping of AGNs, and gravitational wave sirens
have been identified as having the potential to be developed into competitive
probes in the future, and could drive the field in the post-LSST era.

\section{Galaxy Redshift Surveys: Baryon Acoustic Oscillations and Alcock-Paczynski Effect}
\subsection{Executive Summary}

Sound waves propagating in the first 400,000 years after the Big
Bang imprint a characteristic scale in the clustering of matter in
the Universe.  The baryon acoustic oscillations produce a
reasonably sharp peak in the correlation function of galaxies and
other cosmic tracers at a comoving scale of 150 Mpc
\cite{peebles70,sunyaev70,bond84,bond87,jungman96,hu96,hu96a,hu97}.
The length scale of this feature can be accurately predicted from
the simple physics of the early Universe and the measurements of the
CMB anisotropies.  Using this standard ruler, we can measure the 
angular diameter distance and the Hubble parameter as functions of
redshift \cite{tegmark97,goldberg98,efstathiou99,eisenstein98a,eisenstein02,blake03,hu03b,linder03,seo03}.  The method was extensively described in the
recent Weinberg et al. review \cite{Weinberg2012}.

The BAO method has several important advantages.  First, the
simplicity of the physics and the very large physical scale involved
make the method highly robust.  Current theory suggests that the
measurements at $z<3$ can be made to the cosmic variance limit
without being limited by systematic uncertainties.  Second, the method
affords a high level of statistical precision, particularly at
$z>0.5$.  Third, the method allows a direct probe of $H(z)$, further
increasing the leverage at $z>1$.  Fourth, the method allows a 
direct connection to the angular acoustic scale of the CMB, placing
strong constraints on the spatial curvature of the Universe.

The primary challenge of the BAO method is the need for large
redshift surveys.  Surveys to $z=2$ aimed at extracting most of the
BAO information require of order 50 million galaxies.  At $z>2$,
it is likely that Lyman-$\alpha$ forest methods are more advantageous
(e.g. \cite{mcdonald07}).
The BOSS experiment is presently measuring the BAO from 1/4 of the
sky at $z<0.7$, as well as conducting first measurements from the
Lyman-$\alpha$ forest at $z>2$.  Surveys in the coming decade
will extend our view at $z>0.7$ by over an order
of magnitude in cosmic volume.

\subsection{Context}

The acoustic peaks were predicted over 40 years ago but only first
detected in the CMB in 1999-2000.  The first detections in lower redshift
galaxy data took another 5 years \cite{cole05,eisenstein05}.  The large scale of the acoustic peak
means that enormous cosmic volumes are required to detect the signal;
only recent generations of surveys have reached the requisite
volume.  However, the signal has now been detected by several different
groups in six distinct
spectroscopic data sets \cite{cole05,eisenstein05,tegmark06,percival07,
beutler11,blake11c,Anderson2012}
as well as in two photometric redshift data
sets \cite{padmanabhan07,blake07,hutsi10,crocce11,sawangwit11,seo12}.  
The detection in the SDSS-III BOSS Data Release 9 analysis
\cite{Anderson2012} is itself 5$\sigma$; when combined with lower-redshift
SDSS-II data, this reaches 6.5$\sigma$.  The detection of the
acoustic peaks in the CMB anisotopy data is exquisite.  At this point,
there is no question of the existence of the acoustic peak at low
redshift, only the need to improve the measurement of its scale.

The best current BAO data set is that of the SDSS-III BOSS, which has
published measurements of a 1.7\% distance to $z=0.57$ \cite{Anderson2012}.  Improvements
to about 1\% precision are imminent.  At lower redshift, SDSS-II 
produced a 1.9\% distance measurement to $z=0.35$ \cite{padmanabhan12,xu12a}; this measurement
will soon be improved by BOSS due to higher galaxy sampling density and 
somewhat more sky area.  Furthermore, the 6dF Galaxy Survey produced
a 4.5\% measurement at $z=0.1$ \cite{beutler11}; this will improve only slightly in the
future.  At higher redshift, the WiggleZ survey measured the acoustic
peak in a sampling reaching $z=1$, but the aggregate precision is about
4\% at $z=0.6$ \cite{blake11c}, now superceded by BOSS.  

BOSS is also establishing a new view of the BAO using the clustering
of the intergalactic medium at $2<z<3$ as revealed by the Lyman-$\alpha$
forest \cite{white03,mcdonald07}.  
The forest refers to the fluctuating scattering of light
from distant quasars by the neutral hydrogen absorption in the 121.6~nm
transition from the ground to first excited state.  Each quasar provides
a (noisy) map of the density of the intergalactic medium between us
and the quasar.  Using many quasars, BOSS can infer a 3-dimensional 
map of the IGM and study the large-scale clustering within the map.  This
has yielded a first detection of the BAO at $z>1$ and a 3\% measurement
of the Hubble parameter at $z=2.3$ \cite{busca12,slosar13}.

Looking to the coming decade, the BAO method will continue to provide
a precise and accurate measurement of the cosmic distance scale.  

First, regarding precision, the BAO method requires surveys of very
large cosmic volumes with sufficient sampling to detect fluctuations
at wavenumbers of order $0.2h$~Mpc$^{-1}$.
The sampling requirement is modest, typically of order
$10^{-4}$--$10^{-3}h^3$ galaxies per Mpc$^3$, which is well below
the density of galaxies such as the Milky Way.  This means that one
can choose galaxies that are more observationally convenient, e.g.,
those that are brighter, have stronger spectral features, are easier
to pre-select, etc.

However, the volume requirement is ambitious.  Reaching below 1\% precision
requires surveys of a fair fraction of the sky.  Fortunately, 
multiplexed spectrographs on dedicated telescopes make this possible.
Moreover, because there is
a particular size of the observable Universe at a given redshift, there
is a maximum amount of information that can be learned from the BAO
method at a given redshift.  This limitation is called cosmic variance.

\begin{figure}[t]
\begin{center}
\includegraphics[width=0.6\hsize]{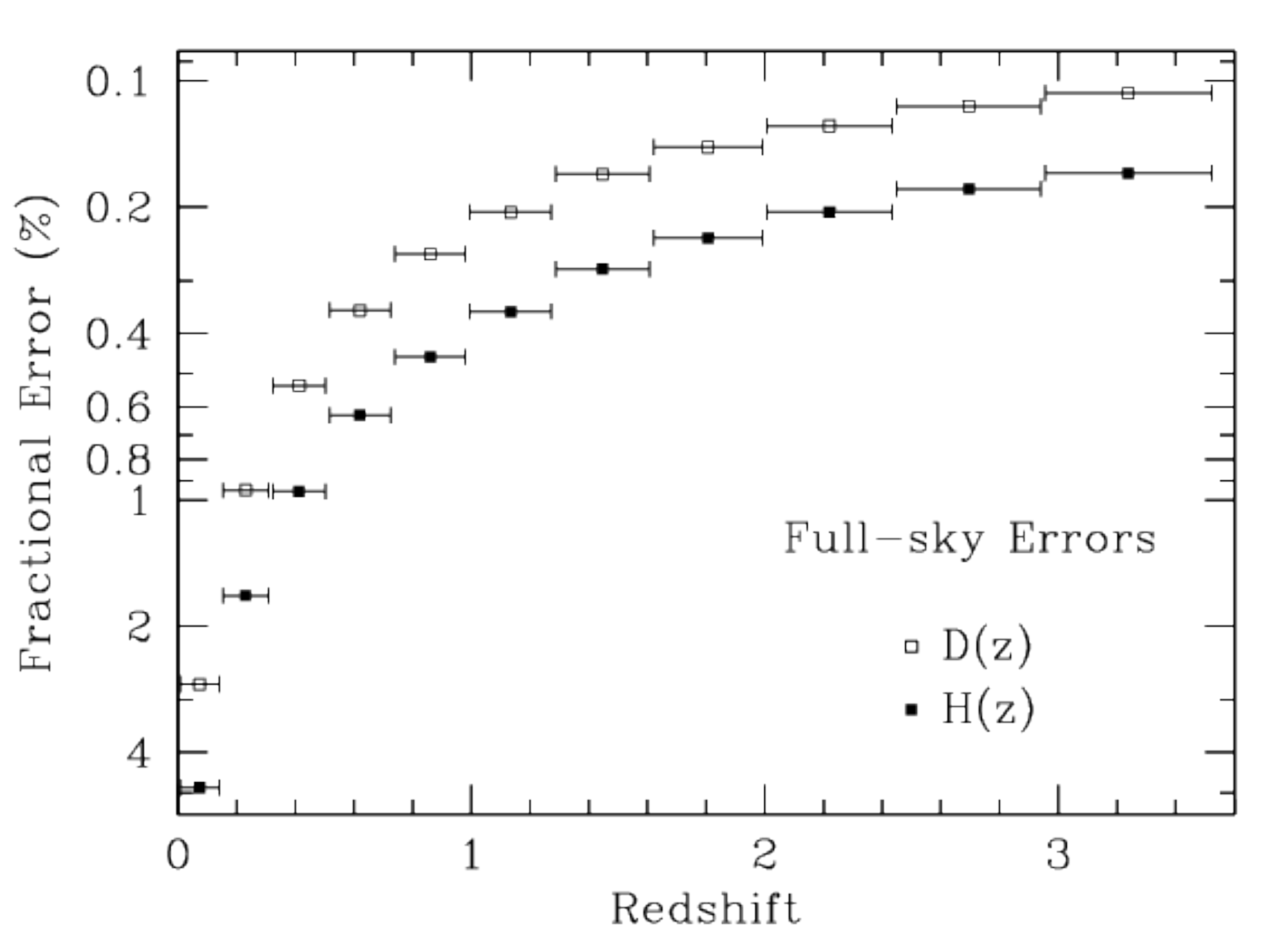}
\caption{
The available precision, reported as the fractional uncertainty on $D_A(z)$
and H(z). This precision assumes a measurement of the BAO for a full-sky survey
with realistic sampling and reconstruction performance, based on Fisher
matrix forecasts from \cite{SeoEisenstein2007}.
The horizontal bars represent the size of each bin for which the fractional
uncertainty is reported.  Uncertainties will increase
as the inverse square root of the fraction of the sky covered by a survey.
One can see that precisions better than 0.2\% in the angular
diameter distance and 0.4\% in the Hubble parameter are available
at $z>1$.  Precisions at $z<1$ degrade because of the smaller cosmic
volumes, but are still better than 1\% at $z>0.3$.  
From \cite{Weinberg2012}; see paper for more explanation.
}
\label{fig:baoforecast}
\end{center}
\end{figure}

Figure~\ref{fig:baoforecast} shows the forecasts for the 
available precision from the measurement of the acoustic peak.
This is assuming realistic sampling and performance from 
reconstruction (to be explained below), with the Fisher matrix 
forecasts from \cite{SeoEisenstein2007} and logarithmic bins in $1+z$.
One can see that precisions better than 0.2\% in the angular
diameter distance and 0.4\% in the Hubble parameter are available
at $z>1$.  Precisions at $z<1$ degrade because of the smaller cosmic
volumes, but are still better than 1\% at $z>0.3$.  Uncertainties will increase
as the inverse square root of the fraction of the sky covered by the survey.

\begin{figure}[t]
\begin{center}
\includegraphics[width=0.6\hsize]{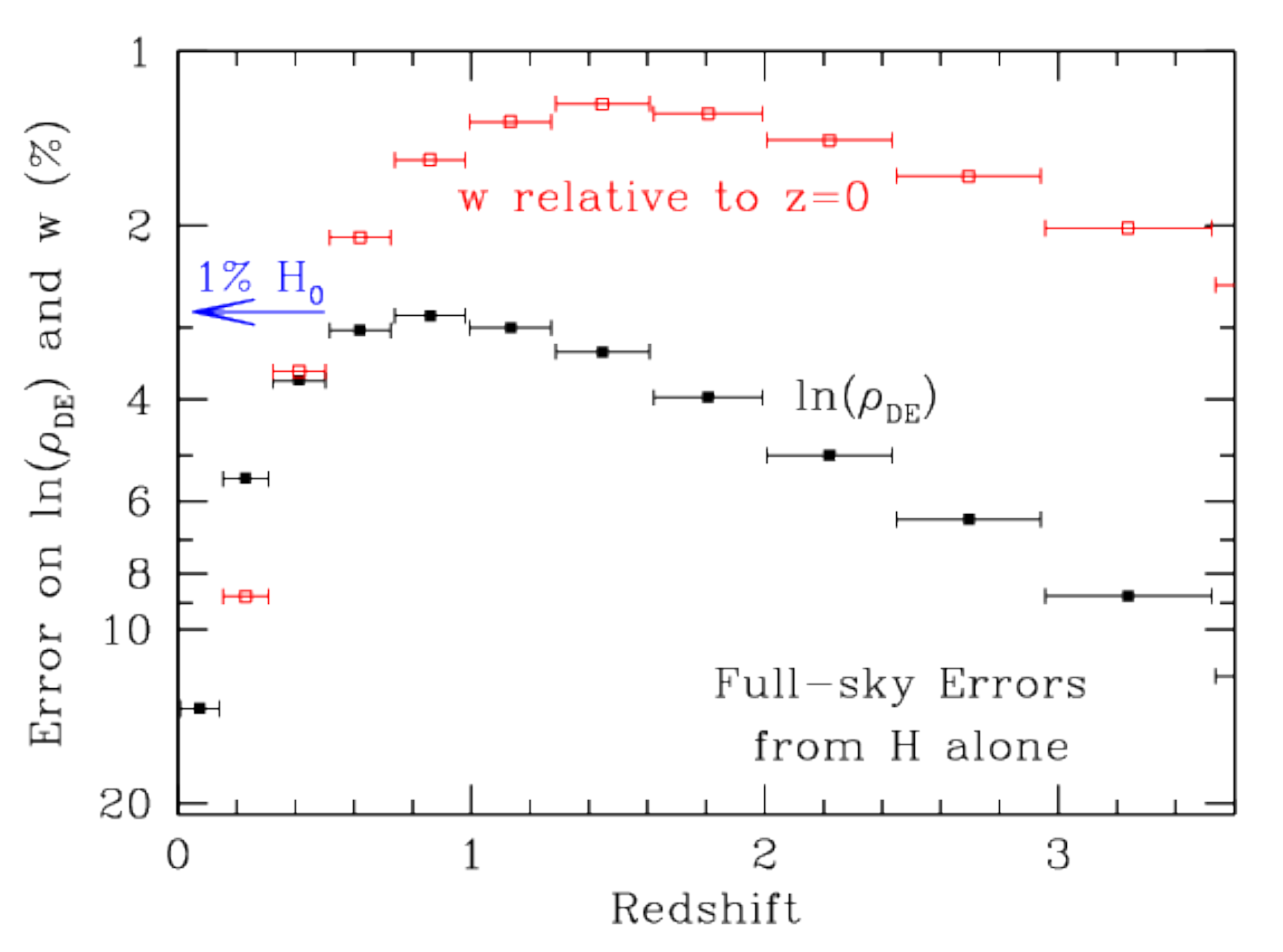}
\caption{
A simple exploration of the impact of measurements of $H(z)$ from the BAO.
In each redshift bin (represented by the horizontal bars) from Figure \ref{fig:baoforecast},
we 
map the measurement of $H(z)$ to a measurement
of the dark energy density, having subtracted off the matter density, which
is assumed to be known from the combination of CMB and lower redshift data.
The black points show the fractional uncertainty on the dark energy density, 
which is better than 5\% even at $z\sim 2$.  
The blue arrow shows the fractional uncertainty on that density that would
result from a hypothetical 1\% measurement of $H_0$.
If we then combine this 
measurement with a known density of dark energy at low redshift, then 
we can infer a power-law evolution, parameterized by the familiar $w$
parameter.  These uncertainties are the red points.  Note that the increasing
lever arm toward higher redshift tends to cancel the decreasing performance,
yielding a broad optimum in the $0.7<z<2.5$ range.
From \cite{Weinberg2012}; see paper for more explanation.
}
\label{fig:density}
\end{center}
\end{figure}

It is commonly assumed that because dark energy is subdominant to
matter at $z>1$, there are diminishing returns in studying it at earlier
epochs.  This is not necessarily true, simply because the precision
of the measurements available to cosmic structure surveys increases 
strongly with redshift.
Figure~\ref{fig:density} explores this in a simplified manner, taking the uncertainties
only from $H(z)$.  Measurements of $H(z)$ are measurements of the
total density of the Universe at that redshift; subtracting off the
matter component reveals the dark energy.  This figure shows that
although the fractional importance of dark energy is dropping at
$z>1$, the available precision on $H(z)$ is still sufficient to
make 5\% measurements of the dark energy density.  Our goal is then
to see whether this density is different from today.  If one compares
each point to a known density today and infers the uncertainty on the
power-law exponent of the evolution (parameterized by the familiar
$w$ choice), then one finds a broad maximum in the performance from
$z=0.7$ to $z=2.5$.  The longer redshift baseline at $z>1$ compensates
for the slowly decreasing precision on the dark energy density.
This calculation is just illustrative; it assumes perfect knowledge 
of the matter density and dark energy density at $z=0$ (but these are
likely measurable to levels to make their uncertainties subdominant in the 
comparison).  It also neglects all of the angular diameter distance
information and doesn't combine more than one redshift bin.

Next, regarding accuracy, the robustness of the BAO method comes primarily from the simplicity
of the early Universe and the relativistic speed of the early sound
waves.  Our theories of the recombination era are exquisitely tested
with the anisotropy measurements of CMB experiments, most notably
the WMAP and Planck satellites and arcminute-scale ground-based
experiments.  From these measurements, we now know the acoustic
scale to better than 1\%, and many possible alterations to the
theory have been sharply ruled out.  Deviations from the adiabatic
cold dark matter theory are now highly constrained!

The relativistic speed of sound before recombination causes the
acoustic scale to be 150 Mpc, which is much larger than the scale
of non-linear gravitational collapse even today.  Large N-body
simulations have found that the shift of the acoustic scale due to
large-scale gravitational flows is only 0.3\% at $z=0$ \cite{seo07,seo10a}; these results
are also found with perturbation theories \cite{padmanabhan09}.  The processes of galaxy
formation occur on scales well below the acoustic scale, but the
small different weighting of the overdense and underdense regions do cause
an additional shift, of order 0.5\% in more extreme cases 
\cite{padmanabhan09,mehta11}.
These shifts are expected to be well predicted by simulations and 
mock catalogs, with uncertainties in the corrections below the statistical
limit of about 0.1\%.

The method of density-field reconstruction aims to sharpen the
acoustic peak by undoing the low-order non-linearity of the large-scale
density field \cite{eisenstein07b,padmanabhan09a}.  Tests in simulations show that reconstruction has
the additional benefit of removing these systematic shifts, providing
another way to suppress systematic uncertainties.

The BAO method is also highly robust as an experimental method.
The core measurements are that of angles and wavelengths, which are
routinely done in astronomy to much higher precision than is required.
The key requirement is the angular homogeneity of the survey, and there
are numerous ways to calibrate this with minimal loss of BAO signal
\cite{vogeley96,tegmark98b,ho08,ross11,ross12}.
Note, however, that the requirements for calibration of the redshift
distortion or Alcock-Paczynski measurements are substantially more 
stringent.

One recently discovered astrophysical effect that can affect the 
acoustic scale peculiarly is the early relative velocity between
baryons and dark matter \cite{tseliakhovich10}.  This velocity modulates
the formation of the smallest early protogalaxies, which may in turn
affect the properties of their larger descendents (although many
practitioners expect the effect at late times to be undetectably
small).  This modulation is due to the same pressure forces that
create the BAO, and the impact could shift the measured acoustic
scale.  However, \cite{Yoo11} shows that this effect also creates
a distinctive three-point clustering signal in the BAO survey data
that allow one to measure the amount of the shift and avoid the systematic
uncertainties.

In summary, we expect that the distance scale information available
in the clustering of matter at $z<3$ can be extracted with sub-dominant
systematic uncertainties even for full-sky ``cosmic-variance-limited''
surveys.  An aggregate precision of 0.1\% allows a remarkable view
of the cosmic distance scale at intermediate redshifts.

\subsection{New projects}

The BOSS project is measuring the BAO at $z<0.7$, with a first look 
at $z>2$.  However, this leaves substantial swaths of cosmic volume
to be surveyed in the coming decade.  New galaxy surveys will focus
on $0.7<z<3$, while a much denser grid of quasars will substantially
improve the constraints from the Lyman-$\alpha$ forest.

The primary DOE-led project is DESI.  This proposes to use a new 
wide-field spectrograph on the Mayall 4-m telescope
at Kitt Peak to perform a massive new survey of 35 million
galaxies over 18,000 square degrees.  This will include a dense
sample of emission-line galaxies at $0.7<z<2$ as well as a sharply
improved sampling of the Lyman-$\alpha$ forest beyond $z=2$.

eBOSS within SDSS-IV will build on the successful SDSS-III BOSS
program.  It will use the existing instrument with new targeting
to provide a first percent-level look at BAO in the $0.7<z<2$ redshift
range, using a sample of quasars and galaxies.  It will also improve the sampling
of the Lyman-$\alpha$ forest.  eBOSS will continue to build the
momentum of BOSS toward DESI science, providing an improved data
set for continued development of the large-scale structure methodology
as well as key training sets for DESI target selection.

The HETDEX project is building a large set of integral-field
spectrographs for the Hobby-Eberly Telescope.  With these, they
will conduct a blank-field search for bright Lyman-$\alpha$ emitting
galaxies at $1.9<z<3.8$.  The initial survey will start in 2014 and
cover 400 square degrees, with a sampling density that is somewhat
sparse but still useful for BAO.  We know of no planned extension of
this technique to $10^4$ square degrees. 
We believe that the Lyman-$\alpha$ forest method is the more efficient
route to the cosmic variance limit at these redshifts.

The PFS project is building a wide-field fiber spectrograph for
Subaru.  It will survey about 1500 square degrees to fainter levels
than DESI.  While this is enough to measure a BAO signal, the major
goals of PFS are in galaxy evolution and smaller-scale clustering.
DESI's much wider survey area is better optimized for BAO and
large-scale clustering.

Euclid is a European Space Agency mission to study dark energy.  It
will perform slitless spectroscopy in the near-infrared over 15,000
square degrees, reaching to $z=2$ but with most signal in the
$0.7<z<1.5$ range.  The satellite is scheduled to launch in 2020 with
a six-year mission lifetime.

WFIRST is a NASA project.  It will provide a wide-field slitless 
spectroscopy capability.  Likely it will aim for denser samples than 
Euclid over somewhat smaller areas.  WFIRST is currently planning
to launch in the early- to mid-2020's.

The major imaging surveys such as DES and LSST will measure BAO
using photometric redshifts.  However, the lack of redshift precision
causes the BAO signal to be blurred out, particularly along the
line-of-sight.  This is a substantial loss to the constraints on
$D_A$ and a total loss of $H(z)$.  The Spanish PAU and JPAS projects
aim to remedy this by using medium-band filters to obtain higher
photometric redshift precision, at a substantial loss in imaging
depth.

An intriguing recent idea is to study the BAO at radio wavelengths
using the 21 cm line.  In most cases, one does not resolve individual
galaxies but instead measures the 3-dimensional large-scale structure
at $>10'$ scale with large interferometric arrays.  This is known
as intensity mapping.  The CHIME project aims to build a 100 meter
square filled interferometer using a cylindrical telescope array
\cite{peterson06} and conduct a lengthy survey at $0.8<z<2.5$.  If
the foregrounds can be adequately controlled, CHIME would be a
powerful demonstrator of the 21 cm method and would yield excellent
cosmological information.  Other projects include the FFT-based
Omniscope \cite{tegmark10} and the Baryon Acoustic Oscillation
Broadband and Broad-beam (BAOBAB) interferometer array \cite{pober1210.2413}.
The challenge in intensity mapping is the removal of other radio emission,
which is about $10^4$ brighter, but expected to be spectrally smooth.

Moving beyond intensity mapping, the SKA could enable an HI-redshift
survey of a billion galaxies, reaching the sample variance limit
over half the sky out to $z=3$ \cite{abdalla05}, which would be a
good approximation to the ultimate BAO experiment.

\subsection{Beyond BAO}

The acoustic peak may be the headliner from these cosmological surveys,
but it is far from the only source of information from redshift
surveys designed to study cosmology and dark energy. 

Modifications of gravity can be tested by the growth of structure as
revealed by the correlation of peculiar velocities and densities
known as redshift-space distortions  \cite{2013arXiv1309.5385H}.  The total mass of neutrinos can
be measured by the tilt of the clustering power spectrum \cite{2013arXiv1309.5383A}.  
Non-Gaussianity from inflation can be measured either in two-point
clustering from the largest scales or in higher-point clustering \cite{2013arXiv1309.5381A}.
And in the largest samples, we may improve upon the CMB measurements
of the matter density and spectral tilt \cite{2013arXiv1309.5381A}.

These items are discussed in contributed white papers.  Here, we discuss an
opportunity that is specific to the cosmic distance scale, namely the 
Alcock-Paczynski (A-P) effect \cite{alcock79}.  This is simply the idea that an intrinsically
spherical object or pattern in the Hubble flow will appear ellipsoidal 
if one assumes the wrong cosmology.  From this, one measures the
product of the angular diameter distance and the Hubble parameter.
As with the BAO, this is of particular interest at higher redshift, where
the $H(z)$ information more directly reveals the density of dark energy.
The BAO method itself is an example of the A-P effect, but with an object
of known size.  

The A-P effect requires the object to be in the Hubble flow, so we
generally focus on large-scale clustering, whether in the two-point
function or non-Gaussian effects such as the shapes of voids
\cite{ryden95,ballinger96,matsubara96,popowski98,hui99,mcdonald99,matsubara01,lavaux10,sutter12}.
In principle, the A-P effect can be measured more precisely than
simply the ellipticity of the BAO ring.  This is because the A-P
effect is a modulation of the broad-band clustering, rather than
the weak oscillation of the BAO, and because the A-P effect can be
measured to higher wavenumbers, where there are lots of modes.

The challenge of the A-P effect is that it is partially degenerate
with the apparent ellipticity of clustering caused by redshift-space
distortions.  The two effects do have distinct dependences on
wavenumber, so with a sufficient range of scales one can separate
them \cite{matsubara04}.  However, the redshift-space distortions are of order unity
ellipticity, whereas the precision of the BAO ellipticity is expected
to be below 1\% and perhaps reach 0.3\%.  Hence, for the broadband
A-P effect to improve on the BAO information, we need to remove the 
redshift-distortion effects to better than one part in 100.  This is
a significant modeling challenge.  There are also observational 
systematics, as the surveys must calibrate the amplitude of their 
radial and transverse clustering measurements.

Nevertheless, the data sets to be taken for BAO measurements will
permit the A-P analyses to proceed.  Furthermore, the modeling
required is the same as for the redshift-distortion measurement of
growth of structure.  If the modeling can succeed, then the A-P
effect can notably improve the $H(z)$ measurements at $z>1$.

\section{Type Ia Supernovae}
\subsection{Executive summary}
Type Ia Supernovae (SNe~Ia) provided the first direct evidence for
the accelerating expansion of the
Universe and today remain a leading probe of cosmology.   Looking toward the new experiments
coming in the next decade,
supernovae will maintain a critical role in the study of dark energy by providing the expansion history
of the Universe to unprecedented accuracy over expanded redshift ranges.
To achieve this goal, we affirm our support of the recommendations given for the upcoming supernova program given
in the  {\it DOE/HEP Dark Energy Science Program: Status and Opportunities} report, outlined in detail in the following subsection.  

Progress requires a coordinated program, including improvements
to low- and high-redshift observing programs, supernova modeling, and analysis procedures;
using knowledge gained in conjunction allows us to suppress the well-understood systematic uncertainties that limit current results.
Low-redshift surveys contribute by providing
hundreds of intensely observed SNe~Ia, the data from which we draw empirical relations and inform theoretical models that predict
luminosities.
The ``cosmology'' workhorses are the Stage IV ground-based experiment LSST and the space-based mission Wide-Field Infrared Survey
Telescope (WFIRST), which will provide unprecedented
numbers of supernovae with exquisite data light curves, over the broad redshift range in which we measure distances.
New technologies must be developed to allow ground-based spectroscopic follow-up to complement the photometric observations
of LSST.  Stage~IV experiments require Stage~IV analysis tools;
advanced approaches to inferring absolute magnitudes from supernova light curves, incorporating training-set data and theoretical models
within the uncertainty band.

Our main findings are as follows:
\begin{itemize}
\item In the next ten years and beyond, Type Ia supernovae will remain a leading probe of dark energy through their measurement of distances
to the far reaches of the Universe.  The improved sensitivity to the dynamical effects of dark energy  is achieved through a coordinated scientific program.
\item Planned (LSST) and prospective (e.g.\ WFIRST) experiments are designed to provide Type Ia supernovae to populate the Hubble
diagram, with
unprecedented numbers, redshift depths, and/or sky coverage.  These experiments provide extended wavelength coverage, increased
signal-to-noise, and improved flux calibration: all important elements needed to reduce limiting systematic uncertainties.
\item A broad program is needed to ensure the reduction of systematic uncertainties to meet the goals of the planned and prospective experiments.
Needed are efforts in low-redshift supernova experiments that study the fundamental properties of SNe~Ia, both physics-based
and empirical high-fidelity supernova modeling, flux calibration programs to establish a primary standard star network, and spectroscopic
follow-up observations that classify transient discoveries.
\item New instrument technologies are being developed that, if successful, will allow order-of-magnitude improvements in data quality
with order-of-magnitude reduction in cost.  It is through technological breakthroughs that giant leaps in scientific discovery become possible;
R\&D today is essential for us to be prepared to attack the new unexpected discoveries of the next decade while maintaining US scientific leadership.
\end{itemize}

\subsection{Context}
Type Ia supernovae are standardizable candles; the luminosity at peak 
brightness of a single supernova can be inferred from the shape and 
wavelength-dependence of its flux evolution. The luminosity distance is 
derived from the ratio between observed flux and inferred luminosity.  
This distance is directly related to the cosmological parameters, including 
the dark energy properties.  Indeed the luminosity distance involves the 
combination of parameters that is the most sensitive current probe 
of the acceleration of the cosmic scale factor, $\ddot a$.  In terms of the 
Hubble parameter $H(z)\equiv \dot{a}/a$, the luminosity distance is 
\begin{eqnarray} 
d_L(z)&=&(1+z)\, \Omega_k^{-1/2} 
\sinh\left[\Omega_k^{1/2}\int_{0}^{z} \frac{dz}{H(z)}\right] \\ 
H(z)&=&H_0\,\left[\Omega_m (1+z)^3+\Omega_{de}\, e^{3\int_0^z d\ln(1+z')[1+w(z')]} + \Omega_k (1+z)^2\right]^{1/2} \,, 
\end{eqnarray} 
where $\Omega_m$ is the matter density in units of the critical density, 
$\Omega_{de}$ the similarly dimensionless dark energy density, 
and $\Omega_k=1-\Omega_m-\Omega_{de}$ 
the curvature density (photon and neutrino contributions
are not included here for concision). A spatially flat universe has $\Omega_k=0$ and 
so the distance is simply given by the integral.  Note 
that $\sinh$ is an analytically complete function, valid for any sign of 
$\Omega_k$.  
The dark energy equation of state function is generally written as 
$w(z)=w_0+w_a z/(1+z)$, shown to accurately approximate exact solutions 
of scalar field dark energy \cite{2003PhRvL..90i1301L} and deliver 0.1\% accuracy 
on cosmological observables such as distances \cite{2008JCAP...10..042D}.  

The expansion factor of the Universe between when the supernova light was 
emitted and today is $a=1/(1+z)$, measured directly from the redshift $z$ 
of the supernova spectral lines or its host galaxy.  Observations of 
luminosity distances and redshifts of a set of supernovae, $d_L(z)$, 
provide the expansion history of the Universe, $a(t)$ -- its relative 
scale as a function of distance or time -- and measure the properties of 
the Universe and the dark energy responsible for its acceleration. 
Relative measurements of  luminosity distance as a function
of redshift, which do not require knowledge of intrinsic supernova luminosities, constrain the energy
densities and equation of state of dark energy; absolute measurements
of luminosity, which do require knowledge of intrinsic  luminosities, constrain the Hubble parameter
$H_0$.

The SNe~Ia method has many strengths as the low-risk, high-reward
probe of dark energy.  It is the most mature probe; indeed
the discovery of dark energy
was made through the accelerated expansion seen in the SN Ia Hubble Diagram
\cite{Riess:1998cb,Perlmutter:1998np}.  SNe~Ia
continue to be a critical contributor to current
measurements of the dark energy equation of state parameters $w_0$ and $w_a$ \cite{2012ApJ...746...85S}
thanks to major cosmological
supernova surveys such as SDSS-II Supernova Survey \cite{2009ApJS..185...32K},
ESSENCE \cite{2007ApJ...666..694W}, Supernova Legacy Survey
(SNLS) \cite{2011ApJ...737..102S},
SCP  \cite{2012ApJ...746...85S}, and the CANDLES and CLASH surveys \cite{2013arXiv1307.0820S}. 

Type Ia supernova cosmology is an active field of research, with current low-redshift surveys such 
as the Nearby Supernova Factory \cite{2002SPIE.4836...61A}, 
Palomar Transient Factory \cite{2009PASP..121.1334R}, and
La Silla/QUEST\cite{2012Msngr.150...34B}, and the high-redshift surveys such as PanSTARRS 1\footnote{\url{http://pan-starrs.ifa.hawaii.edu/public/}}
and the Dark Energy Survey\cite{2012ApJ...753..152B} representing current-generation Stage III experiments.
The power of these surveys is often expressed using a
Figure of Merit (FoM) 
based on a model in which the equation of state of dark energy,
expressed as $P/\rho=w(a)$,
evolves with time as $w(a) = w_0 + w_a(1-a)$.
The projected FoM of these experiments is 100.

Maintaining progress in SN~Ia cosmology first requires the identification of new methodologies
to reduce systematic uncertainties, and then application of that knowledge to high-redshift surveys
with progressively  greater statistical
power.
Current supernova results are limited by fundamental color and flux
calibration uncertainties and not statistical uncertainty;
new experiments must be specifically designed to overcome 
these limiting systematics.
Fortunately, the nature of the systematic uncertainties are understood and can be addressed.
Following this roadmap, future SN~Ia experiments will continue to  provide some
of the strongest individual measures of dark energy, and constrain a unique
complementary phase space
in joint measurements with multiple probes \cite{2012arXiv1208.4012G}.

A strategic SN~Ia program first must have
experiments designed to improve our fundamental understanding
of our distance indicator.  Fruitful efforts have been made in expanding the rest-frame
wavelength range of observational monitoring, at the UV \cite{2008ApJ...674...51E, 2012MNRAS.426.2359M, 2012AJ....143..113F} and
the NIR \cite{2008ApJ...689..377W, 2010AJ....139..120F, 2012MNRAS.425.1007B}.
Optical spectrophotometric time series provide high resolution
information not available in broad-band photometric light curves, such as
equivalent widths, velocities, and ratios of spectroscopic features
\cite{2002SPIE.4836...61A,2012AJ....143..126B,2012MNRAS.425.1789S,2013arXiv1305.6997F}.

From  existing experiments, advances have already been made in
the  determination of an individual SN~Ia's absolute magnitude. Traditionally,
two-parameter supernova models are fit to broad-band optical photometry
\cite{2007A&A...466...11G,2007ApJ...659..122J}.
Now, new light curve analyses have been developed that use more parameters
and NIR data
\cite{2011ApJ...731..120M, 2013ApJ...766...84K} to reduce the residual magnitude
dispersion from 0.15 mag to $<0.10$ mag (less than 5\% distance uncertainty). 
Heterogeneous spectroscopic features have been correlated with color \cite{2011ApJ...742...89F,2011MNRAS.413.3075M}
and absolute magnitude
\cite{2009A&A...500L..17B,2011ApJ...729...55F,2012AJ....143..126B,2011MNRAS.413.3075M,2012MNRAS.425.1889S}.
Some of the unexpected
color corrections ascribed to dust
derived from broad-band color analysis
have now been attributed to a spectral feature whose depth varies
from supernova to supernova \cite{2011A&A...529L...4C}.
Matching observed SN~Ia ``pairs'' with almost identical spectroscopic time series allows one to predict their
absolute magnitudes to 0.08 mag
\cite{Hannah}.
All these reductions of magnitude dispersion are based on new intrinsic parameters
for which population evolution with redshift no longer enters as a source
of systematic uncertainty.

Further studies of fundamental supernova properties are critical for the interpretation 
of data from the next generation
of experiments.  Moreover,  such investigations should continue until a systematic 
floor in the SN~Ia methodology is found, in order
to inform the planning of even more accurate measurements of the
expansion history beyond the next decade.

The other crucial ingredient of a strategic supernova program is improved high-redshift
surveys.  Improvements come in several concrete forms:
access to observables (e.g.\ spectral features, light-curve shapes, colors)
that track SN~Ia diversity, extension of the rest-frame
wavelength range, extension of the redshift range, numbers of supernovae, and the flux calibration
of the optical system.  The ground-based Large Synoptic Survey Telescope \cite{2012arXiv1211.0310L} and space-based
WFIRST \cite{2012arXiv1208.4012G} observatory represent Stage IV missions that offer these improvements.
When considering the improvements brought by new surveys,
it is important to recognize that supernova measurements from new experiments
generally supersede rather than supplement the preexisting
sample;  reductions in the systematic floor from improved experiments
do not extend to old supernovae whose data are already collected.

As we progress through the LSST/WFIRST era, the tightness of constraints on the
dark-energy parameters will set the confidence with which we
will make conclusions about the physics responsible for the accelerating Universe.
We therefore
highlight features of experiments that allow a robust determination of known and as-of-yet unrecognized systematic uncertainties.
The data obtained from low redshift programs that explore new SN~Ia observables
often have quality that exceeds those of high-redshift surveys.
By knowing what information is lost in a degraded
data set, we can quantify the systematic error floor in the high-redshift sample.
Supernovae used for standardization tend to be at lower redshift compared to the cosmology sample,
and so cannot capture all possible evolutionary effects.
It is therefore important that a subset of high-redshift supernovae be
better observed to allow an internal assessment of bias.
For example, a subset of LSST candidates with spectroscopic typing is required to determine
the rate of photometric Type~Ia misclassification.   The generation of a supernova set
that significantly exceeds in number what is needed to reach the expected systematic limit
will allow precise tests of systematic bias in samples divided by supernova and secondary observables.

It is within the context described in this section that 
the {\it DOE/HEP Dark Energy Science Program: Status and Opportunities} report, known familiarly as
Rocky III, identified Key Issues and Opportunities for Supernova Cosmology.  These conclusions remain
operative and so are incoporated by reference
in this document and endorsed by the community.

Key Issues:
\begin{quotation}
\begin{itshape}
For SNe~Ia, the important next steps are aimed at systematic error control. Progress in the DOE Dark Energy program with SNe~Ia will require careful work on several fronts:

1. Calibration instrumentation/studies, and then data collection, for the low-redshift anchor survey(s) and for the upcoming high-redshift surveys (both with major DOE support).

2. Further low-redshift survey work and analysis to identify the key SN~Ia and host features to provide systematics-control of SN~Ia evolution (and dust/color evolution) at the higher redshifts.

3. Study of the observational techniques to carry these calibration and systematics techniques to the higher/highest redshifts. This study will probably require observations in the rest-frame optical and in the near-IR, both photometric and spectroscopic, which are difficult from the ground with present techniques. This situation leads to two possibilities: a space-based near-IR instrument with excellent spectroscopic capabilities, and/or a significant advancement in ground-based IR capabilities.

a) With modest investments in spectroscopic capabilities and a small fraction of mission time, WFIRST-AFTA could be upgraded to provide the detailed supernova measurements at high-redshift to match the low-redshift systematics control, especially with the possibility of using a 2.4m mirror. (WFIRST-AFTA does not currently baseline this precision SN spectroscopy capability.)
{\rm [Editor's note: the most recent WFIRST-AFTA design does include an integral field unit spectrograph for this purpose]
\footnote{The WFIRST-2.4 design released after the
Rocky~III Report does include an integral field unit spectrometer; see \S\ref{wfirst:sec}. \label{WFIRST:ftn}}}
This would be complementary to the EUCLID program.

b) Since such a space mission capability may not be likely to be achieved in the upcoming decade (and may require multi-agency effort for the spectroscopic upgrade), there is a need to explore ground-based alternatives, combining near-IR technology with atmospheric-sky-line suppression and seeing control (e.g., adaptive optics). Sufficient time on 8-meter-class
telescopes would then also be necessary to follow up photometric survey programs such as DES and LSST.

We note that these three elements together make a comprehensive DOE SN program, with a well- sequenced combination of R\&D, construction, operations and analysis projects. The DOE SN researchers will be involved in several of these at any given time, since the precision SN cosmology measurement requires an in-depth understanding and use of SN data from all the redshift ranges simultaneously.
\end{itshape}
\end{quotation}

Opportunities:
\begin{quotation}
\begin{itshape}
Several key ingredients will allow the DOE program to build on the photometric survey projects (DES and LSST) so that Stage IV supernova measurements of dark energy can be accomplished. First, the low-redshift searches (e.g., PTF, QUEST), follow-up (e.g., SN Factory), and data analysis projects will continue to build the foundational measurements -- the crucial knowledge to identify and constrain systematics, and the low-redshift statistical sample large enough to compare with the planned high- redshift data sets.

A future Stage IV space-based SNe project would be the simplest way to match, at high redshift, these precision measurements of Type Ia supernovae at low redshift --measurements needed to provide the same systematics control over the entire redshift range from $z \sim 0.01$ to $ z \sim 2$.	With modest investments in spectroscopic capabilities and a small fraction of mission time, WFIRST-AFTA  could be upgraded
{\rm [Editor's note: and has been upgraded in the current baseline; see Footnote~\ref{WFIRST:ftn}]]}
 to become this project, and would be complementary to the lensing programs of LSST/EUCLID. However, given the timescales and many difficulties of a space mission, there is now a need to explore vigorously a ground-based alternative to fill this important missing element in the DOE program. In particular, an
R\&D effort to explore the potential of novel ground-based techniques, combining near-IR technology with OH sky-line suppression, could make it possible to accomplish the precision measurements for SNe from SCP, DES, and LSST, complementing and strengthening these currently approved DOE projects.
\end{itshape}
\end{quotation}

We support the findings of Rocky III, expand upon them, and identify new opportunities that progress the field through the LSST era
and beyond.

\subsection{Flux Calibration}
Accurate flux calibration is essential for placing SNe measured at different redshifts on the same relative distance scale. This systematic uncertainty is likely the largest one currently affecting supernova cosmology \cite{2011ApJS..192....1C, 2012ApJ...746...85S}.
Accurate calibration requires characterization of the telescope+detector, monitoring of the atmosphere, and fundamental astrophysical flux standards.
As more progress is made in instrument calibration (e.g., \cite{Stubbs10,2013A&A...552A.124B}), atmospheric monitoring (e.g., \cite{Burke10,Stubbs12}) and larger samples of SNe become available, our knowledge of flux standards will need to improve as well \cite{Kent09}.
Creating a well-distributed network of astrophysical standards on the sky is key for LSST to reach its target goal of 1\% internal relative flux calibration and 2\% absolute calibration.

Fortunately, calibration projects are now underway \cite{2007ASPC..364..361K, 2009Metro..46S.219S, Saha12} that should yield flux calibration to better than 0.01 magnitudes over a wavelength range of 0.35 to 1.7 $\mu$m. 
However, it is critical that these efforts succeed.  Thus, it is important that active attention and and any necessary supplementation be provided to these programs to ensure the success of these calibration efforts.

The impact of a 0.01~magnitude calibration uncertainty depends on its functional form. For a simulated WFIRST-AFTA supernova dataset (combined with projected Planck constraints), flux-calibration uncertainties at this level will reduce the statistical-only FoM by $\sim 10\%$ (from the original $\sim 700$) if the calibration of each wavelength element is independently uncertain. If the uncertainty takes the form of a secular drift with wavelength, the same 0.01 magnitudes may reduce the FoM by $\sim20\%$. (The true functional form will likely be in between these extremes.) 
Although flux calibration will remain a significant source of systematic uncertainty, it presents no fundamental obstacle to Stage IV supernova programs.

\subsection{Low-Redshift Surveys}
Supernovae at low redshift serve two important roles for cosmology. First,
they provide the statistical reference -- the anchor -- for the SN~Ia
Hubble diagram. Second, they provide much of the information upon which
standardization of SN~Ia luminosities rests. The brightness and the
ability of amateur scientists to discover nearby SNe~Ia belie a number
of complexities important for cosmology.

For instance, in their role as anchor, accurate flux calibration is as
essential for nearby SNe~Ia as for those at high redshift. Furthermore,
uncertainty in the local standard of rest requires coverage over most of
the sky in order to null-out the zeropoint bias of residual bulk flows.
High-redshift surveys frequently monitor specific fields at several
wavelengths, and therefore include multi-color data from the very earliest
light curve phases.  This capability is not present in current low-redshift
searches, and so those searches serve primarily as triggers for follow-up
with different facilities. This limitation
requires discovery soon after the data
are taken, and thus nearby searches of the kind optimal for cosmology
are in many ways more demanding than searches focused on high redshift.

Much is still being learned about standardizing SNe~Ia, so there is a
significant ``science R\&D'' component to this activity. Types of SN
data that previously were scarce -- spectrophotometric times series, NIR
photometry, UV spectroscopy -- have unlocked a great deal of information
about SNe~Ia that is important for their cosmological application. Note
that this information can be critical even if matching data sets at
high redshift cannot be obtained. For instance, information enabling us
to more deeply understand and constrain possible evolutionary effects
and dust corrections is still necessary. Moreover, the effects of not
having such extensive data at high redshift can then be quantified.
Some new standardization techniques demand much higher numbers of
local supernovae than assumed when treating SNe~Ia as a statistically
homogeneous population.  In the inhomogeneity limit, each high-redshift
SNe would be paired with several local SNe, each its counterpart. This
pairing
would require many hundreds to thousands of well-measured local SNe~Ia in order to
optimally utilize the SNe from DES, LSST and WFIRST-AFTA.

Currently nearby SNe experiments oriented towards the cosmological
application of SNe include the Nearby Supernova Factory (SNfactory),
the Palomar Transient Factory, the La Silla-QUEST search, the Carnegie
Supernova Project, the CfA SN program, and a NOAO Survey with WIYN in the NIR.  
Several of these programs
coordinate with one another in order to find and follow nearby SNe~Ia.
These efforts have resulted in major advances in technical capabilities
and scientific knowledge applicable to SN~Ia cosmology. Yet, obtaining
even better data, in the quantity required to properly support DES,
LSST, and WFIRST-AFTA, is a need that should be examined closely. Significant
new instrumentation and facilities (including telescope access) will be required
in order to scale the nearby SN program adequately.  The Zwicky Transient
Factory is now on the drawing broad and is expected to add a major new
transient search capability that could serve as a starting point.

\subsection{LSST}

The Large Synoptic Survey Telescope is an 8-meter class wide-field dedicated survey telescope that will be in operation during the 2020--2030 decade.  With a 10-square-degree field of view, LSST will be able to quickly survey the sky in $ugrizy$ and build both a deep photometric picture of the sky and a time series of the explosive and variable events in the Universe.  With this unique capability to explore the time-domain in the sky, LSST will find several million supernovae during its 10-year mission.  Of these supernovae, on the order of 100,000 will be Type Ia supernovae with well-observed light curves.  This sample will represent a hundred-fold increase in the number of Type Ia supernovae observed to date, and offer new opportunities to probe the expansion of the Universe from $0<z<1$.

The LSST SN~Ia sample will allow for detailed investigations of the behavior of the expansion of the Universe over the past 7 billion years with the ability to probe the homogeneity and isotropy of dark energy.  Each individual SN~Ia makes a distance measurement, and thus one can explore subsets of these SNe~Ia.  
This ability to divide the sample arbitrarily with the very large numbers of LSST SNe~Ia makes possible
the quantification and suppression of systematic limitations due to  subpopulations and their evolution with redshift.
For example, instead of fitting for a luminosity-color-width relationship and applying that to standardize all SNe~Ia to one fiducial SN~Ia, exact analogs can be found at all redshifts.  One can, e.g., compare stretch 1.1 SNe~Ia from $0.1<z<1$, and then separately compare stretch 1.15 SNe~Ia from $0.1<z<1$ and test whether they agree.  
The population evolution of SN~Ia will be well-controlled with this sample, as dividing 100,000 SNe~Ia into $\Delta z=0.01$ bins enables a check of the relative luminosity estimates from different variations of the 1,000 SNe~Ia in that bin (each bin will have the same number of SNe~Ia as the total number in the Hubble-flow available today).  This redshift slicing will also provide a clear picture of the population demographics of supernova properties with redshift.

While most of the SNe~Ia found and studied by LSST will be in the general ``main''-field survey, there will be an additional very-well-studied sample in the ``deep-drilling'' fields.  The ``deep-drilling'' fields are a set of 10--20 fields that LSST will observe every night when they are visible.  This nightly cadence will generate intermediate redshift SN~Ia light curves with the sampling typically only presently achieved with some of the nearby SN~Ia programs.  This sample of 10,000s of SNe will likely be the premier sample with the most photometric information available, which should allow for both robust and reliable photometric classification (including precision sub-typing as described above) and redshifts.  In addition, the host galaxies of these SNe will be studied by intensive spectroscopic follow-up campaigns with wide-field multi-object spectrographs as part of overall spectroscopic studies of these fields.

Accurate and precise measurements of both axes of the Hubble diagram (redshift and distance) are crucial for a successful SN survey:
Accurate distances require a comparison of the observed flux for wavelengths from 400~nm to 1000~nm.  Therefore, {\em accurate flux calibration will be critical} both within the LSST SN~Ia sample and with respect to any external sample used, whether nearby SNe~Ia or very distant SNe~Ia from Euclid or WFIRST-AFTA.
Redshifts are traditionally measured spectroscopically from either narrow galaxy lines or broad SN absorption features, but they can also be estimated photometrically from multi-band light curves.  The uncertainty on any individual photometric redshift is several times larger than a spectroscopic redshift, but unbiased measurements of large SN samples from LSST will still collectively provide competitive cosmology constraints.

Further development of photometric classification and redshift determination techniques will be necessary to realize the SN~Ia cosmology potential of LSST.  The current state-of-the-art in photometric redshits for SN~Ia cosmology results in residual biases in the redshift distribution and classification due to the complex degeneracies between reddening, redshift, and intrinsic SN colors \cite{2010ApJ...717...40K,2011ApJ...738..162S}.  These redshift biases at the moment are prohibitively large.  It is critical to control these biases with both improvements in the analysis techniques and by obtaining spectroscopic redshifts of
a subset of host galaxies for all supernova types.  Without improvements in these techniques and our understanding of supernova properties and dust in other galaxies, realizing the full potential of SN~Ia cosmology with LSST may require spectroscopic redshifts of the SN host galaxies (but without requiring real-time SN confirmation) to construct a Hubble diagram with well-controlled and understood uncertainties. 

The investigation of intrinsic SN color and reddening due to dust will require significant UV, optical, and NIR imaging and spectroscopy combined with modeling and development of analysis techniques in the years up to 2020.

Variations in intrinsic supernova properties likely come from their binary evolution and metallicity.  The metallicity, and potentially binary fraction\slash separation IMF are likely to be functions of the stellar formation history where the supernova is formed.  A central concern is that these distributions and properties may evolve with redshift.  However, the gross nature of this concern should similarly be well-matched by the gross tracing of the properties of the host galaxy.  



In this present decade, as well as contemporaneously with LSST in the next, the nearby  ($z\sim 0.05$) SN~Ia sample will grow to several thousands of SNe~Ia, which will suffice to compare with the different slices through the LSST SN~Ia sample.
The parametrization of the light curves and the ability to do the division into matched subsets as mentioned above will grow more nuanced as the full extent of variation is traced out by this large sample.

{\em Technical Details of Spectroscopy for SN~Ia Cosmology}

We anticipate $R\sim4000$ optical spectroscopy will be required for obtaining host-galaxy redshifts, allowing the [OII] 3726,3729~\AA\ doublet to be used at redshifts where other lines are shifted into the NIR.  
Planned instruments such as DESI on a 4-m telescope would be well suited for studies of SN host galaxies.

Without further improvements in analysis techniques and data on the properties of supernovae from nearby studies, the SN~Ia cosmology
program could potentially require obtaining spectroscopic redshifts of $>$100,000 galaxies (including hosts of non-Ia supernovae).  This would require hundreds of nights with massively-multiplexed spectrographs, which provide the advantage of simultaneous spectroscopy of many host galaxies.  Depth could be built up efficiently by co-adding spectra of fainter galaxies from multiple visits to the same field, while brighter galaxies will only require a one-time fiber placement.  
Observing each galaxy for longer than is necessary to obtain spectroscopic redshifts may be desired to measure spectroscopic properties of SN host-galaxies that could lead to a reduction in the current level of systematic uncertainty in SN cosmology \cite{Childress13}.  This spectroscopic program would not need to begin until several years after the start of the survey, allowing source density on the sky to build up. 

Spectroscopic confirmation will be required for a subset of 2,000--10,000 LSST SNe.
This unbiased sample is necessary to quantify the systematic bias incurred when spectroscopic information is unavailable.
Low-resolution spectra ($R \gtrsim 100$) are sufficient for identifying the large absorption features of SNe, but at any given time the number of SNe bright enough
to be typed  using standard spectroscopic resources per solid angle on the sky is low and single-object followup will generally be required.
To classify a representative sample of live transients, the magnitude limit of such followup must extend down to $i\sim24$~mag, requiring 8--10m class telescopes.  Because the majority of SNe from LSST will be classified based on photometric algorithms, the accuracy of these methods must be tested against a sample with known SN types.  

Usage of high-throughput IFUs may be desired to simultaneously observe SN and host, and to make possible a cleaner subtraction of SN light from its host.  Unlike the spectroscopic host-galaxy program, live-SN followup would be desired from the start of the survey, though there is no requirement for the spectroscopic sample to be distributed evenly over the duration of the survey.

\subsection{WFIRST-AFTA}
\label{wfirst:sec}
Type Ia supernovae were used to discover the accelerated expansion of our Universe.  In order to study the source of this acceleration future measurements will have to extend to higher redshifts (well above $z=1.5$) in order to observe the presence (or absence) of the transition from acceleration to deceleration at high redshifts predicted by General Relativity. To extend this method to higher redshifts currently
requires observations from space. The reason for this is twofold. At such high redshifts supernova are extremely faint and the superb resolution (0.1 arcsec) and the lack of atmospheric distortions in space, especially for spectroscopy, are crucial. Second, at these redshifts the supernova light is redshifted into the infrared, and the infrared background in space is four orders of magnitude smaller than from ground based telescopes. WFIRST-AFTA, the highest priority large space mission recommended by the last Decadal Survey, is motivated in part by a supernova survey to study the nature of Dark Energy.

The current Design Reference Mission, WFIRST-2.4, uses a 2.4m diameter mirror with  a 0.28 square degree imager consisting of 18 H4RG infrared detectors, which have a wavelength  sensitivity from 0.6 to 2 $\mu$m, and with 4Kx4K 10 $\mu$m pixels. The pixel scale is 0.11 arcseconds, and there is a filter wheel that accommodates 4 filters. 
The baseline design also includes an integral field spectrometer unit (IFU) for the supernova survey to allow the light curves to be obtained via spectrophotometry. The resolution of the IFU is designed to be $R=75$, a solution that balances the degradation of measured supernova spectral features with a reduction of
non-negligible detector noise.
The technical readiness of all of the essential components of the design is very high. The telescope assembly exists and has been qualified as flight ready. The only components that  still require R\&D are the H4RG detectors, which are expected not to be a real problem.

The plan is to run the supernova survey for two calendar years with a 30-hour supernova observation every fifth day for a total observing time of 183 days. The survey is expected to produce 2700 well-measured Type Ia supernovae in a redshift range of 0.2 to 1.7. The imager will be used to discover supernovae with repeated scans of the same sky area with a 5-day cadence. 
There 
are three tiers to the survey: a shallow survey over 
27.44 deg$^2$ for SNe at $z < 0.4$, a medium survey over 
8.96 deg$^2$ for SNe at $z < 0.8$, and a deep survey over 
5.04 deg$^2$ for SNe out to $z = 1.7$.
The light curves will be obtained using spectrophotometry with an IFU intended for that purpose. The IFU is needed to be able to exploit the superior statistics possible with the 2.4m mirror by reducing the systematic uncertainties on the light curves to match the reduced statistical uncertainty
 allowed by the large mirror. Generating the light curves from spectroscopy eliminates the need for interpolating between observer frame filter bands to obtain supernova magnitudes in rest-frame filter bands (K-corrections), thus eliminating one of the most significant systematic uncertainties.
 Furthermore the high efficiency of the IFU spectrometer allows a very deep spectrum (signal to noise better then 15 per resolution element at a resolution of $R=75$) of the supernova to be taken near maximum light, which will make spectral features, and ratios of spectral features, available to further reduce the spread of intrinsic luminosities of the supernovae.

The calculations indicate that this sample of supernovae will yield a precision of 0.5\% around $z=0.5$ to 1\% at $z= 1.7$ in the supernova distance measurements in each 0.1 wide redshift bin  \cite{2012arXiv1208.4012G} . In terms of the DETF Figure of Merit an FoM=580 is expected from the supernova survey alone, assuming a sample of 800 ground-based nearby supernovae and the usual Stage III priors.

There are several other major projects intended to study the acceleration of the Universe and thus the nature of dark energy. LSST, a ground based telescope, will have difficulty in reaching high redshifts for a supernova survey. The European space mission, EUCLID, will not have an IFU spectrometer, will have a considerably smaller diameter telescope mirror, and is not planning to carry out a significant supernova survey. Thus WFIRST-2.4 will  enable a unique space-based supernova survey.

\subsection{Hubble Constant}
\label{hubble:sec}
A measurement of the Hubble constant to 1\% precision would provide an outstanding addition to the set of cosmological constraints used to measure dark energy.  Progress in the last few years using a simplified distance ladder, geometry to Cepheids to SNe Ia, indicates this goal may not be far off.

The extended reach of new HST instruments, ACS and WFC3, has been used to build  a new Cepheid bridge between NGC 4258, the maser host with a geometric distance good to 3\% by \cite{2013arXiv1307.6031H}, and the hosts of recent SNe Ia observed with CCDs to 40 Mpc (SHOES Team; \cite{2011ApJ...730..119R}).  This new bridge reduced systematic uncertainties of the prior by acquiring Cepheids of similar metallicity and period at both ends and by observing both samples with the same camera to eliminate the use of uncertain flux zeropoints.   In addition, the Cepheids were all observed in the NIR (with NICMOS to reach 5\% uncertainty by 2009, with WFC3-IR to approach 3\% in 2011) to mitigate variations in host dust and remaining chemical sensitivity.  The factor of 8 increase in volume reached by the new bridge provided a sample of 8 recent, nearby SNe Ia with the same high quality CCD photometry used to measure the expanding Universe to a few Gpc, about 25 times farther than other secondary distance indicators.  The SHOES Team is now doubling the size of this rate-limiting sample. 
Trigonometric parallaxes to Cepheids in the Milky Way can, in principle, anchor a distance ladder to reach 1\% precision.  To maintain the high level of precision across the distance ladder,  Cepheids with high quality parallax measurements in the  Milky Way must have similar long periods and be observed with the same, near or far infrared instrument as those at the far end of the Cepheid bridge.  A new capability, spatial scanning with WFC3, has begun providing the needed flux measurements of the bright Cepheids with little or no saturation.  A program by the SHOES Team has begun in Cycle 20 to use the enhanced sampling of spatial scanning with WFC3 to measure astrometry to 40 microarcseconds and the parallaxes of the less common but crucial, longer period Cepheids which live between 1 to 3 kpc.  By the end of this decade the ESA GAIA mission will also provide the needed Cepheid parallaxes out to to 10 kpc.

On another front, the Carnegie Hubble Program \cite{2011AJ....142..192F} is obtaining a Cepheid calibration in the mid-infrared using HST parallaxes and SPITZER 3.6 
$\mu$m photometry, where the effects of total line-of-sight reddening are greatly reduced and where atmospheric metallicity effects are predicted to be minimal. In a totally independent but parallel effort, the Carnegie RR Lyrae Program (CRRP) is building an independent path to the Hubble constant that is completely decoupled in its methods, calibrations and systematics from the Cepheid path. Using HST parallaxes for Galactic RR Lyrae variables and SPITZER 3.6 $\mu$m photometry, the RR Lyrae Period-Luminosity relation is being calibrated and applied to Local Group galaxies where large populations of these variables have been previously found from the ground and by HST in space. The goal is to firmly calibrate the tip of the red giant branch luminosity by using the RR Lyraes in nearby galaxies and then calibrate Type Ia supernovae in more distant galaxies where the brightest red giants can be detected and measured by HST.

Other techniques such as strong lensing and distant masers are improving as well and are poised to yield powerful, competitive constraints in the next few years.

\subsection{Technology R\&D}

The goal of bringing precision measurements of Type Ia supernovae to the DETF Stage IV level requires three new elements in the SNe program:  1) photometry capable of comparing supernovae at the same rest-frame wavelength as we move from low to high redshifts,  2) spectrophotometry capable of measuring SN evolution at high redshift with high signal-to-noise, and 3) controlling dust systematics for high-redshift supernovae with well-calibrated rest-frame near-IR photometry.   All three of these require a renewed effort for precision high-redshift supernovae measurements in the near-IR range.   Although earlier DOE efforts at the high redshifts focused on a space-based program, achieving such a capability may not be likely in the upcoming decade.   Now, there are new efforts aimed at developing novel ground-based techniques to combine near-IR technology, suppression of the interfering bright sky lines, and seeing control (e.g., adaptive optics) to follow up photometry survey programs such as DES and LSST.
  
Ground-based wide-field adaptive optics with OH-line suppression in the infrared would create fundamentally new opportunities to pursue high-redshift cosmology from ground-based telescopes.  Observations through the Earth's atmosphere in the infrared are strongly affected by emission and absorption from the sky.  These effects currently place fundamental limits on how far in redshift we can observe Type Ia supernovae in the rest-frame near-infrared (1--3 $\mu$m).  Technologies that can suppress the effect of these lines through (1) spatial and (2) spectroscopic discrimination could open up a new redshift regime to ground-based telescopes.

Adaptive optics systems allow telescopes and instruments to compensate for the blurring caused by small time-scale changes in the Earth's atmosphere.  By reducing the width of the point-spread function of the joint $\mbox{atmosphere}+\mbox{telescope}+\mbox{detector}$ optical system, adaptive optic systems can yield both (a) substantially improved angular resolution along with significant improvements in the signal-to-noise ratio.  By using multiple standard point sources (either natural stars or laser spots), multi-conjugate adaptive optics systems can correct the point-spread-function for multiple layers in the atmosphere and thus provide an improved point-spread function over larger fields of view.  The Gemini Multi-conjugate adaptive optics system (GeMS) is just finishing science verification and is achieving near-diffraction-limited imaging with Strehl ratios of 30\% in $K$ over a 85-arcsecond diameter field of view.\footnote{\url{http://www.gemini.edu/?q=node/11728}}
This performance results in a 1-magnitude improvement in sensitivity over the Hubble Space Telescope in the $K$ band.
But a major current challenge in wide-field adaptive optics is obtaining reliable photometric calibration to the 1\% requirements needed for measure distances with Type Ia supernovae.  Further efforts in photometric calibration are necessary to enable Type Ia supernova cosmology to take advantage of these capabilities.

\begin{figure}[!t]
\centering
\includegraphics[width=0.9\textwidth]{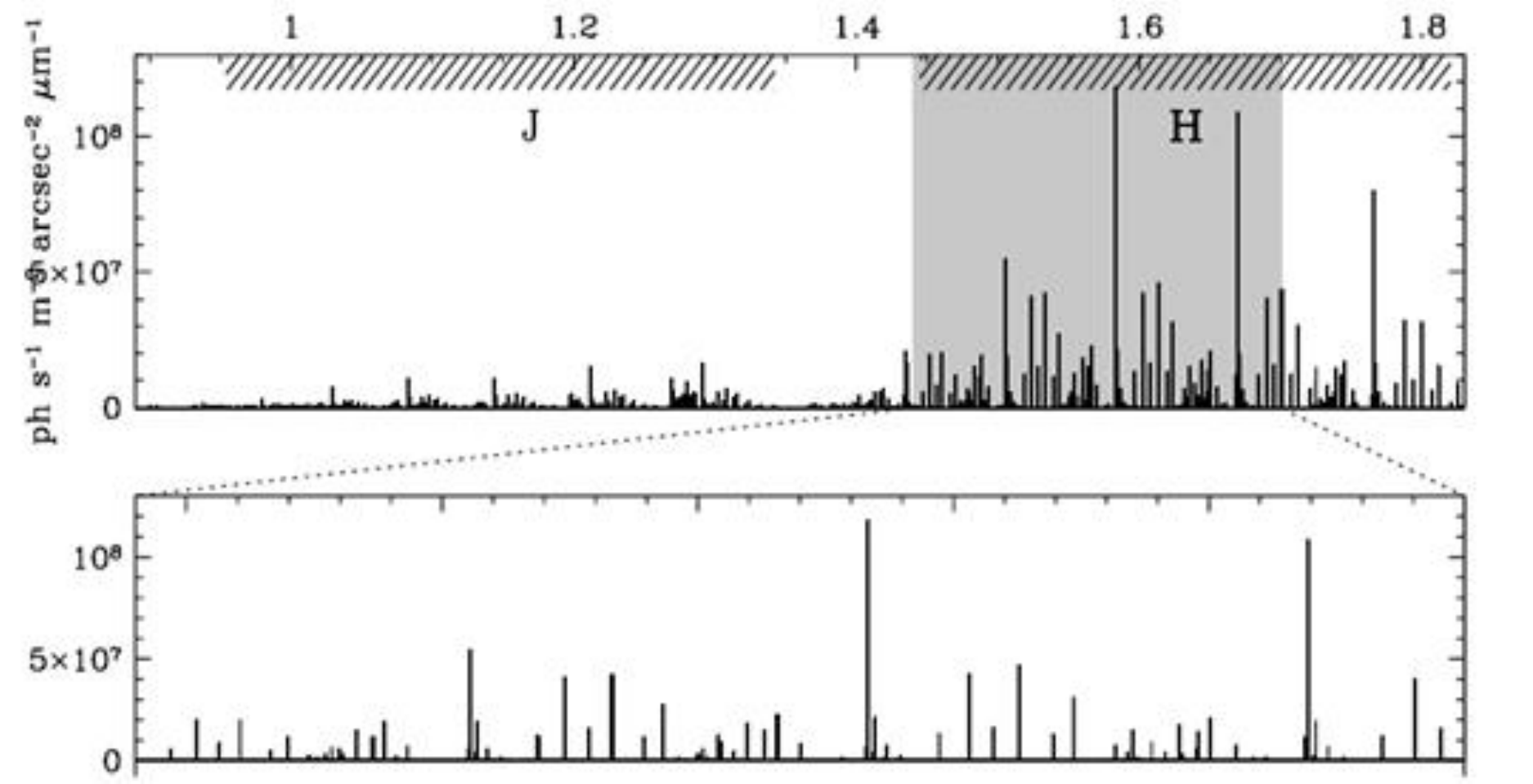} 
\caption{Top: measured sky spectrum of the NIR wavelengths, as well as the range of the broadband J and H filters that are commonly used in infrared astronomy.  The spikes are the emission lines from OH.  Bottom: expanded view of the H-band region from 1.44 -- 1.7 $\mu$m.\label{oh:fig}}
\end{figure}

Ground-based astronomy in the near-infrared region, wavelength range 0.8--2.3 $\mu$m, is plagued by the emission lines of the hydroxyl molecule OH in Earth's upper atmosphere \cite{2008MNRAS.386...47E}.   Figure~\ref{oh:fig} (top) shows a measured spectrum of the NIR wavelengths, as well as the range of the broadband J and H filters that are commonly used in infrared astronomy.  The spikes are the emission lines from OH.  Figure~\ref{oh:fig} (bottom) shows an expanded view of the H-band region from 1.44 -- 1.7 $\mu$m.   In the H-band range the predicted background sky brightness is improved by 8.2 magnitudes without the OH lines  \cite{2008MNRAS.386...47E}.  There are many dozens of techniques to suppress narrow wavelength lines, many from the telecommunications community.  A prototype system using optical fiber devices called fiber Bragg gratings \cite{2012MNRAS.425.1682E} has been tested on a limited number of lines,  but will be difficult to scale up to suppress $\sim 1000$ OH lines or to mass produce.  R\&D on scalable technologies, such as special material-science techniques like metamaterials or silicon-based photonics, is needed to provide an effective general-purpose OH suppression system.

The spectroscopic follow-up of supernovae presents a technical challenge.  The telescope's slit or fibers must be accurately and precisely directed toward 
the coordinates of freshly discovered active supernovae.  The exposure times needed to resolve spectral features are significantly longer
than those needed for broadband photometric light curves; the simultaneous observation of many supernovae (multiplexing)
is needed to keep spectroscopic follow-up times comparable with the associated photometric survey time.  In addition, spectrophotometry can be
required, precluding fiber-fed spectroscopy that does not allow robust subtraction of the polluting light from the background host galaxy.
The same triggering/multiplexing challenge for photometric observations
was resolved through wide-field imagers covering $>20' \times 20'$.  In a ``rolling survey,''
cadenced observations of a fixed field generate light curves of 10--100 supernovae that happen to lie within the field.
Supernova surveys would be revolutionized by using large-format arrays of pixels that provide wavelength (energy) information as well as position for each of
its detected photons: a passive rolling
survey with no live-time requirements can generate all the data needed for a SN-cosmology analysis.  Such detectors must be sensitive from
optical through near-infrared wavelengths, permit photometric flux extraction, and give a resolution of $R \gtrsim 75$ per two resolution elements.
We advocate R\&D on emerging detector technologies that will lead to wide-field imagers that meet these requirements.

\subsection{Next Generation Analyses}
The main computational challenges for a LSST supernova program can be
broken down into three distinct processes: (1) Near real-time
discovery and classification, (2) the automated triggering and
subsequent reduction of follow-up resources and (3) the creation of a
SN~Ia Hubble diagram incorporating all the photometric uncertainties
and correlations along with the best possible understanding of SNe~Ia
at low and high-redshift.

Current wide-field nearby surveys for transients (e.g.\ The Palomar
Transient Factory, The La Silla Supernova Search, etc.) process ~100
GB of raw data covering 1000--2000 sq.\ deg.\ of survey area 2--3 times a
night. These data generate another 1 TB of processed data encompassing
reference images, subtraction images and catalogs from each. If one
decides to search the entire LSST data set of $\sim 20$ TB a night, 
it will be over an order of magnitude more data. Near real-time
processing requires one core per 2k $\times$ 4k image given a one minute
turn-around. LSST would necessitate $\sim 400$ cores to keep up with the
data in real-time. The real-time pipeline includes processing incoming
data, building references, performing subtractions, and running of the
real-bogus classification codes. Furthermore the codes compare
historical images (answering the question is this a new transient or
one we have seen before), calculate orbits of known asteroids to
eliminate them from the subtraction images as well as compare to known
lists of stars and galaxies.

Triggering on nearby events is a challenging task. The huge potential
for scientific gain (discovering a nearby supernova within minutes of
explosion, a GRB orphan afterglow, optical counterpart to a
gravitational wave detection, etc.) must be balanced with the
available follow-up resources and the quality of a particular
candidate. While current surveys like PTF generate one to two such
candidates a night --- and can be followed by 4--10m class telescopes,
LSST has the potential for finding hundreds of candidates
to feed telescopes of these apertures per
night. Simulating this effort prior to first light will be mandatory
to maximizing LSST science in this area and understanding the
computational requirements for such an undertaking.

The success of the program requires support of supernova-modeling
research that extends beyond the scope of the experimental projects.
The development of new supernova models to determine  distances from
photometric and spectral data is a critical line of research that spans supernova astrophysics,
statistics, computer learning, and high-performance computing.
The current supernova models used to determine distances are starting
to prove inadequate for the improved accuracies targeted by next-generation
experiments.  Evidence for this inadequacy
includes observed correlations of Hubble residuals
with host-galaxy properties, recognition that more than two parameters are
needed to describe the diversity of light-curve shapes and colors, and the heterogeneous
spectral features within supernovae with similar light curves.
New supernova models will be informed with data, empirical relations, and physics-based simulations.
They will be more complex, described
by a higher-dimensionality parameter space and responsible for providing a self-consistent estimate of its own statistical uncertainty.
Finally, methods for determining 
simultaneous supernova classification and redshift from photometric data only, a critical component of
both DES and LSST supernova plans, are only in their infancy and have not yet been shown
to achieve the  accuracies needed for precision Stage~IV cosmology \cite{2010PASP..122.1415K, 2011ApJ...738..162S}.
The extra demands on the model will make its training computationally intensive.

Physics-based simulations of SNe~Ia can help identify and limit
cosmological systematic uncertainties associated with evolution in the
progenitor population and its environment.  There is general agreement
that SNe~Ia result from the thermonuclear explosion of carbon-oxygen
white dwarfs, but the nature and mass(es) of the progenitor system
(single or double white dwarf) and the physics of ignition and
propagation of nuclear burning are still debated.  The model space is
highly constrained by observations, and several groups have developed
the ability to calculate not only the hydrodynamics of the explosions,
but also their spectra and multi-band light curves over all angles
and times, e.g., \cite{2008ApJ...681.1448J,2009Natur.460..869K,2010Natur.463...61P,
2012JPhCS.402a2023C,2013FrPhy...8..116H,2011ApJ...734...38W}.  
Given sufficient computational resources, a grid of models
sampling the full space of proposed progenitor systems, triggering
mechanisms, and physics parameters can be constructed.  Comparing
synthetic light curves and spectra to observed ones would identify the
subset of models providing the best representation of SNe~Ia.  Such a
validated model grid would provide a map between the underlying (and
typically observationally inaccessible) physical parameters (e.g., white
dwarf mass, carbon/oxygen ratio, metallicity) and the key observables
(e.g., peak brightness, light curve width, colors, spectral features).
Simulation ``data'' can also have arbitrarily dense time and wavelength
sampling, greater signal-to-noise, and can sample a greater range of
intrinsic diversity.  Comparison of existing simulations to SN~Ia observations
have already proven useful  for studying sources of systematic uncertainty, e.g., 
\cite{2010MNRAS.406..782S, 2011ApJ...729...55F, 2011MNRAS.417.1280B, 2013arXiv1303.1168D, 2013ApJ...764...48K}

For current methods of SN cosmology, the potential benefits of
preparing such a model grid are numerous.  (1) The model grid may be
used as a control sample for testing how well empirical light curve
fitters are able to infer intrinsic luminosity over a diverse and
evolving SN sample.  (2) Detailed models can help improve the current
purely empirical spectral surface templates used for color and
K-corrections.  In particular, validated models can extrapolate behavior
where high-quality observations are scarce or difficult to obtain (e.g.,
at very early times, or in ultraviolet/infrared bands) while maintaining
self-consistency.  (3) Models can be used to determine which observables
divide SNe into physically meaningful subgroups; the cosmological
parameters can then be calculated for each subgroup independently,
permitting a physically motivated ``like-to-like'' comparison.  (4) The
models can be used to confirm and physically explain secondary
parameters (e.g.  colors, spectral features) that provide luminosity
corrections beyond the light curve shape.  Note that these applications
do not require that we have converged on a settled theory for the origin
of SNe~Ia.  Rather, it is only necessary that the validated simulations
capture the relevant correlations between physical inputs and
observables, and that the simulation grid spans as broad a range of
possibilities as Nature.

Carefully constructed and calibrated SN simulations also have the
potential to open new pathways for SN cosmology in the future.  In
principle, high-fidelity models could at least supplement today's more
empirically parameterized light curve fitters.  Fitting or performing
Bayesian inference on observed light curves and spectra with simulations
would yield physics-motivated ``parameters'' like progenitor metallicity
that could be monitored for drift as a function of redshift.  Detailed
stellar explosion models can also provide synthetic observables that can
be sampled in Monte Carlo simulations of future missions, or in a Markov
chain Monte Carlo for forward modeling of the scene of low signal-to-noise SN
targets.  Part of the challenge is in making better models, and that
includes not only simulating the physics correctly but also properly quantifying
uncertainties in the physics.  The new challenge is to efficiently
utilize computationally expensive simulations to make meaningful
inferences from data: Direct application of simulations for SN~Ia
distance estimation will remain computationally impractical for the
forseeable future, even into the LSST era.  Emulators based on
simulations that have been calibrated with high-quality low-redshift
data sets provide a new way forward.  Application of sparse sampling
techniques for deciding which simulations to run, Gaussian processes for
emulating simulations between inputs, and Bayesian calibration of models
against high-quality data sets must become areas of active exploration
to make this a reality \cite{RSSB:RSSB294,sant:will:notz:2003}.  Such
efforts can leverage more developed work already underway in the study
of large-scale structure simulations \cite{2009ApJ...705..156H}.

\section{BAO and Supernova Complementarity}
It is important to stress the deep complementarity between the BAO
and SNe methods.  In practical terms, SNe have no relevant cosmic
variance limit; if we wait, we can build up arbitrarily large samples.
This allows us to measure relative distances even at $z<0.5$,
where the cosmic volume sharply limits BAO measurements.  However, the BAO
offer comparable and even higher precision as redshifts
climb above $z>1$ and provide an absolute calibration 
that ties to the CMB at $z=1000$.  A combined analysis of CMB, BAO,
and SNe allows us to build a calibrated distance scale from $z=0$
to $z=1000$, in principle with sub-percent precision and accuracy.
This is where the most precise current tests of dark energy come from,
and we believe that this opportunity will remain strong in the coming
decade.

\section{Prospective Distance Probes}
There is a suite of other probes that
are striving to match and/or exceed the probative power and robustness of Type~Ia
supernovae and BAO distance measurements.  Important reasons
to explore new distance probes include the following: sensitivity to different systematic uncertainties
allows consistency checks across probes and increased sensitivity in joint
measurements; different measures of distance have different functional dependence
on dark energy and so have complementary intersections in parameter confidence
regions; new probes may have fundamentally lower statistical and systematic uncertainty limits
than supernova and BAO.  This section summarizes three probes 
that generated a response as part of the Snowmass process.

In the late 1980's, the first high-redshift Type~Ia supernova search
was initiated to measure the deceleration parameter
$q_0$.  This early search was met with heavy skepticism; it
was only through hard work in detailed supernova characterization, refinement of
observing strategies for high-redshift supernova discovery and follow-up, and the advent of wide-field
cameras
that have established supernovae and their tracing of cosmic
acceleration as a pillar of modern cosmology.
There is no guarantee that the probes discussed in this section will overcome their scientific
and observational challenges to match the success of Type~Ia supernovae.
Still, we must recognize that exciting new opportunities may be spontaneously initialized
from within the community

\subsection{Clusters}

Complementing other cosmological tests based on the mass
function and clustering of galaxy clusters (e.g.\ 
\cite{Allen1103.4829,Weinberg1201.2434}), nature offers two
independent ways of using clusters to measure cosmic distances.  The
first uses measurements of the X-ray emitting gas mass fraction in the
largest clusters, which is an approximately standard quantity,
independent of mass and redshift. The second uses combined mm and
X-ray measurements of cluster pressure profiles.

\subsubsection{Distance measurements from the cluster 
X-ray gas mass fraction}

\paragraph{Overview and current status}

The largest clusters of galaxies provide approximately fair
samples of the matter content of the Universe. This enables X-ray
measurements of their baryonic mass fraction (the baryonic mass is
dominated by the X-ray emitting gas) to provide robust, and
essentially model-independent, constraints on the mean matter density
of the Universe \cite{White93}.  Additionally, measurements of the
apparent evolution of the cluster X-ray gas mass fraction, hereafter
$f_{\rm gas}$, can be used to probe the acceleration of the Universe
(\cite{Allen0405340,Allen0706.0033}).  This latter constraint
originates from the distance dependence of the $f_{\rm gas}$
measurements, which derive from the observed X-ray gas temperature and
density profiles, on the assumed distances to the clusters, $f_{\rm
gas} \propto d^{1.5}$.

\fgas{} measurements can be determined from X-ray observations under
the assumptions of spherical symmetry and hydrostatic equilibrium. To
ensure that these assumptions are as accurate as possible, it is
essential to limit the analysis to the most dynamically relaxed
clusters.  A further restriction to the hottest, most X-ray luminous
systems simplifies the cosmological analysis and minimizes the required
exposure times. The state of the art for this work \cite{Mantz13}
employs Chandra measurements for a sample of 40 of the hottest
($kT_{2500}> 5$\,keV), most dynamically relaxed clusters known. The
selection on dynamical state is applied automatically, based on X-ray
images.  The \fgas{} measurements are made within a spherical shell
spanning the radial range 0.8--1.2\,$r_{2500}$, for a given reference
cosmology (angular range 0.8--1.2\,$\theta^{\rm ref}_{2500}$).

Combining the two aspects discussed above (fair sample and distance
dependence) the cosmological model fitted to the $f_{\rm gas}(z)$ data
typically has the form

\begin{equation}
  f^{\rm ref}_{\rm gas}(z;\theta^{\rm ref}_{2500}) = 
K\, A\, \Upsilon_{2500}\,\left(\frac{\Omega_{\rm b}}{\Omega_{\rm m}}\right)
\left(\frac{d^{\rm ref}_{\rm A}}{d_{\rm A}}\right)^{3/2}\,.
\label{eq:fgas3}
\end{equation}

Here $K$ encompasses the main systematic uncertainties,
associated with instrumental calibration and the accuracy of the
hydrostatic mass measurements. Fortunately, $K$ can be constrained
robustly through combination with independent weak lensing mass
measurements \cite{Applegate13b}. The angular correction factor, $A$,
accounts for the fact that $\theta^{\rm ref}_{2500}$ for the reference
cosmology and $\theta^{\rm }_{2500}$ for a given trial cosmology are
not identical. $\Upsilon_{2500}$ is the gas depletion parameter, the
average ratio of the cluster gas mass fraction to the cosmic mean
baryon fraction at the measurement radii, as predicted by
hydrodynamical simulations
(e.g. \cite{Planelles1209.5058,Battaglia1209.4082}).  Priors on the
Hubble parameter, $h$, and the mean baryon density $\Omega_bh^2$ are
also required to constrain cosmology from \fgas{} data alone.

\begin{figure}[t] \centering
  \begin{minipage}[c]{0.33\textwidth}
    \includegraphics[width=\textwidth]{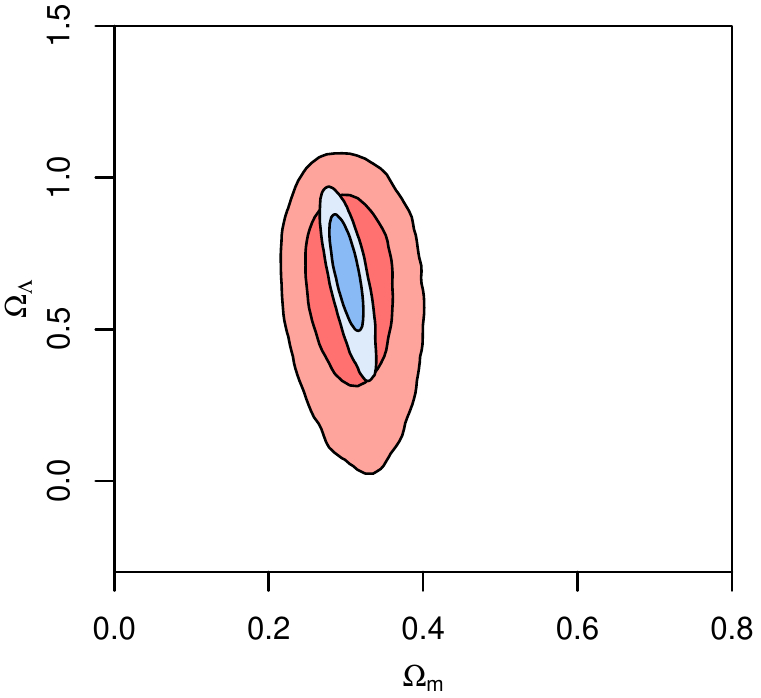}
  \end{minipage}%
  \begin{minipage}[c]{0.33\textwidth}
    \includegraphics[width=\textwidth]{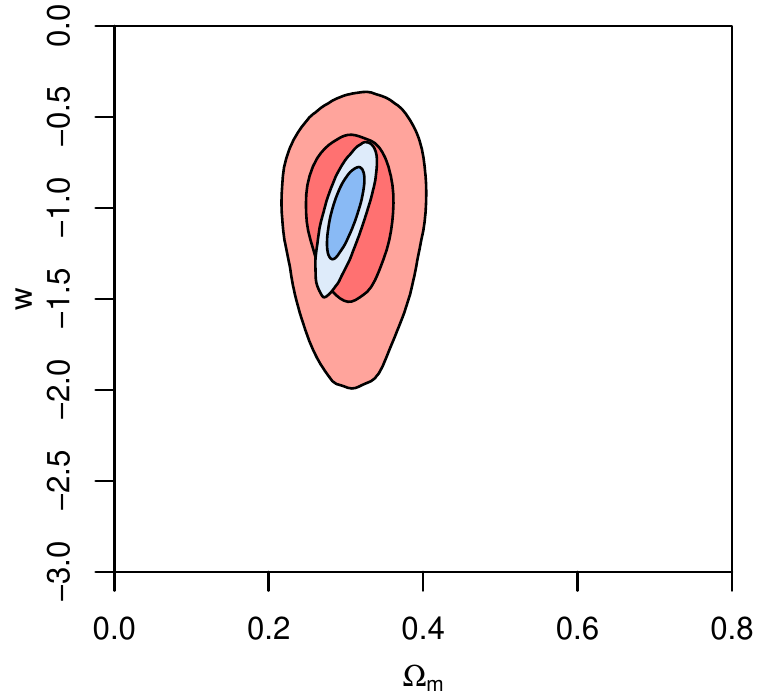}
  \end{minipage}%
  \begin{minipage}[c]{0.33\textwidth}
    \includegraphics[width=\textwidth]{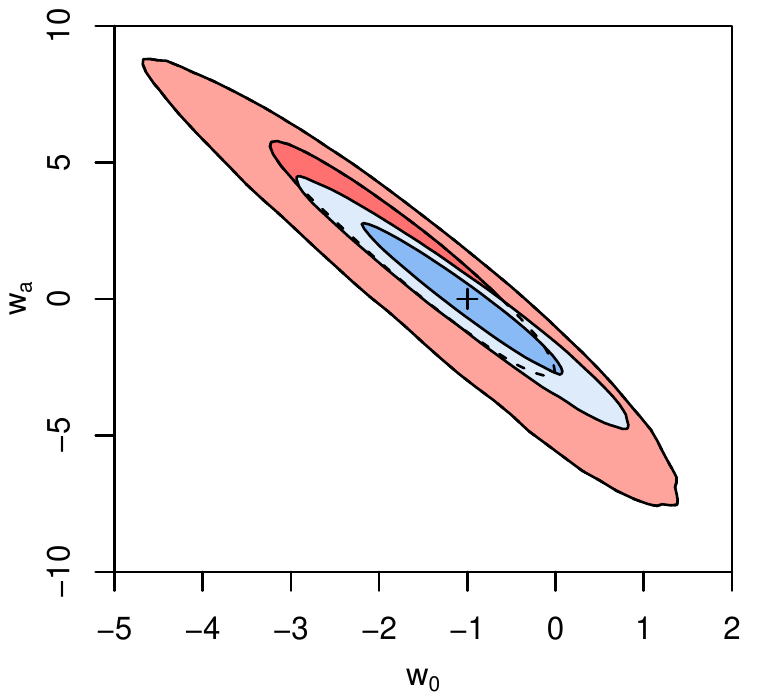}
  \end{minipage}%
  \caption{Joint 68.3\% and 95.4\% confidence constraints from the
\fgas{} method for non-flat \LCDM{} (left), flat $w$CDM (center) and
flat evolving-$w$ (right) models. The red contours show the
constraints from current \fgas{} data (\cite{Mantz13}; also
employing standard priors on $\Omegab h^2$ and $h$). The blue contours
show the improved constraints expected with the addition of an extra
10\,Ms of Chandra observing time, together with modest improvements in
external priors \cite{2013arXiv1307.8152A}.}
\label{fig:simres}
\end{figure}

Fig.~\ref{fig:simres} shows the cosmological constraints from current
\fgas{} data (red curves; \cite{Mantz13}) for non-flat \LCDM{};
flat, constant-$w$; and evolving $w$ models. In all cases the
results are marginalized over conservative systematic uncertainties.
The \fgas{} data provide comparable constraints on dark energy
to current SNIa measurements \cite{2012ApJ...746...85S}, and an
impressively tight constraint on $\Omegam$ independent of the
cosmological model assumed \cite{Mantz13}.

\paragraph{Prospects for improvement}

New optical, X-ray and mm-wave surveys over the next
decade will find in excess of 100,000 clusters, including thousands of
hot, massive systems out to high redshifts. The hottest, most X-ray
luminous and most dynamically relaxed of these will be the targets for
further, deeper observations designed to enable the \fgas{} (and XSZ;
see below) experiments.

The blue curves in Fig.~\ref{fig:simres} show the expected
improvements in the cosmological constraints from the \fgas{} method
over the next decade, using current techniques and assuming that a
further 10\,Ms of Chandra observing time ($\sim 5$\% of the available
total) will be invested in this work. (These predictions also assume
modest improvements in the external priors on $K$, $\Upsilon_{2500}$,
$\Omega_bh^2$ and $h$; see \cite{2013arXiv1307.8152A} for details.) Observations
of this type are also likely to be interesting for a broad range of
other astrophysical and cosmological studies.

The availability of a new X-ray observatory with comparable spatial
resolution to Chandra and a collecting area $\sim 30$ times larger
would likely enable a reduction in the area of the
confidence contours shown in Fig.~\ref{fig:simres}(c) by a further
factor of $\sim 7$ (for details see
\cite{2013arXiv1307.8152A}). Possibilities include the SMART-X
mission (http://hea-www.cfa.harvard.edu/SMARTX/) and Athena+
\cite{Nandra1306.2307}.\footnote{The Athena+ science
theme “The hot and energetic Universe” was recently selected for the
second Large-class mission in ESA’s Cosmic Vision science program.}

\paragraph{Key challenges}

The key challenges to realizing the prospects described above will be
securing the required observing time on flagship X-ray observatories,
and delivering the continued, expected improvements in $K$ and
$\Upsilon_{2500}$ from weak lensing studies and hydrodynamical
simulations, respectively \cite{2013arXiv1307.8152A}.

\subsubsection{Distance measurements from SZ and X-ray pressure profiles}

\paragraph{Overview and current status}

Cosmic microwave background (CMB) photons passing through a galaxy
cluster have a non-negligible chance to inverse Compton scatter off
the hot, X-ray emitting gas.  This scattering boosts the photon energy
and gives rise to a small but significant frequency-dependent shift in
the CMB spectrum observed through the cluster known as the thermal
Sunyaev-Zel'dovich (hereafter SZ) effect \cite{Sunyaev72}.

It has been noted \cite{WhiteSilk78} that X-ray and SZ measurements can be
combined to determine distances to galaxy clusters.  The spectral
shift to the CMB due to the SZ effect can be written in terms of the
Compton $y$-parameter, which is a measure of the integrated electron
pressure along the line of sight ($y \propto \int n_{\rm e}\, T\,
dl$). This shift, $y_{\rm SZ}$, is
independent of the cosmology assumed.  A second, independent
measure of the $y$-parameter can also be obtained from X-ray data
where, for a given reference cosmology, the X-ray measurement, $y_{\rm
X}^{\rm ref}$, depends on the square root of the angular diameter
distance to the cluster. Combining the results, we obtain

\begin{equation}
{y_{\rm X}^{\rm ref}} = {y_{\rm SZ}^{\rm }}\, k(z)\left( \frac{d_{\rm A}^{\rm ref}}{d_{\rm A}^{\rm }}\right)^{1/2}\,.
\label{eq:ycompton}
\end{equation}

Due to the modest distance dependence of this method, and the
limitations of the available data, this test (sometimes referred to as
the XSZ or SZX test) has to date only been used to constrain $h$, with
all other cosmological parameters fixed.  Assuming spatial flatness
and fixing $\Omega_{\rm m} = 0.3$, \cite{Bonamente0512349} applied
this method to 38 X-ray luminous clusters, finding $h =
0.77^{+0.11}_{-0.09}$.

\paragraph{Prospects for improvement}

\begin{figure}[t] \centering
  \begin{minipage}[c]{0.49\textwidth}
    \includegraphics[width=0.9\textwidth]{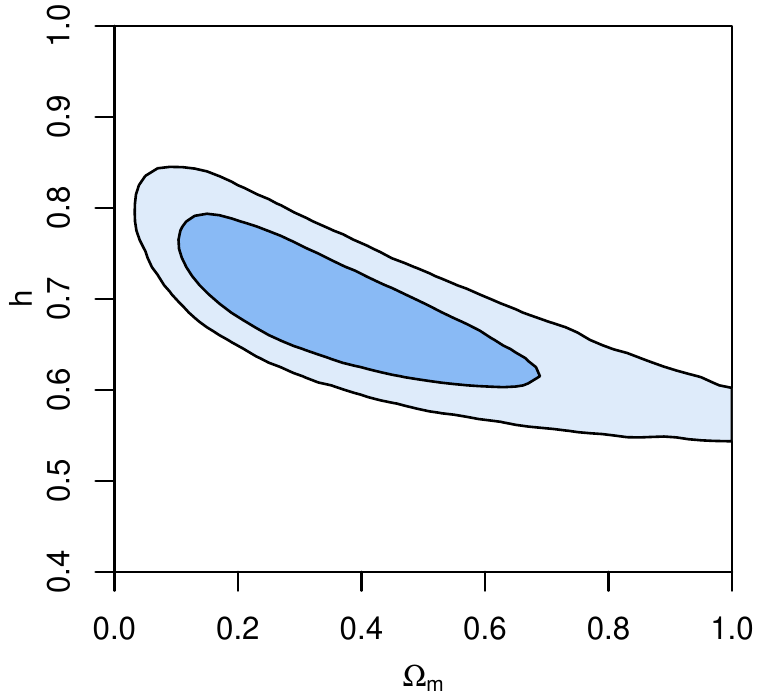}
  \end{minipage}%
  \caption{Joint 68.3\% and 95.4\% confidence constraints on $\Omegam$
and $h$ for a flat \LCDM{} model using simulated XSZ data for 500
clusters, as described in the text.
}
\label{fig:xszsim}
\end{figure}

Over the next decade, utilizing X-ray measurements for $\sim 500$
clusters from the eROSITA satellite (measurements of a type that will
be gathered by default by that satellite for its central science
goals), in combination with follow-up SZ measurements from e.g.
CARMA\footnote{http://www.mmarray.org/} or
CCAT\footnote{http://www.ccatobservatory.org/}, the XSZ technique can
be expected to provide cosmological constraints similar to those shown
in Fig.~\ref{fig:xszsim}. With this expanded data set, and assuming
plausible levels of uncertainty in the X-ray and SZ instrument
calibration, interesting constraints on $h$ should be achievable for
non-flat \LCDM{} and $w$CDM models, solving simultaneously for
$\Omegam$ and $\Omegal$ and/or $w$ \cite{2013arXiv1307.8152A}.

\paragraph{Key challenge}

The key challenge to realizing the prospects for the XSZ experiment
described above will likely lie in providing precise, robust absolute
calibrations for the X-ray and SZ $y$-parameter measurements
\cite{2013arXiv1307.8152A}.

\subsection{Strong Lensing Time Delays}
Strong lensing time delay distances measure the geometric ratio of the 
distance between the observer and source, observer and deflecting lens, and 
lens and source, effectively the focal length of a gravitational lens: 
\begin{equation} 
D_{\Delta t}=(1+z_l)\frac{d_l d_s}{d_{ls}} \ , 
\end{equation} 
where the angular distances $d$ are functions of the matter density, dark 
energy density, and equation of state. 

This dimensionful quantity is observed through the time delays between 
the multiple images created of the source (typically a time varying quasar) 
by the lens (typically a galaxy).  By its nature, the time delay distance 
has two key powerful and unique properties: 1) Being dimensionful, it is 
directly sensitive to the Hubble constant $H_0$, and 2) Being a distance 
ratio, it has different covariances between cosmological parameters than 
luminosity (e.g.\ supernovae) or angular (e.g.\ BAO) distances, and is highly 
complementary for cosmological constraints on dark energy \cite{lin11}. 

Moreover, the strong lensing probe does not require an elaborate 
independent observing program, for the most part building on planned wide 
field and time domain surveys, and so is highly cost effective.  The 
lensed, multiply imaged quasars are detected in wide field surveys, 
then followed up in time domain surveys over a period of several years. 
Redshifts of the images and lenses are measured through spectroscopy, in a 
non time critical way, and 
high resolution imaging of the lens system is obtained through space telescopes 
or future ground adaptive optics systems (radio/submillimeter telescopes such as 
ALMA may play a role as well).
Out of the thousands 
of strong lens systems expected to be measured, we are free to choose the 
cleanest for distance measurements. 

As a new geometric method of exploring dark energy, strong lensing time 
delay distances are a valuable addition to 
standard techniques.  Already the method has matured to the level where 
single lensing systems have delivered 6\% distance measurements (including 
systematics), and programs are underway to increase the quantity and quality 
of measurements.  

In this decade, wide field imaging surveys such as Dark Energy Survey should 
discover $\sim 10^3$ lensed quasars, enabling the detailed study of 
$\sim$100 of 
these systems and resulting in substantial gains in the dark energy figure 
of merit.  For example this would increase the FOM from supernovae 
in combination (hence from purely geometric probes) by a factor of almost 5, 
due to 
degeneracy breaking.  In the next decade, a further order of magnitude 
improvement will be possible with the $10^4$ systems expected to be detected 
and measured with LSST and Euclid. To fully exploit these gains, three 
priorities are to: 
\begin{itemize} 
\item Support the development of techniques required for the accurate 
analysis of the data. 
\item Transform small robotic telescopes (1-4m in diameter) into a high 
cadence, long term dark energy monitoring experiment (non-exclusive use) 
to keep up with the discoveries. 
\item Support high resolution imaging capabilities, such as those 
enabled by the James Webb Space Telescope and next generation adaptive 
optics systems on large ground based telescopes. 
\end{itemize} 

\begin{figure}[!h]
\includegraphics[width=0.4\textwidth]{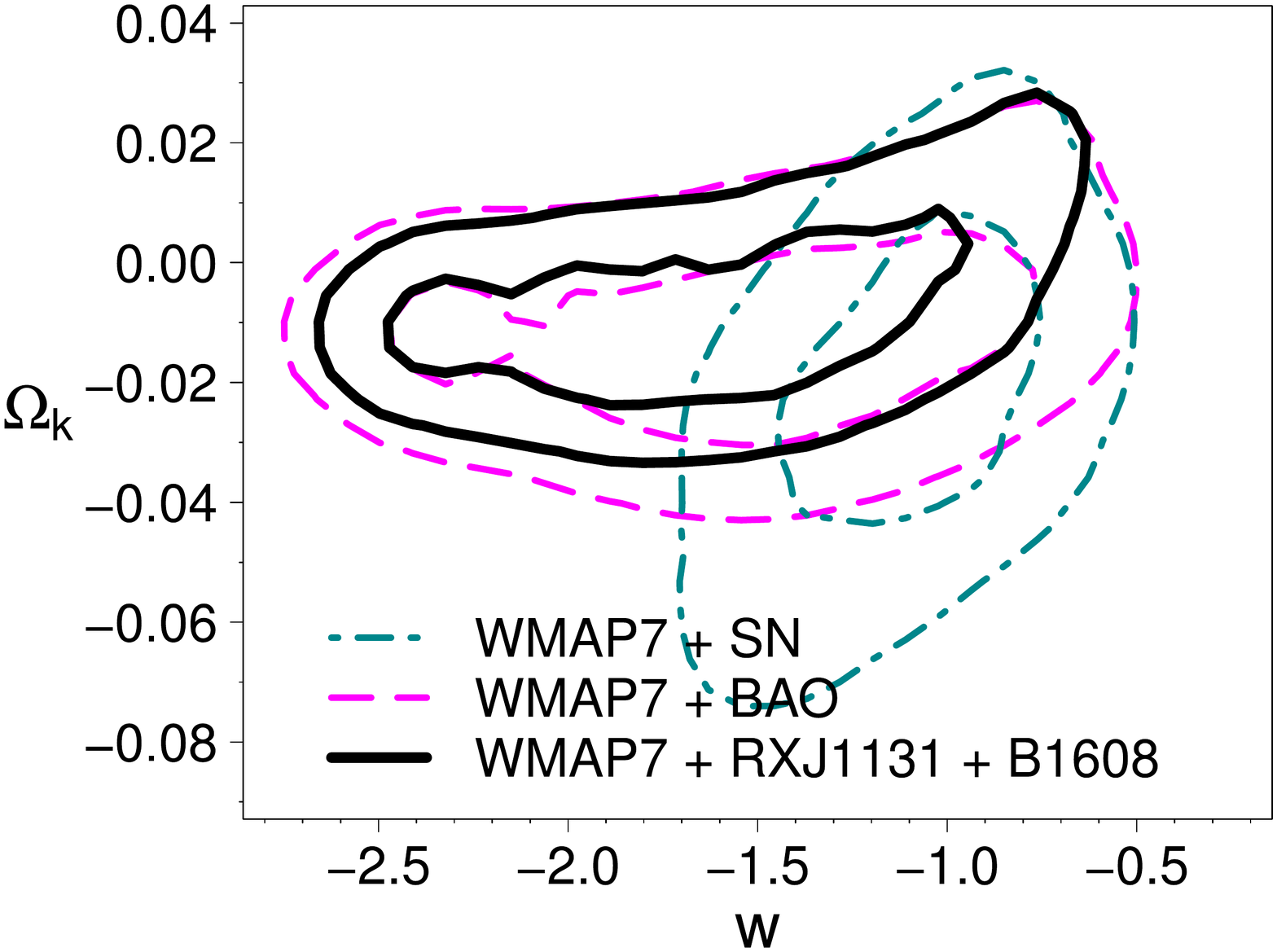} 
  \includegraphics[trim=0 0 100 520, clip,width=0.59\textwidth]{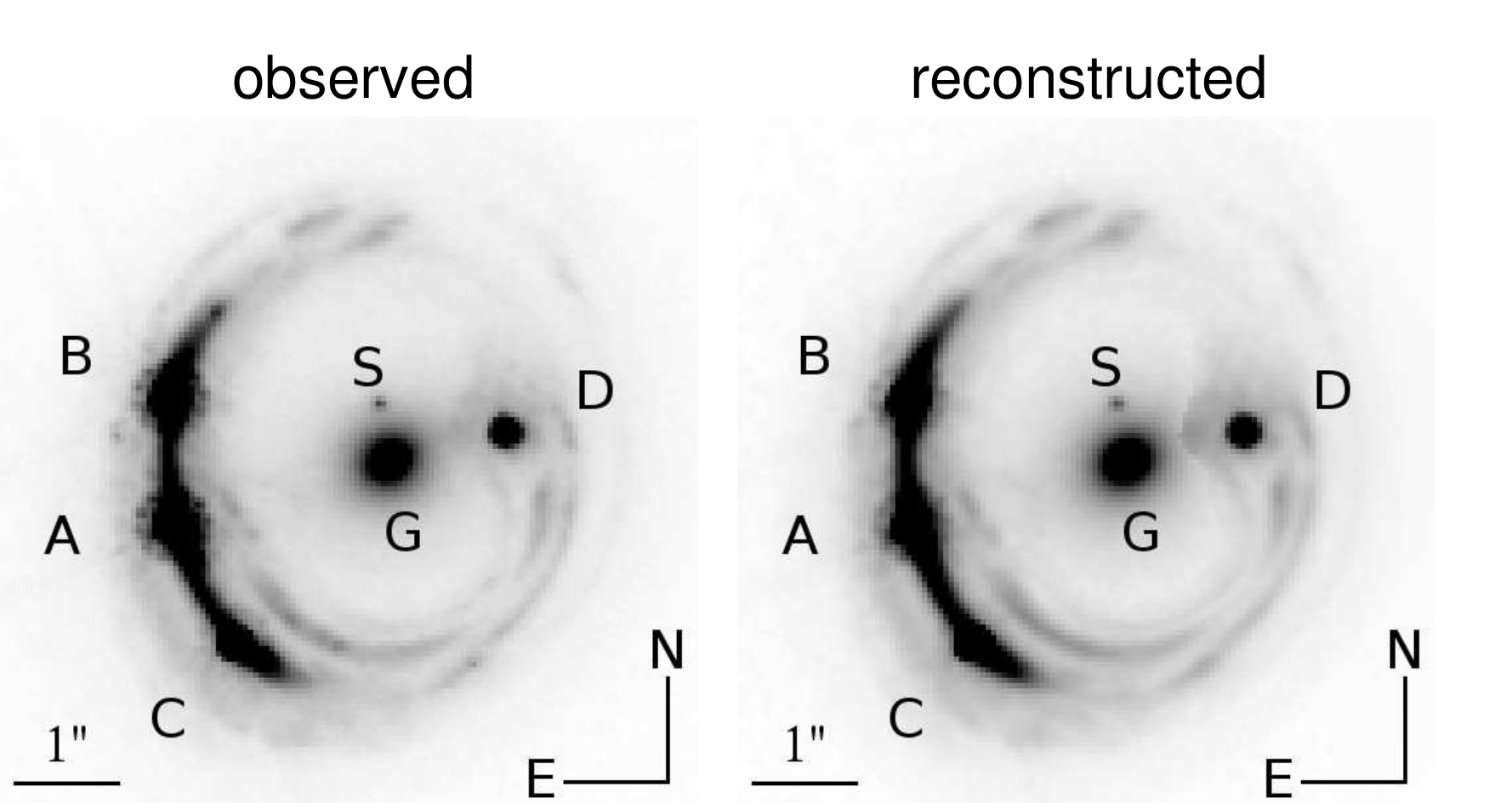}
  \caption{[Left panel] 68\% and 95\% confidence limits on the dark energy 
equation of state $w$ and the curvature $\Omega_k$ for WMAP7 combined with 
just two time delay distances, compared to other probes, showing the high 
complementarity of strong lensing.  
Observed HST [center panel] and reconstructed 
[right panel] image of a strong lens.  The lens model reproduces to high 
  fidelity tens of thousands of data points providing extremely tight 
  constraints on the mass model of the deflector and thus on 
  cosmological parameters.  Figures from \cite{suyu13a,suyu13b}. 
} 
\label{fig:cosmolens} 
\end{figure}

Current blind analyses of time delay distances \cite{suyu10,suyu13a,komatsu11} 
demonstrate the maturing power of strong lensing as a dark energy probe.  
Figure~\ref{fig:cosmolens} illustrates the current cosmological leverage 
from a mere two time delay distances (left panel) and the maturity of 
modeling the gravitational lens (center and right panels).  
Moreover, the observed time delays can give signatures of microlensing from 
dark matter substructure along the line of sight, making this method a 
probe of dark matter properties as well.  The Snowmass white paper 
{\it Dark Energy with Gravitational Time Delays\/} \cite{snowsl} gives 
greater detail on this new dark energy probe.

The main challenge currently for strong lensing as a cosmological probe 
is the small sample size; this will be greatly ameliorated by DES and LSST, 
and in that era the scarcity will be in follow up with high resolution 
imaging. On the systematics side, once we move below 5\% precision then 
accurately accounting for the projected mass, e.g.\ through using large 
ray tracing simulations to calibrate the convergence for a field with the 
observed galaxy density, will require careful attention.

\subsection{Gravitational Wave Sirens}
We are on the verge of a new era of gravitational wave (GW) astronomy.  It is
widely expected that the coming few years will witness the first direct
detection of GWs. A network of ground-based observatories, composed of the LIGO
and Virgo detectors, are currently being upgraded to ``advanced''
sensitivity. Once operational, these detectors are expected to observe a
significant stellar mass compact binary merger rate (perhaps dozens per year;
e.g., \cite{2010CQGra..27q3001A,2013ApJ...779...72D}).  All being well, the second half of the
current decade should see routine ground-based GW observations of binary
coalescences.  The launch of a space based GW antenna would extend
the GW window to low frequencies and probe processes involving supermassive ($M
\gtrsim 10^5\,M_\odot$) black holes.  GWs were recently selected as part of the ESA
cosmic vision science program, with a nominal launch date of a mission such as
eLISA in 2034.

Standard sirens are gravitational wave sources for which the absolute
luminosity distance and redshift can be determined, and are thus the GW analog
to standard
candles~\cite{1986Natur.323..310S,2002luml.conf..207S,2005ApJ...629...15H,2006PhRvD..74f3006D,2009PhRvD..80j4009C,2010PhRvD..81l4046H,2010ApJ...725..496N,2011ApJ...739...99N,tempref}.
Standard sirens are of particular interest because they take advantage of the
simplicity of black holes (which are understood from a basic physics
perspective, and are {\em fully}\/ described by mass, spin, and charge) to
provide an absolute distance, allowing measurements to cosmological scales
without the use of distance-ladders or phenomenological scaling relations.

As first pointed out by~\cite{1986Natur.323..310S,2002luml.conf..207S}, by measuring
the gravitational waveform during the inspiral and merger of a binary
it is possible to make a {\em direct and absolute}\/ measurement of
the luminosity distance to a source. 
This is because the physics underlying the inspiral of a binary due to
GW emission is well described and understood in
general relativity.
These sources thus offer an entirely
independent and complementary way to measure the evolution history
of our Universe. Standard sirens are physical, not astrophysical,
measures of distance.

Multi-messenger astronomy is crucial to unlock the power of these
binary sources. An important limitation of GW binaries
is that they do not provide the redshift to the source; the
redshift is degenerate with the binaries' intrinsic parameters.
However, if an electromagnetic counterpart to a binary
inspiral event is identified, it is then possible to measure the
redshift independently.\footnote{As
emphasized by \cite{1986Natur.323..310S,2012PhRvD..86d3011D}, it is also possible
to do this in a statistical fashion, rather than identifying
redshifts of individual sources.}
In addition, by determining the exact location of the
source on the sky, the fit to distance is significantly improved.  In
short, gravitational waves provide absolute distance while
electromagnetic measurements provide redshift, and the combination makes it possible to
put a very accurate point on the distance--redshift curve.  A binary
inspiral source coupled with an electromagnetic counterpart would
therefore constitute an exceedingly good cosmological standard
siren.

There are three standard siren cases of particular interest: 1. stellar mass
compact binaries detected by ground-based GW observatories, 2. supermassive
binary black holes detected by space-based GW observatories, and 3. stellar mass
compact binaries detected by space-based GW observatories. We discuss them
briefly below:

{\bf 1.}
The advanced LIGO/Virgo network is expected to begin operating within the next
few years. One of the most promising sources for
these ground-based observatories is the inspiral and merger of
binaries consisting of neutron stars and/or black holes. These sources will be
detected to $\sim350\,\mbox{Mpc}$, and the event rates are thought to be in the
range of $0.1$--$1000/\mbox{year}$ for both double neutron star and neutron
star-black hole systems~\cite{2010ApJ...715L.138B,
  2010ApJ...708..117B,Abadieetal:2010,2012ApJ...749...91F,2012ApJ...757...91B,
  2012ApJ...759...52D,aasi:2013}. There has been much activity recently exploring the
possibility of electromagnetic counterparts to these events. For example, it is
conceivable that the radioactive powered ejecta of r-process elements will
produce an isotropic optical/infrared counterpart to the merger (a
``macronovae'' or ``kilonovae'').

A particularly exciting multi-messenger source for ground-based GW observatories is the
inspiral and merger of a stellar-mass compact binary associated with a
short/hard gamma-ray burst (GRB). We know these GRBs exist and that
they occur in the local Universe. 
In addition, there is growing evidence that they
are associated with binary inspiral (i.e., double neutron star or
neutron star--black hole systems), and if this is the case, then we know that they would
be detectable by a network of ground-based GW detectors with advanced
LIGO sensitivity \cite{2013arXiv1307.6586L}.

\begin{figure}
\centering 
\includegraphics[width=0.98\columnwidth]{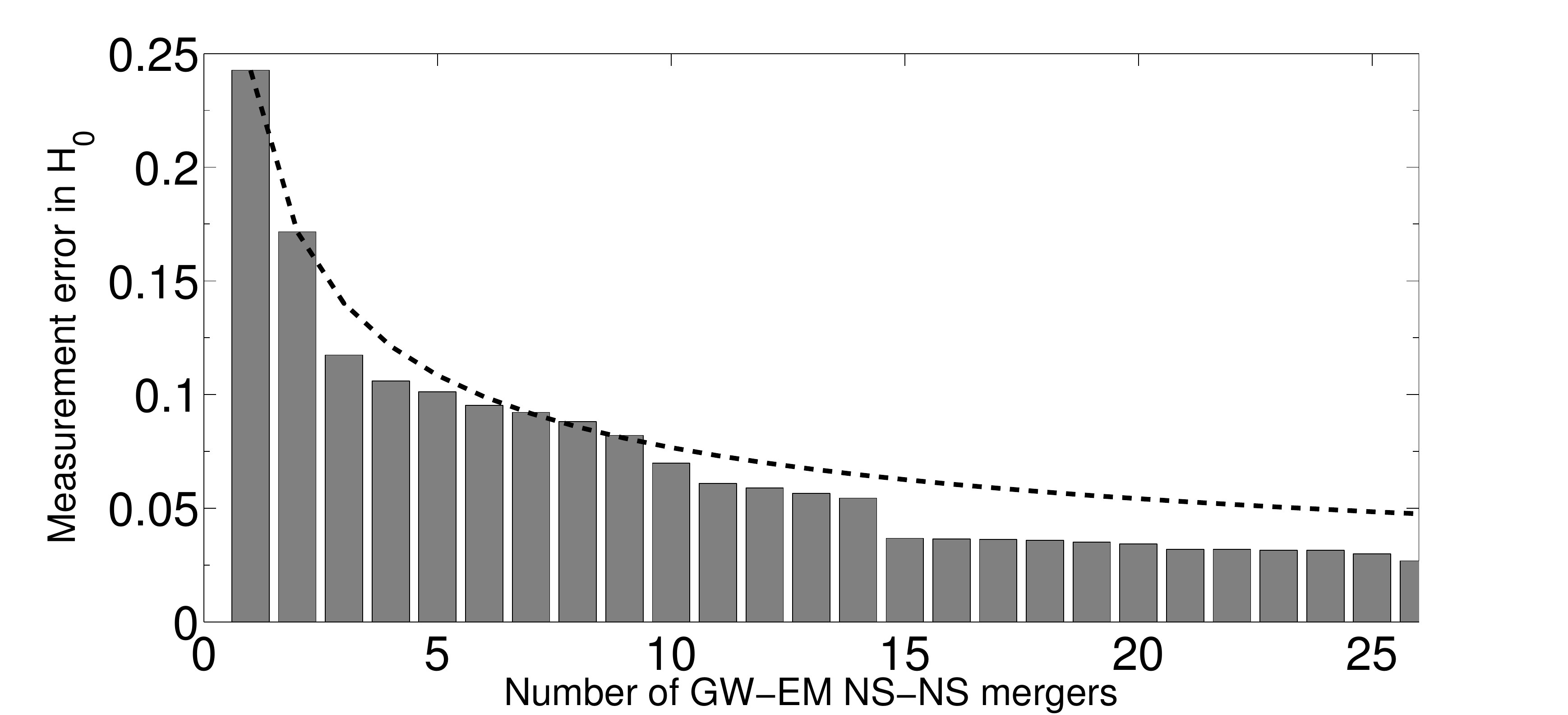}
\caption{$H_0$ measurement uncertainty as a function of the number of multi-messenger
  (GW+EM) double neutron star merger events observed by an advanced LIGO-Virgo
  network. The dashed line shows Gaussian convergence.}
\label{fig:H0}
\end{figure} 

If these sources are observed in both the GW and the EM spectra, we can ask how
well they can be used as standard candles~\cite{2006PhRvD..74f3006D,2010ApJ...725..496N,tempref}.
Fig.~\ref{fig:H0}, taken from~\cite{tempref}, shows that
20 events would provide a 3\% measurement of $H_0$. This improves to 1\% if the 
sources are beamed, as would be expected for GRBs.
Although, as was shown in~\cite{2010ApJ...725..496N},
the distance-inclination degeneracy compromises the utility of individual
sources, the posterior probability distribution function for an ensemble of
sources sharply mitigates this. This arises precisely because of the strong non-Gaussianity in the
distance uncertainties due to the degeneracy.
This measurement of $H_0$ would
provide one of the best, and of greater import, one of the cleanest and most
direct estimates of the age of the
Universe.

{\bf 2.}
Space-based gravitational wave observatories
are sensitive to supermassive binary black hole coalescences out to
very high redshift ($z\sim10$). If we are able to independently
determine redshift, we could use these systems as standard sirens to
provide a powerful complementary cosmological
probe~\cite{2005ApJ...629...15H}. Weak lensing compromises the utility of
individual sources at high redshift, but for sufficient statistics the lensing
effects average out, and precision cosmology is possible.

{\bf 3.} Extrapolating to the distant
future, a space-based decihertz gravitational-wave mission, such as
the Big Bang Observer, would constitute a truly revolutionary
cosmology mission, measuring the equation-of-state of the dark energy
and the growth of structure to a precision comparable to that of all other
proposed dark energy/cosmology missions {\em
  combined}~\cite{2009PhRvD..80j4009C}.

\subsection{Active Galactic Nuclei Radius-Luminosity Relationship}

\subsubsection{Overview and current status} 
The most luminous Active Galactic Nuclei (AGNs) can be detected across much 
of the observable Universe, and the spectral properties of 
most are remarkably uniform. These two properties have motivated many 
attempts to find ways to use them as standard candles \cite{1977ApJ...214..679B,1999MNRAS.302L..24C,2002ApJ...581L..67E,wang13}.
One promising technique
that has been shown to yield a reliable luminosity distance indicator for AGN \cite{2011ApJ...740L..49W}
is to utilize the relationship between
the radius of the AGN broad line region (BLR) and 
the continuum luminosity \cite{2000ApJ...533..631K,2013ApJ...767..149B}. The BLR consists of 
high-velocity gas clouds that surround the supermassive black hole at the 
center of the AGN. These gas clouds are photoionized by radiation from 
the immediate vicinity of the black hole, and thus the distances at which 
these gas clouds produce emission in various lines are determined by the 
intensity of the ionizing continuum. The naive physical expectation is that 
the radius $R$ for a particular broad emission line should be proportional to 
the square root of the luminosity $L^{1/2}$, since the flux declines as the 
inverse square of the distance from the center. 

While the BLR is spatially unresolved, the radius of the BLR can be measured 
with the technique of reverberation mapping \cite{1982ApJ...255..419B,1993PASP..105..247P}. 
Reverberation mapping takes advantage of the intrinsic continuum variability 
of AGNs and that this continuum photoionizes the BLR. As the continuum varies 
in luminosity, the luminosities of the broad emission lines in the BLR also 
change, and the time lag $\tau$ between the continuum and line variations is 
due to the light travel time to the BLR. Measurement of this time lag 
provides the radius $R = c\tau$ of the BLR, and this in turn can be used to 
predict the intrinsic luminosity from the radius-luminosity relation. 

Time lags for $\sim 50$ moderate-luminosity AGNs with $z<0.3$ have been made to 
date (see \cite{2013ApJ...773...90G} for a recent reanalysis.) These lag measurements are 
typically based on continuum and spectroscopic observations on at least 30 
epochs, a cadence that is on order five times smaller than the lag or shorter, 
and few percent flux uncertainties in the line and continuum. With a careful 
consideration of all of the observational uncertainties, as well as accurate 
subtraction of the host galaxy starlight, the best fit for the
radius-luminosity relation has a power-law slope of $0.533^{+0.035}_{-0.033}$ 
\cite{2013ApJ...767..149B}. This result is based on AGNs that span four orders of 
magnitude in luminosity and is in good agreement with the expectation that the 
BLR is photoionized by the continuum. 

The intrinsic scatter in the radius-luminosity relationship for the best 
reverberation data is $\sim 0.11$ dex, which is not much larger than the 
typical uncertainty of $\sim 0.09$ dex in the data \cite{2010IAUS..267..151P}. 
The current uncertainties in an ``AGN Hubble Diagram'' for a larger number of 
AGNs have a root mean square scatter of 0.13 dex after two outliers are 
excluded \cite{2011ApJ...740L..49W,2013ApJ...767..149B}. Based on a reduced $\chi^2$ analysis, nearly 
half of the total scatter appears to be due to observational uncertainty. 
These estimates suggest that the scatter can be reduced to 0.08 dex or 
0.20 mag in distance modulus \cite{2011ApJ...740L..49W}. 

\subsubsection{Prospects for improvement}
One important, next step is to demonstrate that the uncertainties can be 
substantially reduced for a larger sample of nearby objects. The main 
requirement is more observations to produce better-sampled light curves, which 
would help eliminate incorrect lags due to aliasing or long-term variations. 
Other areas for improvement include reliable extinction corrections, better 
distance estimates for the nearest AGNs, more precise flux calibrations, and 
models to relate the variation in the observed continuum to the variation in 
the ionizing continuum. Many month-long campaigns at 1-m to 3-m class 
telescopes over the last few years \cite{2010ApJ...721..715D,2011ApJ...743L...4B} have gradually 
increased the number of AGNs with higher-quality observational data and begin 
to provide a dataset with which observational uncertainties can be minimized. 
A key advantage of AGNs is that since they do not substantially dim with time
over human timescales, 
these long-term studies can eliminate AGNs with uncertain extinction 
corrections, ambiguous lag measurements, and other potential complications. 

The greatest potential for the radius-luminosity relationship is the 
application of these same techniques to higher redshifts because the most 
luminous AGNs are several orders of magnitude brighter than the most luminous 
supernovae. The two main challenges to the application of this relationship 
to higher-redshift AGNs are: 1) The radius-luminosity relationship must be 
measured with different emission lines and calibrated for different continuum 
regions than have been used for low-redshift studies; 2) The most readily 
observable AGNs are the most luminous, and these also have the greatest lags 
(a problem that is compounded by time dilation). Low-redshift studies have 
concentrated on the prominent $H\beta$ emission line at 486~nm, yet by 
$z\sim0.6$ this line has redshifted out of the range of easy observation. 
There are numerous other rest-frame UV lines that are accessible in 
higher-redshift AGNs, but only a few have been studied with reverberation 
mapping \cite{2006ApJ...647..901M,2007ApJ...659..997K}, although microlensing studies of 
gravitationally-lensed QSOs do support the existence of a radius-luminosity 
relationship for the CIV broad emission line at 155~nm \cite{2013ApJ...764..160G}. Substantial, new data 
will be required to demonstrate that any of these other emission lines can 
ultimately lead to comparably small scatter, and that 
evolutionary effects are either negligible or can be corrected. 
Another consideration is that the 
radius-luminosity relationship implies that the time lags could be as long as 
several years for the most luminous objects, particularly those that would be 
easiest to detect at the highest redshifts, and thus observational campaigns 
will similarly require at least several years of data acquisition. Aliasing 
due to seasonal gaps in the data acquisition will also need to be addressed. 
One advantage is that the observing cadence is also slower, so less time is 
required in any one season. Another advantage is that the corrections for host 
galaxy starlight are less significant for the rest-frame UV continuum. A 
final, potentially unique advantage of AGNs is that the same methodology can be 
applied across the entire observable Universe. 

Substantial progress with high-redshift reverberation-mapping campaigns 
appears possible in the next few years thanks to both large, synoptic imaging 
surveys and multi-object spectrographs with sufficient field of view to 
observe many AGNs at once. Two current surveys that aim to study many new AGNs, 
including AGNs at higher redshifts, are DES and BOSS. DES aims to combine 
higher cadence photometric monitoring with lower cadence spectroscopy over the 
course of five years. These data will be used to estimate the radius-luminosity 
relationship for several emission lines in QSOs up to redshift four. BOSS plans 
to use slightly more spectroscopic epochs over a shorter period of time to 
accomplish similar aims, although directed more toward lower redshift and 
lower luminosity AGNs with shorter lag times. A key factor in the success of 
both surveys will be the quality of their data, particularly the relative flux 
calibration of spectroscopic data obtained through fiber optics cables. 
If these surveys meet their data quality goals, the results
should begin to quantify the competitiveness of the radius-luminosity relation 
as a distance indicator at high redshift.

LSST offers unique and exciting opportunities to extend
reverberation\slash mapping 
studies at all redshifts and for a much broader range in luminosity. The LSST 
cadence is well suited to determine time lags for most AGNs of at least 
moderate luminosity, and the fraction of the sky that is nearly circumpolar 
offers an excellent opportunity to avoid aliasing. The combination of excellent 
flux calibration and multi-wavelength photometry, combined with the 
extraordinarily large sample size, also offers the prospect for 
reverberation-mapping with broad-band photometry alone
\cite{2013arXiv1310.6774Z}. Superb, 
multi-wavelength photometry, combined with a subset of AGNs with particularly 
strong emission lines, may lead to robust separation of the continuum and line 
intensity variations with photometric data alone.

 \section*{Acknowledgement}
FNAL is operated by Fermi Research Alliance, LLC under Contract No.\ De-AC02-07CH11359 with the United States Department of Energy.
LBNL is supported by the U.S. Department of Energy, Office of High Energy Physics, under Contract No.\ DE-AC02-05CH11231.
OSU acknowledges support by the National Science Foundation under the  grant AST-1008882.

SWA was supported in part by the U.S. Department of Energy under contract number DE-AC02-76SF00515.
KDD acknowledges support by the National Science Foundation under Award No.\ AST-1302093.
DEH acknowledges support from National Science Foundation CAREER grant PHY-1151836, and support in part by the Kavli Institute for Cosmological Physics at the University of Chicago through NSF grant PHY-1125897 and an endowment from the Kavli Foundation and its founder Fred Kavli.
DK and SEW would like to thank the DOE
HEP Program for support through grant DOE-HEPÐDE-SC00010676.
\bibliography{bib}

\begin{thebibliography}{100}
\expandafter\ifx\csname url\endcsname\relax
  \def\url#1{\texttt{#1}}\fi
\expandafter\ifx\csname urlprefix\endcsname\relax\def\urlprefix{URL }\fi
\expandafter\ifx\csname href\endcsname\relax
  \def\href#1#2{#2} \def\path#1{#1}\fi

\bibitem{2013AJ....145...10D}
K.~S. {Dawson}, D.~J. {Schlegel}, C.~P. {Ahn}, S.~F. {Anderson},
  {\'E}.~{Aubourg}, S.~{Bailey}, R.~H. {Barkhouser}, et~al., {The Baryon
  Oscillation Spectroscopic Survey of SDSS-III}, Astronomical Journal 145
  (2013) 10.
\newblock \href {http://arxiv.org/abs/1208.0022} {\path{arXiv:1208.0022}},
  \href {http://dx.doi.org/10.1088/0004-6256/145/1/10}
  {\path{doi:10.1088/0004-6256/145/1/10}}.

\bibitem{peebles70}
P.~J.~E. {Peebles}, J.~T. {Yu}, {Primeval Adiabatic Perturbation in an
  Expanding Universe}, ApJ 162 (1970) 815--+.
\newblock \href {http://dx.doi.org/10.1086/150713} {\path{doi:10.1086/150713}}.

\bibitem{sunyaev70}
R.~A. {Sunyaev}, Y.~B. {Zeldovich}, {Small-Scale Fluctuations of Relic
  Radiation}, Astrophysics and Space Science 7 (1970) 3--19.
\newblock \href {http://dx.doi.org/10.1007/BF00653471}
  {\path{doi:10.1007/BF00653471}}.

\bibitem{bond84}
J.~R. {Bond}, G.~{Efstathiou}, {Cosmic background radiation anisotropies in
  universes dominated by nonbaryonic dark matter}, ApJL 285 (1984) L45--L48.
\newblock \href {http://dx.doi.org/10.1086/184362} {\path{doi:10.1086/184362}}.

\bibitem{bond87}
J.~R. {Bond}, G.~{Efstathiou}, {The statistics of cosmic background radiation
  fluctuations}, MNRAS 226 (1987) 655--687.

\bibitem{jungman96}
G.~{Jungman}, M.~{Kamionkowski}, A.~{Kosowsky}, D.~N. {Spergel}, {Weighing the
  Universe with the Cosmic Microwave Background}, Phys. Rev. Lett. 76 (1996)
  1007--1010.
\newblock \href {http://arxiv.org/abs/arXiv:astro-ph/9507080}
  {\path{arXiv:arXiv:astro-ph/9507080}}, \href
  {http://dx.doi.org/10.1103/PhysRevLett.76.1007}
  {\path{doi:10.1103/PhysRevLett.76.1007}}.

\bibitem{hu96}
W.~{Hu}, N.~{Sugiyama}, {Small-Scale Cosmological Perturbations: an Analytic
  Approach}, ApJ 471 (1996) 542--+.
\newblock \href {http://arxiv.org/abs/arXiv:astro-ph/9510117}
  {\path{arXiv:arXiv:astro-ph/9510117}}, \href
  {http://dx.doi.org/10.1086/177989} {\path{doi:10.1086/177989}}.

\bibitem{hu96a}
W.~{Hu}, M.~{White}, {Acoustic Signatures in the Cosmic Microwave Background},
  ApJ 471 (1996) 30--+.
\newblock \href {http://arxiv.org/abs/arXiv:astro-ph/9602019}
  {\path{arXiv:arXiv:astro-ph/9602019}}, \href
  {http://dx.doi.org/10.1086/177951} {\path{doi:10.1086/177951}}.

\bibitem{hu97}
W.~{Hu}, N.~{Sugiyama}, J.~{Silk}, {The physics of microwave background
  anisotropies}, Nature 386 (1997) 37--43.
\newblock \href {http://arxiv.org/abs/arXiv:astro-ph/9504057}
  {\path{arXiv:arXiv:astro-ph/9504057}}, \href
  {http://dx.doi.org/10.1038/386037a0} {\path{doi:10.1038/386037a0}}.

\bibitem{tegmark97}
M.~{Tegmark}, {Measuring Cosmological Parameters with Galaxy Surveys}, Phys.
  Rev. Lett. 79 (1997) 3806--3809.
\newblock \href {http://arxiv.org/abs/arXiv:astro-ph/9706198}
  {\path{arXiv:arXiv:astro-ph/9706198}}, \href
  {http://dx.doi.org/10.1103/PhysRevLett.79.3806}
  {\path{doi:10.1103/PhysRevLett.79.3806}}.

\bibitem{goldberg98}
D.~M. {Goldberg}, M.~A. {Strauss}, {Determination of the Baryon Density from
  Large-Scale Galaxy Redshift Surveys}, ApJ 495 (1998) 29--+.
\newblock \href {http://arxiv.org/abs/arXiv:astro-ph/9707209}
  {\path{arXiv:arXiv:astro-ph/9707209}}, \href
  {http://dx.doi.org/10.1086/305284} {\path{doi:10.1086/305284}}.

\bibitem{efstathiou99}
G.~{Efstathiou}, J.~R. {Bond}, {Cosmic confusion: degeneracies among
  cosmological parameters derived from measurements of microwave background
  anisotropies}, MNRAS 304 (1999) 75--97.
\newblock \href {http://arxiv.org/abs/arXiv:astro-ph/9807103}
  {\path{arXiv:arXiv:astro-ph/9807103}}, \href
  {http://dx.doi.org/10.1046/j.1365-8711.1999.02274.x}
  {\path{doi:10.1046/j.1365-8711.1999.02274.x}}.

\bibitem{eisenstein98a}
D.~J. {Eisenstein}, W.~{Hu}, M.~{Tegmark}, {Cosmic Complementarity: $H_0$ and
  $\Omega_M$ from Combining Cosmic Microwave Background Experiments and
  Redshift Surveys}, ApJL 504 (1998) L57+.
\newblock \href {http://arxiv.org/abs/arXiv:astro-ph/9805239}
  {\path{arXiv:arXiv:astro-ph/9805239}}, \href
  {http://dx.doi.org/10.1086/311582} {\path{doi:10.1086/311582}}.

\bibitem{eisenstein02}
D.~{Eisenstein}, {Large-Scale Structure and Future Surveys}, in:
  {M.~J.~I.~Brown \& A.~Dey} (Ed.), Next Generation Wide-Field Multi-Object
  Spectroscopy, Vol. 280 of Astronomical Society of the Pacific Conference
  Series, 2002, pp. 35--+.

\bibitem{blake03}
C.~{Blake}, K.~{Glazebrook}, {Probing Dark Energy Using Baryonic Oscillations
  in the Galaxy Power Spectrum as a Cosmological Ruler}, ApJ 594 (2003)
  665--673.
\newblock \href {http://arxiv.org/abs/arXiv:astro-ph/0301632}
  {\path{arXiv:arXiv:astro-ph/0301632}}, \href
  {http://dx.doi.org/10.1086/376983} {\path{doi:10.1086/376983}}.

\bibitem{hu03b}
W.~{Hu}, Z.~{Haiman}, {Redshifting rings of power}, Physical Review D 68~(6)
  (2003) 063004.
\newblock \href {http://arxiv.org/abs/arXiv:astro-ph/0306053}
  {\path{arXiv:arXiv:astro-ph/0306053}}, \href
  {http://dx.doi.org/10.1103/PhysRevD.68.063004}
  {\path{doi:10.1103/PhysRevD.68.063004}}.

\bibitem{linder03}
E.~V. {Linder}, {Baryon oscillations as a cosmological probe}, Physical Review
  D 68~(8) (2003) 083504--+.
\newblock \href {http://arxiv.org/abs/arXiv:astro-ph/0304001}
  {\path{arXiv:arXiv:astro-ph/0304001}}, \href
  {http://dx.doi.org/10.1103/PhysRevD.68.083504}
  {\path{doi:10.1103/PhysRevD.68.083504}}.

\bibitem{seo03}
H.-J. {Seo}, D.~J. {Eisenstein}, {Probing Dark Energy with Baryonic Acoustic
  Oscillations from Future Large Galaxy Redshift Surveys}, ApJ 598 (2003)
  720--740.
\newblock \href {http://arxiv.org/abs/arXiv:astro-ph/0307460}
  {\path{arXiv:arXiv:astro-ph/0307460}}, \href
  {http://dx.doi.org/10.1086/379122} {\path{doi:10.1086/379122}}.

\bibitem{Weinberg2012}
D.~H. {Weinberg}, M.~J. {Mortonson}, D.~J. {Eisenstein}, C.~{Hirata}, A.~G.
  {Riess}, E.~{Rozo}, {Observational Probes of Cosmic Acceleration}, ArXiv
  e-prints\href {http://arxiv.org/abs/1201.2434} {\path{arXiv:1201.2434}}.

\bibitem{mcdonald07}
P.~{McDonald}, D.~J. {Eisenstein}, {Dark energy and curvature from a future
  baryonic acoustic oscillation survey using the Lyman-{$\alpha$} forest},
  Physical Review D 76~(6) (2007) 063009--+.
\newblock \href {http://arxiv.org/abs/arXiv:astro-ph/0607122}
  {\path{arXiv:arXiv:astro-ph/0607122}}, \href
  {http://dx.doi.org/10.1103/PhysRevD.76.063009}
  {\path{doi:10.1103/PhysRevD.76.063009}}.

\bibitem{cole05}
S.~{Cole}, W.~J. {Percival}, J.~A. {Peacock}, P.~{Norberg}, C.~M. {Baugh},
  C.~S. {Frenk}, et~al., {The 2dF Galaxy Redshift Survey: power-spectrum
  analysis of the final data set and cosmological implications}, MNRAS 362
  (2005) 505--534.
\newblock \href {http://arxiv.org/abs/arXiv:astro-ph/0501174}
  {\path{arXiv:arXiv:astro-ph/0501174}}, \href
  {http://dx.doi.org/10.1111/j.1365-2966.2005.09318.x}
  {\path{doi:10.1111/j.1365-2966.2005.09318.x}}.

\bibitem{eisenstein05}
D.~J. {Eisenstein}, I.~{Zehavi}, D.~W. {Hogg}, R.~{Scoccimarro}, M.~R.
  {Blanton}, R.~C. {Nichol}, R.~{Scranton}, et~al., {Detection of the Baryon
  Acoustic Peak in the Large-Scale Correlation Function of SDSS Luminous Red
  Galaxies}, \apj 633 (2005) 560--574.
\newblock \href {http://arxiv.org/abs/arXiv:astro-ph/0501171}
  {\path{arXiv:arXiv:astro-ph/0501171}}, \href
  {http://dx.doi.org/10.1086/466512} {\path{doi:10.1086/466512}}.

\bibitem{tegmark06}
M.~{Tegmark}, D.~J. {Eisenstein}, M.~A. {Strauss}, D.~H. {Weinberg}, M.~R.
  {Blanton}, J.~A. {Frieman}, et~al., {Cosmological constraints from the SDSS
  luminous red galaxies}, Physical Review D 74~(12) (2006) 123507--+.
\newblock \href {http://arxiv.org/abs/arXiv:astro-ph/0608632}
  {\path{arXiv:arXiv:astro-ph/0608632}}, \href
  {http://dx.doi.org/10.1103/PhysRevD.74.123507}
  {\path{doi:10.1103/PhysRevD.74.123507}}.

\bibitem{percival07}
W.~J. {Percival}, S.~{Cole}, D.~J. {Eisenstein}, R.~C. {Nichol}, J.~A.
  {Peacock}, A.~C. {Pope}, A.~S. {Szalay}, {Measuring the Baryon Acoustic
  Oscillation scale using the Sloan Digital Sky Survey and 2dF Galaxy Redshift
  Survey}, MNRAS 381 (2007) 1053--1066.
\newblock \href {http://arxiv.org/abs/0705.3323} {\path{arXiv:0705.3323}},
  \href {http://dx.doi.org/10.1111/j.1365-2966.2007.12268.x}
  {\path{doi:10.1111/j.1365-2966.2007.12268.x}}.

\bibitem{beutler11}
F.~{Beutler}, C.~{Blake}, M.~{Colless}, D.~H. {Jones}, L.~{Staveley-Smith},
  L.~{Campbell}, Q.~{Parker}, et~al., {The 6dF Galaxy Survey: baryon acoustic
  oscillations and the local Hubble constant}, MNRAS 416 (2011) 3017--3032.
\newblock \href {http://arxiv.org/abs/1106.3366} {\path{arXiv:1106.3366}},
  \href {http://dx.doi.org/10.1111/j.1365-2966.2011.19250.x}
  {\path{doi:10.1111/j.1365-2966.2011.19250.x}}.

\bibitem{blake11c}
C.~{Blake}, E.~A. {Kazin}, F.~{Beutler}, T.~M. {Davis}, D.~{Parkinson},
  S.~{Brough}, et~al., {The WiggleZ Dark Energy Survey: mapping the
  distance-redshift relation with baryon acoustic oscillations}, MNRAS (2011)
  1598--+\href {http://arxiv.org/abs/1108.2635} {\path{arXiv:1108.2635}}, \href
  {http://dx.doi.org/10.1111/j.1365-2966.2011.19592.x}
  {\path{doi:10.1111/j.1365-2966.2011.19592.x}}.

\bibitem{Anderson2012}
L.~{Anderson}, E.~{Aubourg}, S.~{Bailey}, D.~{Bizyaev}, M.~{Blanton}, A.~S.
  {Bolton}, J.~{Brinkmann}, et~al., {The clustering of galaxies in the SDSS-III
  Baryon Oscillation Spectroscopic Survey: baryon acoustic oscillations in the
  Data Release 9 spectroscopic galaxy sample}, MNRAS 427 (2012) 3435--3467.
\newblock \href {http://arxiv.org/abs/1203.6594} {\path{arXiv:1203.6594}},
  \href {http://dx.doi.org/10.1111/j.1365-2966.2012.22066.x}
  {\path{doi:10.1111/j.1365-2966.2012.22066.x}}.

\bibitem{padmanabhan07}
N.~{Padmanabhan}, D.~J. {Schlegel}, U.~{Seljak}, A.~{Makarov}, N.~A. {Bahcall},
  M.~R. {Blanton}, et~al., {The clustering of luminous red galaxies in the
  Sloan Digital Sky Survey imaging data}, MNRAS 378 (2007) 852--872.
\newblock \href {http://arxiv.org/abs/arXiv:astro-ph/0605302}
  {\path{arXiv:arXiv:astro-ph/0605302}}, \href
  {http://dx.doi.org/10.1111/j.1365-2966.2007.11593.x}
  {\path{doi:10.1111/j.1365-2966.2007.11593.x}}.

\bibitem{blake07}
C.~{Blake}, A.~{Collister}, S.~{Bridle}, O.~{Lahav}, {Cosmological baryonic and
  matter densities from 600000 SDSS luminous red galaxies with photometric
  redshifts}, MNRAS 374 (2007) 1527--1548.
\newblock \href {http://arxiv.org/abs/arXiv:astro-ph/0605303}
  {\path{arXiv:arXiv:astro-ph/0605303}}, \href
  {http://dx.doi.org/10.1111/j.1365-2966.2006.11263.x}
  {\path{doi:10.1111/j.1365-2966.2006.11263.x}}.

\bibitem{hutsi10}
G.~{H{\"u}tsi}, {Power spectrum of the maxBCG sample: detection of acoustic
  oscillations using galaxy clusters}, MNRAS 401 (2010) 2477--2489.
\newblock \href {http://arxiv.org/abs/0910.0492} {\path{arXiv:0910.0492}},
  \href {http://dx.doi.org/10.1111/j.1365-2966.2009.15824.x}
  {\path{doi:10.1111/j.1365-2966.2009.15824.x}}.

\bibitem{crocce11}
M.~{Crocce}, E.~{Gazta{\~n}aga}, A.~{Cabr{\'e}}, A.~{Carnero},
  E.~{S{\'a}nchez}, {Clustering of photometric luminous red galaxies - I.
  Growth of structure and baryon acoustic feature}, MNRAS 417 (2011)
  2577--2591.
\newblock \href {http://arxiv.org/abs/1104.5236} {\path{arXiv:1104.5236}},
  \href {http://dx.doi.org/10.1111/j.1365-2966.2011.19425.x}
  {\path{doi:10.1111/j.1365-2966.2011.19425.x}}.

\bibitem{sawangwit11}
U.~{Sawangwit}, T.~{Shanks}, F.~B. {Abdalla}, R.~D. {Cannon}, S.~M. {Croom},
  A.~C. {Edge}, N.~P. {Ross}, D.~A. {Wake}, {Angular correlation function of
  1.5 million luminous red galaxies: clustering evolution and a search for
  baryon acoustic oscillations}, MNRAS 416 (2011) 3033--3056.
\newblock \href {http://arxiv.org/abs/0912.0511} {\path{arXiv:0912.0511}},
  \href {http://dx.doi.org/10.1111/j.1365-2966.2011.19251.x}
  {\path{doi:10.1111/j.1365-2966.2011.19251.x}}.

\bibitem{seo12}
H.-J. {Seo}, S.~{Ho}, M.~{White}, A.~J. {Cuesta}, A.~J. {Ross}, S.~{Saito},
  B.~{Reid}, et~al., {Acoustic Scale from the Angular Power Spectra of SDSS-III
  DR8 Photometric Luminous Galaxies}, \apj 761 (2012) 13.
\newblock \href {http://arxiv.org/abs/1201.2172} {\path{arXiv:1201.2172}},
  \href {http://dx.doi.org/10.1088/0004-637X/761/1/13}
  {\path{doi:10.1088/0004-637X/761/1/13}}.

\bibitem{padmanabhan12}
N.~{Padmanabhan}, X.~{Xu}, D.~J. {Eisenstein}, R.~{Scalzo}, A.~J. {Cuesta},
  K.~T. {Mehta}, E.~{Kazin}, {A 2 per cent distance to z = 0.35 by
  reconstructing baryon acoustic oscillations - I. Methods and application to
  the Sloan Digital Sky Survey}, MNRAS 427 (2012) 2132--2145.
\newblock \href {http://arxiv.org/abs/1202.0090} {\path{arXiv:1202.0090}},
  \href {http://dx.doi.org/10.1111/j.1365-2966.2012.21888.x}
  {\path{doi:10.1111/j.1365-2966.2012.21888.x}}.

\bibitem{xu12a}
X.~{Xu}, N.~{Padmanabhan}, D.~J. {Eisenstein}, K.~T. {Mehta}, A.~J. {Cuesta},
  {A 2 per cent distance to z = 0.35 by reconstructing baryon acoustic
  oscillations - II. Fitting techniques}, MNRAS 427 (2012) 2146--2167.
\newblock \href {http://arxiv.org/abs/1202.0091} {\path{arXiv:1202.0091}},
  \href {http://dx.doi.org/10.1111/j.1365-2966.2012.21573.x}
  {\path{doi:10.1111/j.1365-2966.2012.21573.x}}.

\bibitem{white03}
M.~{White}, {The Ly-a forest}, in: The Davis Meeting On Cosmic Inflation,
  arXiv:astro-ph/0305474, 2003.
\newblock \href {http://arxiv.org/abs/arXiv:astro-ph/0305474}
  {\path{arXiv:arXiv:astro-ph/0305474}}.

\bibitem{busca12}
N.~G. {Busca}, T.~{Delubac}, J.~{Rich}, S.~{Bailey}, A.~{Font-Ribera},
  D.~{Kirkby}, J.-M. {Le Goff}, et~al., {Baryon acoustic oscillations in the
  Ly{$\alpha$} forest of BOSS quasars}, Astronomy and Astrophysics 552 (2013)
  A96.
\newblock \href {http://arxiv.org/abs/1211.2616} {\path{arXiv:1211.2616}},
  \href {http://dx.doi.org/10.1051/0004-6361/201220724}
  {\path{doi:10.1051/0004-6361/201220724}}.

\bibitem{slosar13}
A.~{Slosar}, V.~{Ir{\v s}i{\v c}}, D.~{Kirkby}, S.~{Bailey}, N.~G. {Busca},
  T.~{Delubac}, J.~{Rich}, et~al., {Measurement of baryon acoustic oscillations
  in the Lyman-{$\alpha$} forest fluctuations in BOSS data release 9}, JCAP 4
  (2013) 26.
\newblock \href {http://arxiv.org/abs/1301.3459} {\path{arXiv:1301.3459}},
  \href {http://dx.doi.org/10.1088/1475-7516/2013/04/026}
  {\path{doi:10.1088/1475-7516/2013/04/026}}.

\bibitem{SeoEisenstein2007}
H.-J. {Seo}, D.~J. {Eisenstein}, {Improved Forecasts for the Baryon Acoustic
  Oscillations and Cosmological Distance Scale}, APJ 665 (2007) 14--24.
\newblock \href {http://arxiv.org/abs/arXiv:astro-ph/0701079}
  {\path{arXiv:arXiv:astro-ph/0701079}}, \href
  {http://dx.doi.org/10.1086/519549} {\path{doi:10.1086/519549}}.

\bibitem{seo07}
H.-J. Seo, D.~J. Eisenstein, {Improved forecasts for the baryon acoustic
  oscillations and cosmological distance scale}, {ApJ} 665 (2007) 14--24.
\newblock \href {http://arxiv.org/abs/astro-ph/0701079}
  {\path{arXiv:astro-ph/0701079}}, \href {http://dx.doi.org/10.1086/519549}
  {\path{doi:10.1086/519549}}.

\bibitem{seo10a}
H.-J. {Seo}, J.~{Eckel}, D.~J. {Eisenstein}, K.~{Mehta}, M.~{Metchnik},
  N.~{Padmanabhan}, P.~{Pinto}, et~al., {High-precision Predictions for the
  Acoustic Scale in the Nonlinear Regime}, ApJ 720 (2010) 1650--1667.
\newblock \href {http://arxiv.org/abs/0910.5005} {\path{arXiv:0910.5005}},
  \href {http://dx.doi.org/10.1088/0004-637X/720/2/1650}
  {\path{doi:10.1088/0004-637X/720/2/1650}}.

\bibitem{padmanabhan09}
N.~{Padmanabhan}, M.~{White}, {Calibrating the baryon oscillation ruler for
  matter and halos}, Physical Review D 80~(6) (2009) 063508--+.
\newblock \href {http://arxiv.org/abs/0906.1198} {\path{arXiv:0906.1198}},
  \href {http://dx.doi.org/10.1103/PhysRevD.80.063508}
  {\path{doi:10.1103/PhysRevD.80.063508}}.

\bibitem{mehta11}
K.~T. {Mehta}, H.-J. {Seo}, J.~{Eckel}, D.~J. {Eisenstein}, M.~{Metchnik},
  P.~{Pinto}, X.~{Xu}, {Galaxy Bias and Its Effects on the Baryon Acoustic
  Oscillation Measurements}, ApJ 734 (2011) 94--+.
\newblock \href {http://arxiv.org/abs/1104.1178} {\path{arXiv:1104.1178}},
  \href {http://dx.doi.org/10.1088/0004-637X/734/2/94}
  {\path{doi:10.1088/0004-637X/734/2/94}}.

\bibitem{eisenstein07b}
D.~J. {Eisenstein}, H.-J. {Seo}, E.~{Sirko}, D.~N. {Spergel}, {Improving
  Cosmological Distance Measurements by Reconstruction of the Baryon Acoustic
  Peak}, ApJ 664 (2007) 675--679.
\newblock \href {http://arxiv.org/abs/arXiv:astro-ph/0604362}
  {\path{arXiv:arXiv:astro-ph/0604362}}, \href
  {http://dx.doi.org/10.1086/518712} {\path{doi:10.1086/518712}}.

\bibitem{padmanabhan09a}
N.~{Padmanabhan}, M.~{White}, J.~D. {Cohn}, {Reconstructing baryon
  oscillations: A Lagrangian theory perspective}, Physical Review D 79~(6)
  (2009) 063523--+.
\newblock \href {http://arxiv.org/abs/0812.2905} {\path{arXiv:0812.2905}},
  \href {http://dx.doi.org/10.1103/PhysRevD.79.063523}
  {\path{doi:10.1103/PhysRevD.79.063523}}.

\bibitem{vogeley96}
M.~S. {Vogeley}, A.~S. {Szalay}, {Eigenmode Analysis of Galaxy Redshift
  Surveys. I. Theory and Methods}, ApJ 465 (1996) 34--+.
\newblock \href {http://arxiv.org/abs/arXiv:astro-ph/9601185}
  {\path{arXiv:arXiv:astro-ph/9601185}}, \href
  {http://dx.doi.org/10.1086/177399} {\path{doi:10.1086/177399}}.

\bibitem{tegmark98b}
M.~{Tegmark}, A.~J.~S. {Hamilton}, M.~A. {Strauss}, M.~S. {Vogeley}, A.~S.
  {Szalay}, {Measuring the Galaxy Power Spectrum with Future Redshift Surveys},
  ApJ 499 (1998) 555--+.
\newblock \href {http://arxiv.org/abs/arXiv:astro-ph/9708020}
  {\path{arXiv:arXiv:astro-ph/9708020}}, \href
  {http://dx.doi.org/10.1086/305663} {\path{doi:10.1086/305663}}.

\bibitem{ho08}
S.~{Ho}, C.~{Hirata}, N.~{Padmanabhan}, U.~{Seljak}, N.~{Bahcall}, {Correlation
  of CMB with large-scale structure. I. Integrated Sachs-Wolfe tomography and
  cosmological implications}, Physical Review D 78~(4) (2008) 043519.
\newblock \href {http://arxiv.org/abs/0801.0642} {\path{arXiv:0801.0642}},
  \href {http://dx.doi.org/10.1103/PhysRevD.78.043519}
  {\path{doi:10.1103/PhysRevD.78.043519}}.

\bibitem{ross11}
A.~J. {Ross}, S.~{Ho}, A.~J. {Cuesta}, R.~{Tojeiro}, W.~J. {Percival},
  D.~{Wake}, et~al., {Ameliorating systematic uncertainties in the angular
  clustering of galaxies: a study using the SDSS-III}, MNRAS 417 (2011)
  1350--1373.
\newblock \href {http://arxiv.org/abs/1105.2320} {\path{arXiv:1105.2320}},
  \href {http://dx.doi.org/10.1111/j.1365-2966.2011.19351.x}
  {\path{doi:10.1111/j.1365-2966.2011.19351.x}}.

\bibitem{ross12}
A.~J. {Ross}, W.~J. {Percival}, A.~G. {S{\'a}nchez}, L.~{Samushia}, S.~{Ho},
  E.~{Kazin}, et~al., {The clustering of galaxies in the SDSS-III Baryon
  Oscillation Spectroscopic Survey: analysis of potential systematics}, MNRAS
  424 (2012) 564--590.
\newblock \href {http://arxiv.org/abs/1203.6499} {\path{arXiv:1203.6499}},
  \href {http://dx.doi.org/10.1111/j.1365-2966.2012.21235.x}
  {\path{doi:10.1111/j.1365-2966.2012.21235.x}}.

\bibitem{tseliakhovich10}
D.~{Tseliakhovich}, C.~{Hirata}, {Relative velocity of dark matter and baryonic
  fluids and the formation of the first structures}, Physical Review D 82~(8)
  (2010) 083520.
\newblock \href {http://arxiv.org/abs/1005.2416} {\path{arXiv:1005.2416}},
  \href {http://dx.doi.org/10.1103/PhysRevD.82.083520}
  {\path{doi:10.1103/PhysRevD.82.083520}}.

\bibitem{Yoo11}
J.~{Yoo}, N.~{Dalal}, U.~{Seljak}, {Supersonic relative velocity effect on the
  baryonic acoustic oscillation measurements}, JCAP 7 (2011) 18.
\newblock \href {http://arxiv.org/abs/1105.3732} {\path{arXiv:1105.3732}},
  \href {http://dx.doi.org/10.1088/1475-7516/2011/07/018}
  {\path{doi:10.1088/1475-7516/2011/07/018}}.

\bibitem{peterson06}
J.~B. {Peterson}, K.~{Bandura}, U.~L. {Pen}, {The Hubble Sphere Hydrogen
  Survey}, arXiv:astro-ph/0606104\href
  {http://arxiv.org/abs/arXiv:astro-ph/0606104}
  {\path{arXiv:arXiv:astro-ph/0606104}}.

\bibitem{tegmark10}
M.~{Tegmark}, M.~{Zaldarriaga}, {Omniscopes: Large area telescope arrays with
  only NlogN computational cost}, Physical Review D 82~(10) (2010) 103501--+.
\newblock \href {http://arxiv.org/abs/0909.0001} {\path{arXiv:0909.0001}},
  \href {http://dx.doi.org/10.1103/PhysRevD.82.103501}
  {\path{doi:10.1103/PhysRevD.82.103501}}.

\bibitem{pober1210.2413}
J.~C. {Pober}, A.~R. {Parsons}, D.~R. {DeBoer}, P.~{McDonald}, M.~{McQuinn},
  J.~E. {Aguirre}, Z.~{Ali}, et~al., {The Baryon Acoustic Oscillation Broadband
  and Broad-beam Array: Design Overview and Sensitivity Forecasts},
  Astronomical Journal 145 (2013) 65.
\newblock \href {http://arxiv.org/abs/1210.2413} {\path{arXiv:1210.2413}},
  \href {http://dx.doi.org/10.1088/0004-6256/145/3/65}
  {\path{doi:10.1088/0004-6256/145/3/65}}.

\bibitem{abdalla05}
F.~B. {Abdalla}, S.~{Rawlings}, {Probing dark energy with baryonic oscillations
  and future radio surveys of neutral hydrogen}, MNRAS 360 (2005) 27--40.
\newblock \href {http://arxiv.org/abs/arXiv:astro-ph/0411342}
  {\path{arXiv:arXiv:astro-ph/0411342}}, \href
  {http://dx.doi.org/10.1111/j.1365-2966.2005.08650.x}
  {\path{doi:10.1111/j.1365-2966.2005.08650.x}}.

\bibitem{2013arXiv1309.5385H}
D.~{Huterer}, D.~{Kirkby}, R.~{Bean}, A.~{Connolly}, K.~{Dawson},
  S.~{Dodelson}, et~al., {Growth of Cosmic Structure: Probing Dark Energy
  Beyond Expansion}, ArXiv e-prints\href {http://arxiv.org/abs/1309.5385}
  {\path{arXiv:1309.5385}}.

\bibitem{2013arXiv1309.5383A}
K.~N. {Abazajian}, K.~{Arnold}, J.~{Austermann}, B.~A. {Benson}, C.~{Bischoff},
  J.~{Bock}, et~al., {Neutrino Physics from the Cosmic Microwave Background and
  Large Scale Structure}, ArXiv e-prints\href {http://arxiv.org/abs/1309.5383}
  {\path{arXiv:1309.5383}}.

\bibitem{2013arXiv1309.5381A}
K.~N. {Abazajian}, K.~{Arnold}, J.~{Austermann}, B.~A. {Benson}, C.~{Bischoff},
  J.~{Bock}, et~al., {Inflation Physics from the Cosmic Microwave Background
  and Large Scale Structure}, ArXiv e-prints\href
  {http://arxiv.org/abs/1309.5381} {\path{arXiv:1309.5381}}.

\bibitem{alcock79}
C.~Alcock, B.~Paczynski, {An evolution free test for non-zero cosmological
  constant}, {Nature} 281 (1979) 358--359.

\bibitem{ryden95}
B.~S. {Ryden}, {Measuring $q_0$ from the Distortion of Voids in Redshift
  Space}, ApJ 452 (1995) 25.
\newblock \href {http://arxiv.org/abs/arXiv:astro-ph/9506028}
  {\path{arXiv:arXiv:astro-ph/9506028}}, \href
  {http://dx.doi.org/10.1086/176277} {\path{doi:10.1086/176277}}.

\bibitem{ballinger96}
W.~E. Ballinger, J.~A. Peacock, A.~F. Heavens, {Measuring the cosmological
  constant with redshift surveys}, {MNRAS} 282 (1996) 877--888.
\newblock \href {http://arxiv.org/abs/astro-ph/9605017}
  {\path{arXiv:astro-ph/9605017}}.

\bibitem{matsubara96}
T.~Matsubara, Y.~Suto, {Cosmological redshift distortion of correlation
  functions as a probe of the density parameter and the cosmological constant},
  {ApJ} 470 (1996) L1--L5.
\newblock \href {http://arxiv.org/abs/astro-ph/9604142}
  {\path{arXiv:astro-ph/9604142}}, \href {http://dx.doi.org/10.1086/310290}
  {\path{doi:10.1086/310290}}.

\bibitem{popowski98}
P.~A. {Popowski}, D.~H. {Weinberg}, B.~S. {Ryden}, P.~S. {Osmer}, {Quasar
  Clustering and Spacetime Geometry}, ApJ 498 (1998) 11.
\newblock \href {http://arxiv.org/abs/arXiv:astro-ph/9707175}
  {\path{arXiv:arXiv:astro-ph/9707175}}, \href
  {http://dx.doi.org/10.1086/305528} {\path{doi:10.1086/305528}}.

\bibitem{hui99}
L.~Hui, A.~Stebbins, S.~Burles, {A Geometrical Test of the Cosmological Energy
  Contents Using the Lyman-alpha Forest}, {ApJ} 511 (1999) L5--9.
\newblock \href {http://arxiv.org/abs/astro-ph/9807190}
  {\path{arXiv:astro-ph/9807190}}, \href {http://dx.doi.org/10.1086/311826}
  {\path{doi:10.1086/311826}}.

\bibitem{mcdonald99}
P.~{McDonald}, J.~{Miralda-Escud{\'e}}, {Measuring the Cosmological Geometry
  from the Lyman-alpha Forest along Parallel Lines of Sight}, ApJ 518 (1999)
  24--31.
\newblock \href {http://arxiv.org/abs/arXiv:astro-ph/9807137}
  {\path{arXiv:arXiv:astro-ph/9807137}}, \href
  {http://dx.doi.org/10.1086/307264} {\path{doi:10.1086/307264}}.

\bibitem{matsubara01}
T.~{Matsubara}, A.~S. {Szalay}, {Constraining the Cosmological Constant from
  Large-Scale Redshift-Space Clustering}, ApJL 556 (2001) L67--L70.
\newblock \href {http://arxiv.org/abs/arXiv:astro-ph/0105493}
  {\path{arXiv:arXiv:astro-ph/0105493}}, \href
  {http://dx.doi.org/10.1086/322268} {\path{doi:10.1086/322268}}.

\bibitem{lavaux10}
G.~Lavaux, B.~D. Wandelt, {Precision cosmology with voids: definition, methods,
  dynamics}, {MNRAS} 403 (2010) 1392--1408.
\newblock \href {http://arxiv.org/abs/0906.4101} {\path{arXiv:0906.4101}},
  \href {http://dx.doi.org/10.1111/j.1365-2966.2010.16197.x}
  {\path{doi:10.1111/j.1365-2966.2010.16197.x}}.

\bibitem{sutter12}
P.~M. {Sutter}, G.~{Lavaux}, B.~D. {Wandelt}, D.~H. {Weinberg}, {A First
  Application of the Alcock-Paczynski Test to Stacked Cosmic Voids}, ApJ 761
  (2012) 187.
\newblock \href {http://arxiv.org/abs/1208.1058} {\path{arXiv:1208.1058}},
  \href {http://dx.doi.org/10.1088/0004-637X/761/2/187}
  {\path{doi:10.1088/0004-637X/761/2/187}}.

\bibitem{matsubara04}
T.~Matsubara, {Correlation Function in Deep Redshift Space as a Cosmological
  Probe}, {ApJ} 615 (2004) 573--585.
\newblock \href {http://arxiv.org/abs/astro-ph/0408349}
  {\path{arXiv:astro-ph/0408349}}, \href {http://dx.doi.org/10.1086/424561}
  {\path{doi:10.1086/424561}}.

\bibitem{2003PhRvL..90i1301L}
E.~V. {Linder}, {Exploring the Expansion History of the Universe}, Physical
  Review Letters 90~(9) (2003) 091301.
\newblock \href {http://arxiv.org/abs/arXiv:astro-ph/0208512}
  {\path{arXiv:arXiv:astro-ph/0208512}}, \href
  {http://dx.doi.org/10.1103/PhysRevLett.90.091301}
  {\path{doi:10.1103/PhysRevLett.90.091301}}.

\bibitem{2008JCAP...10..042D}
R.~{de Putter}, E.~V. {Linder}, {Calibrating dark energy}, JCAP 10 (2008) 42.
\newblock \href {http://arxiv.org/abs/0808.0189} {\path{arXiv:0808.0189}},
  \href {http://dx.doi.org/10.1088/1475-7516/2008/10/042}
  {\path{doi:10.1088/1475-7516/2008/10/042}}.

\bibitem{Riess:1998cb}
A.~G. {Riess}, A.~V. {Filippenko}, P.~{Challis}, A.~{Clocchiatti},
  A.~{Diercks}, P.~M. {Garnavich}, R.~L. {Gilliland}, et~al., {Observational
  Evidence from Supernovae for an Accelerating Universe and a Cosmological
  Constant}, AJ 116 (1998) 1009--1038.
\newblock \href {http://arxiv.org/abs/arXiv:astro-ph/9805201}
  {\path{arXiv:arXiv:astro-ph/9805201}}, \href
  {http://dx.doi.org/10.1086/300499} {\path{doi:10.1086/300499}}.

\bibitem{Perlmutter:1998np}
S.~{Perlmutter}, G.~{Aldering}, G.~{Goldhaber}, R.~A. {Knop}, P.~{Nugent},
  P.~G. {Castro}, S.~{Deustua}, et~al., {Measurements of Omega and Lambda from
  42 High-Redshift Supernovae}, \apj 517 (1999) 565--586.
\newblock \href {http://arxiv.org/abs/arXiv:astro-ph/9812133}
  {\path{arXiv:arXiv:astro-ph/9812133}}, \href
  {http://dx.doi.org/10.1086/307221} {\path{doi:10.1086/307221}}.

\bibitem{2012ApJ...746...85S}
N.~{Suzuki}, D.~{Rubin}, C.~{Lidman}, G.~{Aldering}, R.~{Amanullah},
  K.~{Barbary}, L.~F. {Barrientos}, et~al., {The Hubble Space Telescope Cluster
  Supernova Survey. V. Improving the Dark-energy Constraints above $z > 1$ and
  Building an Early-type-hosted Supernova Sample}, ApJ 746 (2012) 85.
\newblock \href {http://arxiv.org/abs/1105.3470} {\path{arXiv:1105.3470}},
  \href {http://dx.doi.org/10.1088/0004-637X/746/1/85}
  {\path{doi:10.1088/0004-637X/746/1/85}}.

\bibitem{2009ApJS..185...32K}
R.~{Kessler}, A.~C. {Becker}, D.~{Cinabro}, J.~{Vanderplas}, J.~A. {Frieman},
  J.~{Marriner}, T.~M. {Davis}, et~al., {First-Year Sloan Digital Sky Survey-II
  Supernova Results: Hubble Diagram and Cosmological Parameters}, ApJS 185
  (2009) 32--84.
\newblock \href {http://arxiv.org/abs/0908.4274} {\path{arXiv:0908.4274}},
  \href {http://dx.doi.org/10.1088/0067-0049/185/1/32}
  {\path{doi:10.1088/0067-0049/185/1/32}}.

\bibitem{2007ApJ...666..694W}
W.~M. {Wood-Vasey}, G.~{Miknaitis}, C.~W. {Stubbs}, S.~{Jha}, A.~G. {Riess},
  P.~M. {Garnavich}, R.~P. {Kirshner}, et~al., {Observational Constraints on
  the Nature of Dark Energy: First Cosmological Results from the ESSENCE
  Supernova Survey}, ApJ 666 (2007) 694--715.
\newblock \href {http://arxiv.org/abs/arXiv:astro-ph/0701041}
  {\path{arXiv:arXiv:astro-ph/0701041}}, \href
  {http://dx.doi.org/10.1086/518642} {\path{doi:10.1086/518642}}.

\bibitem{2011ApJ...737..102S}
M.~{Sullivan}, J.~{Guy}, A.~{Conley}, N.~{Regnault}, P.~{Astier}, C.~{Balland},
  S.~{Basa}, et~al., {SNLS3: Constraints on Dark Energy Combining the Supernova
  Legacy Survey Three-year Data with Other Probes}, ApJ 737 (2011) 102.
\newblock \href {http://arxiv.org/abs/1104.1444} {\path{arXiv:1104.1444}},
  \href {http://dx.doi.org/10.1088/0004-637X/737/2/102}
  {\path{doi:10.1088/0004-637X/737/2/102}}.

\bibitem{2013arXiv1307.0820S}
V.~{Salzano}, S.~A. {Rodney}, I.~{Sendra}, R.~{Lazkoz}, A.~G. {Riess},
  M.~{Postman}, T.~{Broadhurst}, D.~{Coe}, {Improving Dark Energy Constraints
  with High Redshift Type Ia Supernovae from CANDELS and CLASH}, ArXiv
  e-prints\href {http://arxiv.org/abs/1307.0820} {\path{arXiv:1307.0820}}.

\bibitem{2002SPIE.4836...61A}
G.~{Aldering}, G.~{Adam}, P.~{Antilogus}, P.~{Astier}, R.~{Bacon},
  S.~{Bongard}, C.~{Bonnaud}, et~al., {Overview of the Nearby Supernova
  Factory}, in: J.~A. {Tyson}, S.~{Wolff} (Eds.), Society of Photo-Optical
  Instrumentation Engineers (SPIE) Conference Series, Vol. 4836 of Society of
  Photo-Optical Instrumentation Engineers (SPIE) Conference Series, 2002, pp.
  61--72.
\newblock \href {http://dx.doi.org/10.1117/12.458107}
  {\path{doi:10.1117/12.458107}}.

\bibitem{2009PASP..121.1334R}
A.~{Rau}, S.~R. {Kulkarni}, N.~M. {Law}, J.~S. {Bloom}, D.~{Ciardi}, G.~S.
  {Djorgovski}, D.~B. {Fox}, et~al., {Exploring the Optical Transient Sky with
  the Palomar Transient Factory}, PASP 121 (2009) 1334--1351.
\newblock \href {http://arxiv.org/abs/0906.5355} {\path{arXiv:0906.5355}},
  \href {http://dx.doi.org/10.1086/605911} {\path{doi:10.1086/605911}}.

\bibitem{2012Msngr.150...34B}
C.~{Baltay}, D.~{Rabinowitz}, E.~{Hadjiyska}, M.~{Schwamb}, N.~{Ellman},
  R.~{Zinn}, S.~{Tourtellotte}, et~al., {The La Silla-QUEST Southern Hemisphere
  Variability Survey}, The Messenger 150 (2012) 34--38.

\bibitem{2012ApJ...753..152B}
J.~P. {Bernstein}, R.~{Kessler}, S.~{Kuhlmann}, R.~{Biswas}, E.~{Kovacs},
  G.~{Aldering}, I.~{Crane}, et~al., {Supernova Simulations and Strategies for
  the Dark Energy Survey}, \apj 753 (2012) 152.
\newblock \href {http://arxiv.org/abs/1111.1969} {\path{arXiv:1111.1969}},
  \href {http://dx.doi.org/10.1088/0004-637X/753/2/152}
  {\path{doi:10.1088/0004-637X/753/2/152}}.

\bibitem{2012arXiv1208.4012G}
J.~{Green}, P.~{Schechter}, C.~{Baltay}, R.~{Bean}, D.~{Bennett}, R.~{Brown},
  C.~{Conselice}, et~al., {Wide-Field InfraRed Survey Telescope (WFIRST) Final
  Report}, ArXiv e-prints\href {http://arxiv.org/abs/1208.4012}
  {\path{arXiv:1208.4012}}.

\bibitem{2008ApJ...674...51E}
R.~S. {Ellis}, M.~{Sullivan}, P.~E. {Nugent}, D.~A. {Howell}, A.~{Gal-Yam},
  P.~{Astier}, D.~{Balam}, et~al., {Verifying the Cosmological Utility of Type
  Ia Supernovae: Implications of a Dispersion in the Ultraviolet Spectra}, ApJ
  674 (2008) 51--69.
\newblock \href {http://arxiv.org/abs/0710.3896} {\path{arXiv:0710.3896}},
  \href {http://dx.doi.org/10.1086/524981} {\path{doi:10.1086/524981}}.

\bibitem{2012MNRAS.426.2359M}
K.~{Maguire}, M.~{Sullivan}, R.~S. {Ellis}, P.~E. {Nugent}, D.~A. {Howell},
  A.~{Gal-Yam}, J.~{Cooke}, et~al., {Hubble Space Telescope studies of
  low-redshift Type Ia supernovae: evolution with redshift and ultraviolet
  spectral trends}, MNRAS 426 (2012) 2359--2379.
\newblock \href {http://arxiv.org/abs/1205.7040} {\path{arXiv:1205.7040}},
  \href {http://dx.doi.org/10.1111/j.1365-2966.2012.21909.x}
  {\path{doi:10.1111/j.1365-2966.2012.21909.x}}.

\bibitem{2012AJ....143..113F}
R.~J. {Foley}, A.~V. {Filippenko}, R.~{Kessler}, B.~{Bassett}, J.~A. {Frieman},
  P.~M. {Garnavich}, S.~W. {Jha}, et~al., {A Mismatch in the Ultraviolet
  Spectra between Low-redshift and Intermediate-redshift Type Ia Supernovae as
  a Possible Systematic Uncertainty for Supernova Cosmology}, AJ 143 (2012)
  113.
\newblock \href {http://arxiv.org/abs/1010.2749} {\path{arXiv:1010.2749}},
  \href {http://dx.doi.org/10.1088/0004-6256/143/5/113}
  {\path{doi:10.1088/0004-6256/143/5/113}}.

\bibitem{2008ApJ...689..377W}
W.~M. {Wood-Vasey}, A.~S. {Friedman}, J.~S. {Bloom}, M.~{Hicken}, M.~{Modjaz},
  R.~P. {Kirshner}, D.~L. {Starr}, et~al., {Type Ia Supernovae Are Good
  Standard Candles in the Near Infrared: Evidence from PAIRITEL}, ApJ 689
  (2008) 377--390.
\newblock \href {http://arxiv.org/abs/0711.2068} {\path{arXiv:0711.2068}},
  \href {http://dx.doi.org/10.1086/592374} {\path{doi:10.1086/592374}}.

\bibitem{2010AJ....139..120F}
G.~{Folatelli}, M.~M. {Phillips}, C.~R. {Burns}, C.~{Contreras}, M.~{Hamuy},
  W.~L. {Freedman}, S.~E. {Persson}, M.~{Stritzinger}, N.~B. {Suntzeff},
  K.~{Krisciunas}, L.~{Boldt}, S.~{Gonz{\'a}lez}, W.~{Krzeminski},
  N.~{Morrell}, M.~{Roth}, F.~{Salgado}, B.~F. {Madore}, D.~{Murphy},
  P.~{Wyatt}, W.~{Li}, A.~V. {Filippenko}, N.~{Miller}, {The Carnegie Supernova
  Project: Analysis of the First Sample of Low-Redshift Type-Ia Supernovae}, AJ
  139 (2010) 120--144.
\newblock \href {http://arxiv.org/abs/0910.3317} {\path{arXiv:0910.3317}},
  \href {http://dx.doi.org/10.1088/0004-6256/139/1/120}
  {\path{doi:10.1088/0004-6256/139/1/120}}.

\bibitem{2012MNRAS.425.1007B}
R.~L. {Barone-Nugent}, C.~{Lidman}, J.~S.~B. {Wyithe}, J.~{Mould}, D.~A.
  {Howell}, I.~M. {Hook}, M.~{Sullivan}, et~al., {Near-infrared observations of
  Type Ia supernovae: the best known standard candle for cosmology}, MNRAS 425
  (2012) 1007--1012.
\newblock \href {http://arxiv.org/abs/1204.2308} {\path{arXiv:1204.2308}},
  \href {http://dx.doi.org/10.1111/j.1365-2966.2012.21412.x}
  {\path{doi:10.1111/j.1365-2966.2012.21412.x}}.

\bibitem{2012AJ....143..126B}
S.~{Blondin}, T.~{Matheson}, R.~P. {Kirshner}, K.~S. {Mandel}, P.~{Berlind},
  M.~{Calkins}, P.~{Challis}, et~al., {The Spectroscopic Diversity of Type Ia
  Supernovae}, AJ 143 (2012) 126.
\newblock \href {http://arxiv.org/abs/1203.4832} {\path{arXiv:1203.4832}},
  \href {http://dx.doi.org/10.1088/0004-6256/143/5/126}
  {\path{doi:10.1088/0004-6256/143/5/126}}.

\bibitem{2012MNRAS.425.1789S}
J.~M. {Silverman}, R.~J. {Foley}, A.~V. {Filippenko}, M.~{Ganeshalingam}, A.~J.
  {Barth}, R.~{Chornock}, C.~V. {Griffith}, et~al., {Berkeley Supernova Ia
  Program - I. Observations, data reduction and spectroscopic sample of 582
  low-redshift Type Ia supernovae}, MNRAS 425 (2012) 1789--1818.
\newblock \href {http://arxiv.org/abs/1202.2128} {\path{arXiv:1202.2128}},
  \href {http://dx.doi.org/10.1111/j.1365-2966.2012.21270.x}
  {\path{doi:10.1111/j.1365-2966.2012.21270.x}}.

\bibitem{2013arXiv1305.6997F}
G.~{Folatelli}, N.~{Morrell}, M.~M. {Phillips}, E.~{Hsiao}, A.~{Campillay},
  C.~{Contreras}, S.~{Castell{\'o}n}, et~al., {Spectroscopy of Type Ia
  Supernovae by the Carnegie Supernova Project}, ArXiv e-prints\href
  {http://arxiv.org/abs/1305.6997} {\path{arXiv:1305.6997}}.

\bibitem{2007A&A...466...11G}
J.~{Guy}, P.~{Astier}, S.~{Baumont}, D.~{Hardin}, R.~{Pain}, N.~{Regnault},
  S.~{Basa}, et~al., {SALT2: using distant supernovae to improve the use of
  type Ia supernovae as distance indicators}, A\&A 466 (2007) 11--21.
\newblock \href {http://arxiv.org/abs/arXiv:astro-ph/0701828}
  {\path{arXiv:arXiv:astro-ph/0701828}}, \href
  {http://dx.doi.org/10.1051/0004-6361:20066930}
  {\path{doi:10.1051/0004-6361:20066930}}.

\bibitem{2007ApJ...659..122J}
S.~{Jha}, A.~G. {Riess}, R.~P. {Kirshner}, {Improved Distances to Type Ia
  Supernovae with Multicolor Light-Curve Shapes: MLCS2k2}, ApJ 659 (2007)
  122--148.
\newblock \href {http://arxiv.org/abs/arXiv:astro-ph/0612666}
  {\path{arXiv:arXiv:astro-ph/0612666}}, \href
  {http://dx.doi.org/10.1086/512054} {\path{doi:10.1086/512054}}.

\bibitem{2011ApJ...731..120M}
K.~S. {Mandel}, G.~{Narayan}, R.~P. {Kirshner}, {Type Ia Supernova Light Curve
  Inference: Hierarchical Models in the Optical and Near-infrared}, ApJ 731
  (2011) 120.
\newblock \href {http://arxiv.org/abs/1011.5910} {\path{arXiv:1011.5910}},
  \href {http://dx.doi.org/10.1088/0004-637X/731/2/120}
  {\path{doi:10.1088/0004-637X/731/2/120}}.

\bibitem{2013ApJ...766...84K}
A.~G. {Kim}, R.~C. {Thomas}, G.~{Aldering}, P.~{Antilogus}, C.~{Aragon},
  S.~{Bailey}, C.~{Baltay}, et~al., {Standardizing Type Ia Supernova Absolute
  Magnitudes Using Gaussian Process Data Regression}, ApJ 766 (2013) 84.
\newblock \href {http://arxiv.org/abs/1302.2925} {\path{arXiv:1302.2925}},
  \href {http://dx.doi.org/10.1088/0004-637X/766/2/84}
  {\path{doi:10.1088/0004-637X/766/2/84}}.

\bibitem{2011ApJ...742...89F}
R.~J. {Foley}, N.~E. {Sanders}, R.~P. {Kirshner}, {Velocity Evolution and the
  Intrinsic Color of Type Ia Supernovae}, ApJ 742 (2011) 89.
\newblock \href {http://arxiv.org/abs/1107.3555} {\path{arXiv:1107.3555}},
  \href {http://dx.doi.org/10.1088/0004-637X/742/2/89}
  {\path{doi:10.1088/0004-637X/742/2/89}}.

\bibitem{2011MNRAS.413.3075M}
K.~{Maeda}, G.~{Leloudas}, S.~{Taubenberger}, M.~{Stritzinger}, J.~{Sollerman},
  N.~{Elias-Rosa}, S.~{Benetti}, et~al., {Effects of the explosion asymmetry
  and viewing angle on the Type Ia supernova colour and luminosity
  calibration}, MNRAS 413 (2011) 3075--3094.
\newblock \href {http://arxiv.org/abs/1101.3935} {\path{arXiv:1101.3935}},
  \href {http://dx.doi.org/10.1111/j.1365-2966.2011.18381.x}
  {\path{doi:10.1111/j.1365-2966.2011.18381.x}}.

\bibitem{2009A&A...500L..17B}
S.~{Bailey}, G.~{Aldering}, P.~{Antilogus}, C.~{Aragon}, C.~{Baltay},
  S.~{Bongard}, C.~{Buton}, et~al., {Using spectral flux ratios to standardize
  SN Ia luminosities}, A\&A 500 (2009) L17--L20.
\newblock \href {http://arxiv.org/abs/0905.0340} {\path{arXiv:0905.0340}},
  \href {http://dx.doi.org/10.1051/0004-6361/200911973}
  {\path{doi:10.1051/0004-6361/200911973}}.

\bibitem{2011ApJ...729...55F}
R.~J. {Foley}, D.~{Kasen}, {Measuring Ejecta Velocity Improves Type Ia
  Supernova Distances}, ApJ 729 (2011) 55.
\newblock \href {http://arxiv.org/abs/1011.4517} {\path{arXiv:1011.4517}},
  \href {http://dx.doi.org/10.1088/0004-637X/729/1/55}
  {\path{doi:10.1088/0004-637X/729/1/55}}.

\bibitem{2012MNRAS.425.1889S}
J.~M. {Silverman}, M.~{Ganeshalingam}, W.~{Li}, A.~V. {Filippenko}, {Berkeley
  Supernova Ia Program - III. Spectra near maximum brightness improve the
  accuracy of derived distances to Type Ia supernovae}, MNRAS 425 (2012)
  1889--1916.
\newblock \href {http://arxiv.org/abs/1202.2130} {\path{arXiv:1202.2130}},
  \href {http://dx.doi.org/10.1111/j.1365-2966.2012.21526.x}
  {\path{doi:10.1111/j.1365-2966.2012.21526.x}}.

\bibitem{2011A&A...529L...4C}
N.~{Chotard}, E.~{Gangler}, G.~{Aldering}, P.~{Antilogus}, C.~{Aragon},
  S.~{Bailey}, C.~{Baltay}, et~al., {The reddening law of type Ia supernovae:
  separating intrinsic variability from dust using equivalent widths}, A\&A 529
  (2011) L4.
\newblock \href {http://arxiv.org/abs/1103.5300} {\path{arXiv:1103.5300}},
  \href {http://dx.doi.org/10.1051/0004-6361/201116723}
  {\path{doi:10.1051/0004-6361/201116723}}.

\bibitem{Hannah}
H.~Fakhouri, {Supernova Ia Spectra and Spectrophotometric Time Series:
  Recognizing Twins and the Consequences for Cosmological Distance
  Measurements}, Ph.D. thesis, University of California, Berkeley (2013).

\bibitem{2012arXiv1211.0310L}
{LSST Dark Energy Science Collaboration}, {Large Synoptic Survey Telescope:
  Dark Energy Science Collaboration}, ArXiv e-prints\href
  {http://arxiv.org/abs/1211.0310} {\path{arXiv:1211.0310}}.

\bibitem{2011ApJS..192....1C}
A.~{Conley}, J.~{Guy}, M.~{Sullivan}, N.~{Regnault}, P.~{Astier}, C.~{Balland},
  S.~{Basa}, et~al., {Supernova Constraints and Systematic Uncertainties from
  the First Three Years of the Supernova Legacy Survey}, ApJS 192 (2011) 1.
\newblock \href {http://arxiv.org/abs/1104.1443} {\path{arXiv:1104.1443}},
  \href {http://dx.doi.org/10.1088/0067-0049/192/1/1}
  {\path{doi:10.1088/0067-0049/192/1/1}}.

\bibitem{Stubbs10}
C.~W. {Stubbs}, P.~{Doherty}, C.~{Cramer}, G.~{Narayan}, Y.~J. {Brown}, K.~R.
  {Lykke}, J.~T. {Woodward}, J.~L. {Tonry}, {Precise Throughput Determination
  of the PanSTARRS Telescope and the Gigapixel Imager Using a Calibrated
  Silicon Photodiode and a Tunable Laser: Initial Results}, \apjs 191 (2010)
  376--388.
\newblock \href {http://arxiv.org/abs/1003.3465} {\path{arXiv:1003.3465}},
  \href {http://dx.doi.org/10.1088/0067-0049/191/2/376}
  {\path{doi:10.1088/0067-0049/191/2/376}}.

\bibitem{2013A&A...552A.124B}
M.~{Betoule}, J.~{Marriner}, N.~{Regnault}, J.-C. {Cuillandre}, P.~{Astier},
  J.~{Guy}, C.~{Balland}, et~al., {Improved photometric calibration of the SNLS
  and the SDSS supernova surveys}, A\&A 552 (2013) A124.
\newblock \href {http://arxiv.org/abs/1212.4864} {\path{arXiv:1212.4864}},
  \href {http://dx.doi.org/10.1051/0004-6361/201220610}
  {\path{doi:10.1051/0004-6361/201220610}}.

\bibitem{Burke10}
D.~L. {Burke}, T.~{Axelrod}, S.~{Blondin}, C.~{Claver}, {\v Z}.~{Ivezi{\'c}},
  L.~{Jones}, A.~{Saha}, et~al., {Precision Determination of Atmospheric
  Extinction at Optical and Near-infrared Wavelengths}, \apj 720 (2010)
  811--823.
\newblock \href {http://dx.doi.org/10.1088/0004-637X/720/1/811}
  {\path{doi:10.1088/0004-637X/720/1/811}}.

\bibitem{Stubbs12}
C.~W. {Stubbs}, J.~L. {Tonry}, {Addressing the Photometric Calibration
  Challenge: Explicit Determination of the Instrumental Response and
  Atmospheric Response Functions, and Tying it All Together}, ArXiv
  e-prints\href {http://arxiv.org/abs/1206.6695} {\path{arXiv:1206.6695}}.

\bibitem{Kent09}
S.~{Kent}, M.~B. {Kaiser}, S.~E. {Deustua}, J.~A. {Smith}, S.~{Adelman},
  S.~{Allam}, B.~{Baptista}, et~al., {Photometric Calibrations for 21st Century
  Science} 2010 (2009) 155.
\newblock \href {http://arxiv.org/abs/0903.2799} {\path{arXiv:0903.2799}}.

\bibitem{2007ASPC..364..361K}
M.~E. {Kaiser}, J.~W. {Kruk}, S.~R. {McCandliss}, D.~J. {Sahnow}, W.~V.
  {Dixon}, R.~C. {Bohlin}, S.~E. {Deustua}, {ACCESS -- Absolute Color
  Calibration Experiment for Standard Stars}, in: C.~{Sterken} (Ed.), The
  Future of Photometric, Spectrophotometric and Polarimetric Standardization,
  Vol. 364 of Astronomical Society of the Pacific Conference Series, 2007, p.
  361.

\bibitem{2009Metro..46S.219S}
A.~W. {Smith}, J.~T. {Woodward}, C.~A. {Jenkins}, S.~W. {Brown}, K.~R. {Lykke},
  {Absolute flux calibration of stars: calibration of the reference telescope},
  Metrologia 46 (2009) 219.
\newblock \href {http://dx.doi.org/10.1088/0026-1394/46/4/S16}
  {\path{doi:10.1088/0026-1394/46/4/S16}}.

\bibitem{Saha12}
A.~{Saha}, S.~E. {Deustua}, R.~C. {Bohlin}, A.~{Rest}, T.~{Axelrod}, R.~L.
  {Gilliland}, J.~B. {Holberg}, et~al.,
  \href{http://www.stsci.edu/cgi-bin/get-proposal-info?id=12967&observatory=HST}{{Establishing
  a Network of DA White Dwarf SED Standards}} (2012).
\newline\urlprefix\url{http://www.stsci.edu/cgi-bin/get-proposal-info?id=12967&observatory=HST}

\bibitem{2010ApJ...717...40K}
R.~{Kessler}, D.~{Cinabro}, B.~{Bassett}, B.~{Dilday}, J.~A. {Frieman}, P.~M.
  {Garnavich}, S.~{Jha}, et~al., {Photometric Estimates of Redshifts and
  Distance Moduli for Type Ia Supernovae}, ApJ 717 (2010) 40--57.
\newblock \href {http://arxiv.org/abs/1001.0738} {\path{arXiv:1001.0738}},
  \href {http://dx.doi.org/10.1088/0004-637X/717/1/40}
  {\path{doi:10.1088/0004-637X/717/1/40}}.

\bibitem{2011ApJ...738..162S}
M.~{Sako}, B.~{Bassett}, B.~{Connolly}, B.~{Dilday}, H.~{Cambell}, J.~A.
  {Frieman}, L.~{Gladney}, et~al., {Photometric Type Ia Supernova Candidates
  from the Three-year SDSS-II SN Survey Data}, ApJ 738 (2011) 162.
\newblock \href {http://arxiv.org/abs/1107.5106} {\path{arXiv:1107.5106}},
  \href {http://dx.doi.org/10.1088/0004-637X/738/2/162}
  {\path{doi:10.1088/0004-637X/738/2/162}}.

\bibitem{Childress13}
M.~{Childress}, G.~{Aldering}, P.~{Antilogus}, C.~{Aragon}, S.~{Bailey},
  C.~{Baltay}, S.~{Bongard}, et~al., {Host Galaxy Properties and Hubble
  Residuals of Type Ia Supernovae from the Nearby Supernova Factory}, \apj 770
  (2013) 108.
\newblock \href {http://arxiv.org/abs/1304.4720} {\path{arXiv:1304.4720}},
  \href {http://dx.doi.org/10.1088/0004-637X/770/2/108}
  {\path{doi:10.1088/0004-637X/770/2/108}}.

\bibitem{2013arXiv1307.6031H}
L.~{Humphreys}, M.~{Reid}, J.~{Moran}, L.~{Greenhill}, A.~{Argon}, {Toward a
  New Geometric Distance to the Active Galaxy NGC 4258. III. Final Results and
  the Hubble Constant}, ArXiv e-prints\href {http://arxiv.org/abs/1307.6031}
  {\path{arXiv:1307.6031}}.

\bibitem{2011ApJ...730..119R}
A.~G. {Riess}, L.~{Macri}, S.~{Casertano}, H.~{Lampeitl}, H.~C. {Ferguson},
  A.~V. {Filippenko}, S.~W. {Jha}, et~al., {A 3\% Solution: Determination of
  the Hubble Constant with the Hubble Space Telescope and Wide Field Camera 3},
  \apj 730 (2011) 119.
\newblock \href {http://arxiv.org/abs/1103.2976} {\path{arXiv:1103.2976}},
  \href {http://dx.doi.org/10.1088/0004-637X/730/2/119}
  {\path{doi:10.1088/0004-637X/730/2/119}}.

\bibitem{2011AJ....142..192F}
W.~L. {Freedman}, B.~F. {Madore}, V.~{Scowcroft}, A.~{Monson}, S.~E. {Persson},
  M.~{Seibert}, J.~R. {Rigby}, L.~{Sturch}, P.~{Stetson}, {The Carnegie Hubble
  Program}, AJ 142 (2011) 192.
\newblock \href {http://arxiv.org/abs/1109.3802} {\path{arXiv:1109.3802}},
  \href {http://dx.doi.org/10.1088/0004-6256/142/6/192}
  {\path{doi:10.1088/0004-6256/142/6/192}}.

\bibitem{2008MNRAS.386...47E}
S.~C. {Ellis}, J.~{Bland-Hawthorn}, {The case for OH suppression at
  near-infrared wavelengths}, MNRAS 386 (2008) 47--64.
\newblock \href {http://arxiv.org/abs/0801.3870} {\path{arXiv:0801.3870}},
  \href {http://dx.doi.org/10.1111/j.1365-2966.2008.13021.x}
  {\path{doi:10.1111/j.1365-2966.2008.13021.x}}.

\bibitem{2012MNRAS.425.1682E}
S.~C. {Ellis}, J.~{Bland-Hawthorn}, J.~{Lawrence}, A.~J. {Horton}, C.~{Trinh},
  S.~G. {Leon-Saval}, K.~{Shortridge}, et~al., {Suppression of the
  near-infrared OH night-sky lines with fibre Bragg gratings - first results},
  MNRAS 425 (2012) 1682--1695.
\newblock \href {http://arxiv.org/abs/1206.6551} {\path{arXiv:1206.6551}},
  \href {http://dx.doi.org/10.1111/j.1365-2966.2012.21602.x}
  {\path{doi:10.1111/j.1365-2966.2012.21602.x}}.

\bibitem{2010PASP..122.1415K}
R.~{Kessler}, B.~{Bassett}, P.~{Belov}, V.~{Bhatnagar}, H.~{Campbell},
  A.~{Conley}, J.~A. {Frieman}, A.~{Glazov}, S.~{Gonz{\'a}lez-Gait{\'a}n},
  R.~{Hlozek}, S.~{Jha}, S.~{Kuhlmann}, M.~{Kunz}, H.~{Lampeitl}, A.~{Mahabal},
  J.~{Newling}, R.~C. {Nichol}, D.~{Parkinson}, N.~S. {Philip}, D.~{Poznanski},
  J.~W. {Richards}, S.~A. {Rodney}, M.~{Sako}, D.~P. {Schneider}, M.~{Smith},
  M.~{Stritzinger}, M.~{Varughese}, {Results from the Supernova Photometric
  Classification Challenge}, PASP 122 (2010) 1415--1431.
\newblock \href {http://arxiv.org/abs/1008.1024} {\path{arXiv:1008.1024}},
  \href {http://dx.doi.org/10.1086/657607} {\path{doi:10.1086/657607}}.

\bibitem{2008ApJ...681.1448J}
G.~C. {Jordan}, IV, R.~T. {Fisher}, D.~M. {Townsley}, A.~C. {Calder},
  C.~{Graziani}, S.~{Asida}, D.~Q. {Lamb}, J.~W. {Truran}, {Three-Dimensional
  Simulations of the Deflagration Phase of the Gravitationally Confined
  Detonation Model of Type Ia Supernovae}, ApJ 681 (2008) 1448--1457.
\newblock \href {http://arxiv.org/abs/arXiv:astro-ph/0703573}
  {\path{arXiv:arXiv:astro-ph/0703573}}, \href
  {http://dx.doi.org/10.1086/588269} {\path{doi:10.1086/588269}}.

\bibitem{2009Natur.460..869K}
D.~{Kasen}, F.~K. {R{\"o}pke}, S.~E. {Woosley}, {The diversity of type Ia
  supernovae from broken symmetries}, Nature 460 (2009) 869--872.
\newblock \href {http://arxiv.org/abs/0907.0708} {\path{arXiv:0907.0708}},
  \href {http://dx.doi.org/10.1038/nature08256}
  {\path{doi:10.1038/nature08256}}.

\bibitem{2010Natur.463...61P}
R.~{Pakmor}, M.~{Kromer}, F.~K. {R{\"o}pke}, S.~A. {Sim}, A.~J. {Ruiter},
  W.~{Hillebrandt}, {Sub-luminous type Ia supernovae from the mergers of
  equal-mass white dwarfs with mass \~{}0.9M$_{solar}$}, Nature 463 (2010)
  61--64.
\newblock \href {http://arxiv.org/abs/0911.0926} {\path{arXiv:0911.0926}},
  \href {http://dx.doi.org/10.1038/nature08642}
  {\path{doi:10.1038/nature08642}}.

\bibitem{2012JPhCS.402a2023C}
A.~C. {Calder}, B.~K. {Krueger}, A.~P. {Jackson}, D.~M. {Townsley}, E.~F.
  {Brown}, F.~X. {Timmes}, {On Simulating Type Ia Supernovae}, Journal of
  Physics Conference Series 402~(1) (2012) 012023.
\newblock \href {http://arxiv.org/abs/1205.0966} {\path{arXiv:1205.0966}},
  \href {http://dx.doi.org/10.1088/1742-6596/402/1/012023}
  {\path{doi:10.1088/1742-6596/402/1/012023}}.

\bibitem{2013FrPhy...8..116H}
W.~{Hillebrandt}, M.~{Kromer}, F.~K. {R{\"o}pke}, A.~J. {Ruiter}, {Towards an
  understanding of Type Ia supernovae from a synthesis of theory and
  observations}, Frontiers of Physics 8 (2013) 116--143.
\newblock \href {http://arxiv.org/abs/1302.6420} {\path{arXiv:1302.6420}},
  \href {http://dx.doi.org/10.1007/s11467-013-0303-2}
  {\path{doi:10.1007/s11467-013-0303-2}}.

\bibitem{2011ApJ...734...38W}
S.~E. {Woosley}, D.~{Kasen}, {Sub-Chandrasekhar Mass Models for Supernovae},
  \apj 734 (2011) 38.
\newblock \href {http://arxiv.org/abs/1010.5292} {\path{arXiv:1010.5292}},
  \href {http://dx.doi.org/10.1088/0004-637X/734/1/38}
  {\path{doi:10.1088/0004-637X/734/1/38}}.

\bibitem{2010MNRAS.406..782S}
M.~{Sullivan}, A.~{Conley}, D.~A. {Howell}, J.~D. {Neill}, P.~{Astier},
  C.~{Balland}, S.~{Basa}, et~al., {The dependence of Type Ia Supernovae
  luminosities on their host galaxies}, MNRAS 406 (2010) 782--802.
\newblock \href {http://arxiv.org/abs/1003.5119} {\path{arXiv:1003.5119}},
  \href {http://dx.doi.org/10.1111/j.1365-2966.2010.16731.x}
  {\path{doi:10.1111/j.1365-2966.2010.16731.x}}.

\bibitem{2011MNRAS.417.1280B}
S.~{Blondin}, D.~{Kasen}, F.~K. {R{\"o}pke}, R.~P. {Kirshner}, K.~S. {Mandel},
  {Confronting 2D delayed-detonation models with light curves and spectra of
  Type Ia supernovae}, MNRAS 417 (2011) 1280--1302.
\newblock \href {http://arxiv.org/abs/1107.0009} {\path{arXiv:1107.0009}},
  \href {http://dx.doi.org/10.1111/j.1365-2966.2011.19345.x}
  {\path{doi:10.1111/j.1365-2966.2011.19345.x}}.

\bibitem{2013arXiv1303.1168D}
B.~{Diemer}, R.~{Kessler}, C.~{Graziani}, G.~C. {Jordan}, IV, D.~Q. {Lamb},
  M.~{Long}, D.~R. {van Rossum}, {Comparing the light curves of simulated Type
  Ia Supernovae with observations using data-driven models}, ArXiv
  e-prints\href {http://arxiv.org/abs/1303.1168} {\path{arXiv:1303.1168}}.

\bibitem{2013ApJ...764...48K}
R.~{Kessler}, J.~{Guy}, J.~{Marriner}, M.~{Betoule}, J.~{Brinkmann},
  D.~{Cinabro}, P.~{El-Hage}, et~al., {Testing Models of Intrinsic Brightness
  Variations in Type Ia Supernovae and Their Impact on Measuring Cosmological
  Parameters}, ApJ 764 (2013) 48.
\newblock \href {http://arxiv.org/abs/1209.2482} {\path{arXiv:1209.2482}},
  \href {http://dx.doi.org/10.1088/0004-637X/764/1/48}
  {\path{doi:10.1088/0004-637X/764/1/48}}.

\bibitem{RSSB:RSSB294}
M.~C. Kennedy, A.~O'Hagan,
  \href{http://dx.doi.org/10.1111/1467-9868.00294}{Bayesian calibration of
  computer models}, Journal of the Royal Statistical Society: Series B
  (Statistical Methodology) 63~(3) (2001) 425--464.
\newblock \href {http://dx.doi.org/10.1111/1467-9868.00294}
  {\path{doi:10.1111/1467-9868.00294}}.
\newline\urlprefix\url{http://dx.doi.org/10.1111/1467-9868.00294}

\bibitem{sant:will:notz:2003}
T.~J. Santner, B.~Williams, W.~Notz, The Design and Analysis of Computer
  Experiments, Springer-Verlag, 2003.

\bibitem{2009ApJ...705..156H}
K.~{Heitmann}, D.~{Higdon}, M.~{White}, S.~{Habib}, B.~J. {Williams},
  E.~{Lawrence}, C.~{Wagner}, {The Coyote Universe. II. Cosmological Models and
  Precision Emulation of the Nonlinear Matter Power Spectrum}, ApJ 705 (2009)
  156--174.
\newblock \href {http://arxiv.org/abs/0902.0429} {\path{arXiv:0902.0429}},
  \href {http://dx.doi.org/10.1088/0004-637X/705/1/156}
  {\path{doi:10.1088/0004-637X/705/1/156}}.

\bibitem{Allen1103.4829}
S.~W. {Allen}, A.~E. {Evrard}, A.~B. {Mantz}, {Cosmological Parameters from
  Observations of Galaxy Clusters}, ARA\&A 49 (2011) 409--470.
\newblock \href {http://arxiv.org/abs/1103.4829} {\path{arXiv:1103.4829}},
  \href {http://dx.doi.org/10.1146/annurev-astro-081710-102514}
  {\path{doi:10.1146/annurev-astro-081710-102514}}.

\bibitem{Weinberg1201.2434}
D.~H. {Weinberg}, M.~J. {Mortonson}, D.~J. {Eisenstein}, C.~{Hirata}, A.~G.
  {Riess}, E.~{Rozo}, {Observational probes of cosmic acceleration}, PhR 530
  (2013) 87--255.
\newblock \href {http://arxiv.org/abs/1201.2434} {\path{arXiv:1201.2434}},
  \href {http://dx.doi.org/10.1016/j.physrep.2013.05.001}
  {\path{doi:10.1016/j.physrep.2013.05.001}}.

\bibitem{White93}
S.~D.~M. {White}, J.~F. {Navarro}, A.~E. {Evrard}, C.~S. {Frenk}, {The baryon
  content of galaxy clusters: a challenge to cosmological orthodoxy}, Nature
  366 (1993) 429--433.
\newblock \href {http://dx.doi.org/10.1038/366429a0}
  {\path{doi:10.1038/366429a0}}.

\bibitem{Allen0405340}
S.~W. {Allen}, R.~W. {Schmidt}, H.~{Ebeling}, A.~C. {Fabian}, L.~{van
  Speybroeck}, {Constraints on dark energy from Chandra observations of the
  largest relaxed galaxy clusters}, MNRAS 353 (2004) 457--467.
\newblock \href {http://arxiv.org/abs/arXiv:astro-ph/0405340}
  {\path{arXiv:arXiv:astro-ph/0405340}}, \href
  {http://dx.doi.org/10.1111/j.1365-2966.2004.08080.x}
  {\path{doi:10.1111/j.1365-2966.2004.08080.x}}.

\bibitem{Allen0706.0033}
S.~W. {Allen}, D.~A. {Rapetti}, R.~W. {Schmidt}, H.~{Ebeling}, R.~G. {Morris},
  A.~C. {Fabian}, {Improved constraints on dark energy from Chandra X-ray
  observations of the largest relaxed galaxy clusters}, MNRAS 383 (2008)
  879--896.
\newblock \href {http://arxiv.org/abs/arXiv:0706.0033}
  {\path{arXiv:arXiv:0706.0033}}, \href
  {http://dx.doi.org/10.1111/j.1365-2966.2007.12610.x}
  {\path{doi:10.1111/j.1365-2966.2007.12610.x}}.

\bibitem{Mantz13}
{Mantz et al.}, \emph{in preparation}.

\bibitem{Applegate13b}
{Applegate et al.}, \emph{in preparation}.

\bibitem{Planelles1209.5058}
S.~{Planelles}, S.~{Borgani}, K.~{Dolag}, S.~{Ettori}, D.~{Fabjan},
  G.~{Murante}, L.~{Tornatore}, {Baryon census in hydrodynamical simulations of
  galaxy clusters}, MNRAS 431 (2013) 1487--1502.
\newblock \href {http://arxiv.org/abs/1209.5058} {\path{arXiv:1209.5058}},
  \href {http://dx.doi.org/10.1093/mnras/stt265}
  {\path{doi:10.1093/mnras/stt265}}.

\bibitem{Battaglia1209.4082}
N.~{Battaglia}, J.~R. {Bond}, C.~{Pfrommer}, J.~L. {Sievers}, {On the Cluster
  Physics of Sunyaev-Zel'dovich and X-ray Surveys III: Measurement Biases and
  Cosmological Evolution of Gas and Stellar Mass Fractions}, ArXiv
  e-prints\href {http://arxiv.org/abs/1209.4082} {\path{arXiv:1209.4082}}.

\bibitem{2013arXiv1307.8152A}
S.~W. {Allen}, A.~B. {Mantz}, R.~G. {Morris}, D.~E. {Applegate}, P.~L. {Kelly},
  A.~{von der Linden}, D.~A. {Rapetti}, R.~W. {Schmidt}, {Measuring cosmic
  distances with galaxy clusters}, ArXiv e-prints\href
  {http://arxiv.org/abs/1307.8152} {\path{arXiv:1307.8152}}.

\bibitem{Nandra1306.2307}
K.~{Nandra}, D.~{Barret}, X.~{Barcons}, A.~{Fabian}, J.-W. {den Herder},
  L.~{Piro}, M.~{Watson}, et~al., {The Hot and Energetic Universe: A White
  Paper presenting the science theme motivating the Athena+ mission},
  arXiv:1306.2307\href {http://arxiv.org/abs/1306.2307}
  {\path{arXiv:1306.2307}}.

\bibitem{Sunyaev72}
R.~A. {Sunyaev}, Y.~B. {Zeldovich}, {The Observations of Relic Radiation as a
  Test of the Nature of X-Ray Radiation from the Clusters of Galaxies},
  Comments on Astrophysics and Space Physics 4 (1972) 173--178.

\bibitem{WhiteSilk78}
J.~{Silk}, S.~D.~M. {White}, {The determination of $Q_{0}$ using X-ray and
  microwave observations of galaxy clusters}, ApJL 226 (1978) L103--L106.
\newblock \href {http://dx.doi.org/10.1086/182841} {\path{doi:10.1086/182841}}.

\bibitem{Bonamente0512349}
M.~{Bonamente}, M.~K. {Joy}, S.~J. {LaRoque}, J.~E. {Carlstrom}, E.~D. {Reese},
  K.~S. {Dawson}, {Determination of the Cosmic Distance Scale from
  Sunyaev-Zel'dovich Effect and Chandra X-Ray Measurements of High-Redshift
  Galaxy Clusters}, \apj 647 (2006) 25--54.
\newblock \href {http://arxiv.org/abs/arXiv:astro-ph/0512349}
  {\path{arXiv:arXiv:astro-ph/0512349}}, \href
  {http://dx.doi.org/10.1086/505291} {\path{doi:10.1086/505291}}.

\bibitem{lin11}
E.~V. {Linder}, {Lensing time delays and cosmological complementarity}, \prd
  84~(12) (2011) 123529.
\newblock \href {http://arxiv.org/abs/1109.2592} {\path{arXiv:1109.2592}},
  \href {http://dx.doi.org/10.1103/PhysRevD.84.123529}
  {\path{doi:10.1103/PhysRevD.84.123529}}.

\bibitem{suyu10}
S.~H. {Suyu}, P.~J. {Marshall}, M.~W. {Auger}, S.~{Hilbert}, R.~D. {Blandford},
  L.~V.~E. {Koopmans}, C.~D. {Fassnacht}, T.~{Treu}, {Dissecting the
  Gravitational lens B1608+656. II. Precision Measurements of the Hubble
  Constant, Spatial Curvature, and the Dark Energy Equation of State}, ApJ 711
  (2010) 201--221.
\newblock \href {http://arxiv.org/abs/0910.2773} {\path{arXiv:0910.2773}},
  \href {http://dx.doi.org/10.1088/0004-637X/711/1/201}
  {\path{doi:10.1088/0004-637X/711/1/201}}.

\bibitem{suyu13a}
S.~H. {Suyu}, M.~W. {Auger}, S.~{Hilbert}, P.~J. {Marshall}, M.~{Tewes},
  T.~{Treu}, C.~D. {Fassnacht}, et~al., {Two Accurate Time-delay Distances from
  Strong Lensing: Implications for Cosmology}, ApJ 766 (2013) 70.
\newblock \href {http://arxiv.org/abs/1208.6010} {\path{arXiv:1208.6010}},
  \href {http://dx.doi.org/10.1088/0004-637X/766/2/70}
  {\path{doi:10.1088/0004-637X/766/2/70}}.

\bibitem{komatsu11}
E.~{Komatsu}, K.~M. {Smith}, J.~{Dunkley}, C.~L. {Bennett}, B.~{Gold},
  G.~{Hinshaw}, N.~{Jarosik}, et~al., {Seven-year Wilkinson Microwave
  Anisotropy Probe (WMAP) Observations: Cosmological Interpretation}, ApJS 192
  (2011) 18.
\newblock \href {http://arxiv.org/abs/1001.4538} {\path{arXiv:1001.4538}},
  \href {http://dx.doi.org/10.1088/0067-0049/192/2/18}
  {\path{doi:10.1088/0067-0049/192/2/18}}.

\bibitem{snowsl}
T.~{Treu}, P.~J. {Marshall}, F.-Y. {Cyr-Racine}, C.~D. {Fassnacht}, C.~R.
  {Keeton}, E.~V. {Linder}, L.~A. {Moustakas}, et~al., {Dark energy with
  gravitational lens time delays}, ArXiv e-prints\href
  {http://arxiv.org/abs/1306.1272} {\path{arXiv:1306.1272}}.

\bibitem{suyu13b}
S.~H. {Suyu}, T.~{Treu}, S.~{Hilbert}, A.~{Sonnenfeld}, M.~W. {Auger}, R.~D.
  {Blandford}, T.~{Collett}, et~al., {Cosmology from gravitational lens time
  delays and Planck data}, ArXiv e-prints\href {http://arxiv.org/abs/1306.4732}
  {\path{arXiv:1306.4732}}.

\bibitem{2010CQGra..27q3001A}
J.~{Abadie}, B.~P. {Abbott}, R.~{Abbott}, M.~{Abernathy}, T.~{Accadia},
  F.~{Acernese}, C.~{Adams}, R.~{Adhikari}, P.~{Ajith}, B.~{Allen}, et~al.,
  {TOPICAL REVIEW: Predictions for the rates of compact binary coalescences
  observable by ground-based gravitational-wave detectors}, Classical and
  Quantum Gravity 27~(17) (2010) 173001.
\newblock \href {http://arxiv.org/abs/1003.2480} {\path{arXiv:1003.2480}},
  \href {http://dx.doi.org/10.1088/0264-9381/27/17/173001}
  {\path{doi:10.1088/0264-9381/27/17/173001}}.

\bibitem{2013ApJ...779...72D}
M.~{Dominik}, K.~{Belczynski}, C.~{Fryer}, D.~E. {Holz}, E.~{Berti},
  T.~{Bulik}, I.~{Mandel}, R.~{O'Shaughnessy}, {Double Compact Objects. II.
  Cosmological Merger Rates}, \apj 779 (2013) 72.
\newblock \href {http://arxiv.org/abs/1308.1546} {\path{arXiv:1308.1546}},
  \href {http://dx.doi.org/10.1088/0004-637X/779/1/72}
  {\path{doi:10.1088/0004-637X/779/1/72}}.

\bibitem{1986Natur.323..310S}
B.~F. {Schutz}, {Determining the Hubble constant from gravitational wave
  observations}, \nat 323 (1986) 310.
\newblock \href {http://dx.doi.org/10.1038/323310a0}
  {\path{doi:10.1038/323310a0}}.

\bibitem{2002luml.conf..207S}
B.~F. {Schutz}, {Lighthouses of Gravitational Wave Astronomy}, in:
  {M.~Gilfanov, R.~Sunyeav, \& E.~Churazov} (Ed.), Lighthouses of the Universe:
  The Most Luminous Celestial Objects and Their Use for Cosmology, 2002, p.
  207.
\newblock \href {http://arxiv.org/abs/arXiv:gr-qc/0111095}
  {\path{arXiv:arXiv:gr-qc/0111095}}, \href
  {http://dx.doi.org/10.1007/10856495_29} {\path{doi:10.1007/10856495_29}}.

\bibitem{2005ApJ...629...15H}
D.~E. {Holz}, S.~A. {Hughes}, {Using Gravitational-Wave Standard Sirens}, \apj
  629 (2005) 15--22.
\newblock \href {http://arxiv.org/abs/arXiv:astro-ph/0504616}
  {\path{arXiv:arXiv:astro-ph/0504616}}, \href
  {http://dx.doi.org/10.1086/431341} {\path{doi:10.1086/431341}}.

\bibitem{2006PhRvD..74f3006D}
N.~{Dalal}, D.~E. {Holz}, S.~A. {Hughes}, B.~{Jain}, {Short GRB and binary
  black hole standard sirens as a probe of dark energy}, \prd 74~(6) (2006)
  063006.
\newblock \href {http://arxiv.org/abs/arXiv:astro-ph/0601275}
  {\path{arXiv:arXiv:astro-ph/0601275}}, \href
  {http://dx.doi.org/10.1103/PhysRevD.74.063006}
  {\path{doi:10.1103/PhysRevD.74.063006}}.

\bibitem{2009PhRvD..80j4009C}
C.~{Cutler}, D.~E. {Holz}, {Ultrahigh precision cosmology from gravitational
  waves}, \prd 80~(10) (2009) 104009.
\newblock \href {http://arxiv.org/abs/0906.3752} {\path{arXiv:0906.3752}},
  \href {http://dx.doi.org/10.1103/PhysRevD.80.104009}
  {\path{doi:10.1103/PhysRevD.80.104009}}.

\bibitem{2010PhRvD..81l4046H}
C.~M. {Hirata}, D.~E. {Holz}, C.~{Cutler}, {Reducing the weak lensing noise for
  the gravitational wave Hubble diagram using the non-Gaussianity of the
  magnification distribution}, \prd 81~(12) (2010) 124046.
\newblock \href {http://arxiv.org/abs/1004.3988} {\path{arXiv:1004.3988}},
  \href {http://dx.doi.org/10.1103/PhysRevD.81.124046}
  {\path{doi:10.1103/PhysRevD.81.124046}}.

\bibitem{2010ApJ...725..496N}
S.~{Nissanke}, D.~E. {Holz}, S.~A. {Hughes}, N.~{Dalal}, J.~L. {Sievers},
  {Exploring Short Gamma-ray Bursts as Gravitational-wave Standard Sirens},
  \apj 725 (2010) 496--514.
\newblock \href {http://arxiv.org/abs/0904.1017} {\path{arXiv:0904.1017}},
  \href {http://dx.doi.org/10.1088/0004-637X/725/1/496}
  {\path{doi:10.1088/0004-637X/725/1/496}}.

\bibitem{2011ApJ...739...99N}
S.~{Nissanke}, J.~{Sievers}, N.~{Dalal}, D.~{Holz}, {Localizing Compact Binary
  Inspirals on the Sky Using Ground-based Gravitational Wave Interferometers},
  \apj 739 (2011) 99.
\newblock \href {http://arxiv.org/abs/1105.3184} {\path{arXiv:1105.3184}},
  \href {http://dx.doi.org/10.1088/0004-637X/739/2/99}
  {\path{doi:10.1088/0004-637X/739/2/99}}.

\bibitem{tempref}
S.~{Nissanke}, D.~E. {Holz}, N.~{Dalal}, S.~A. {Hughes}, J.~L. {Sievers}, C.~M.
  {Hirata}, {Determining the Hubble constant from gravitational wave
  observations of merging compact binaries}, ArXiv e-prints\href
  {http://arxiv.org/abs/1307.2638} {\path{arXiv:1307.2638}}.

\bibitem{2012PhRvD..86d3011D}
W.~{Del Pozzo}, {Inference of cosmological parameters from gravitational waves:
  Applications to second generation interferometers}, \prd 86~(4) (2012)
  043011.
\newblock \href {http://arxiv.org/abs/1108.1317} {\path{arXiv:1108.1317}},
  \href {http://dx.doi.org/10.1103/PhysRevD.86.043011}
  {\path{doi:10.1103/PhysRevD.86.043011}}.

\bibitem{2010ApJ...715L.138B}
K.~{Belczynski}, M.~{Dominik}, T.~{Bulik}, R.~{O'Shaughnessy}, C.~{Fryer},
  D.~E. {Holz}, {The Effect of Metallicity on the Detection Prospects for
  Gravitational Waves}, \apjl 715 (2010) L138--L141.
\newblock \href {http://arxiv.org/abs/1004.0386} {\path{arXiv:1004.0386}},
  \href {http://dx.doi.org/10.1088/2041-8205/715/2/L138}
  {\path{doi:10.1088/2041-8205/715/2/L138}}.

\bibitem{2010ApJ...708..117B}
K.~{Belczynski}, D.~E. {Holz}, C.~L. {Fryer}, E.~{Berger}, D.~H. {Hartmann},
  B.~{O'Shea}, {On the Origin of the Highest Redshift Gamma-Ray Bursts}, \apj
  708 (2010) 117--126.
\newblock \href {http://arxiv.org/abs/0812.2470} {\path{arXiv:0812.2470}},
  \href {http://dx.doi.org/10.1088/0004-637X/708/1/117}
  {\path{doi:10.1088/0004-637X/708/1/117}}.

\bibitem{Abadieetal:2010}
J.~{Abadie}, B.~P. {Abbott}, R.~{Abbott}, M.~{Abernathy}, T.~{Accadia},
  F.~{Acernese}, C.~{Adams}, R.~{Adhikari}, P.~{Ajith}, B.~{Allen}, et~al.,
  {TOPICAL REVIEW: Predictions for the rates of compact binary coalescences
  observable by ground-based gravitational-wave detectors}, Classical and
  Quantum Gravity 27~(17) (2010) 173001.
\newblock \href {http://arxiv.org/abs/1003.2480} {\path{arXiv:1003.2480}},
  \href {http://dx.doi.org/10.1088/0264-9381/27/17/173001}
  {\path{doi:10.1088/0264-9381/27/17/173001}}.

\bibitem{2012ApJ...749...91F}
C.~L. {Fryer}, K.~{Belczynski}, G.~{Wiktorowicz}, M.~{Dominik}, V.~{Kalogera},
  D.~E. {Holz}, {Compact Remnant Mass Function: Dependence on the Explosion
  Mechanism and Metallicity}, \apj 749 (2012) 91.
\newblock \href {http://arxiv.org/abs/1110.1726} {\path{arXiv:1110.1726}},
  \href {http://dx.doi.org/10.1088/0004-637X/749/1/91}
  {\path{doi:10.1088/0004-637X/749/1/91}}.

\bibitem{2012ApJ...757...91B}
K.~{Belczynski}, G.~{Wiktorowicz}, C.~L. {Fryer}, D.~E. {Holz}, V.~{Kalogera},
  {Missing Black Holes Unveil the Supernova Explosion Mechanism}, \apj 757
  (2012) 91.
\newblock \href {http://arxiv.org/abs/1110.1635} {\path{arXiv:1110.1635}},
  \href {http://dx.doi.org/10.1088/0004-637X/757/1/91}
  {\path{doi:10.1088/0004-637X/757/1/91}}.

\bibitem{2012ApJ...759...52D}
M.~{Dominik}, K.~{Belczynski}, C.~{Fryer}, D.~E. {Holz}, E.~{Berti},
  T.~{Bulik}, I.~{Mandel}, R.~{O'Shaughnessy}, {Double Compact Objects. I. The
  Significance of the Common Envelope on Merger Rates}, \apj 759 (2012) 52.
\newblock \href {http://arxiv.org/abs/1202.4901} {\path{arXiv:1202.4901}},
  \href {http://dx.doi.org/10.1088/0004-637X/759/1/52}
  {\path{doi:10.1088/0004-637X/759/1/52}}.

\bibitem{aasi:2013}
\textbf{LIGO Scientific Collaboration}, \textbf{Virgo Collaboration},
  J.~{Aasi}, J.~{Abadie}, B.~P. {Abbott}, R.~{Abbott}, T.~D. {Abbott},
  M.~{Abernathy}, T.~{Accadia}, F.~{Acernese}, et~al., {Prospects for
  Localization of Gravitational Wave Transients by the Advanced LIGO and
  Advanced Virgo Observatories}, ArXiv e-prints\href
  {http://arxiv.org/abs/1304.0670} {\path{arXiv:1304.0670}}.

\bibitem{2013arXiv1307.6586L}
T.~{Laskar}, E.~{Berger}, N.~{Tanvir}, B.~{Zauderer}, R.~{Margutti},
  A.~{Levan}, D.~{Perley}, et~al., {GRB 120521C at z\~{}6 and the Properties of
  High-redshift GRBs}, ArXiv e-prints\href {http://arxiv.org/abs/1307.6586}
  {\path{arXiv:1307.6586}}.

\bibitem{1977ApJ...214..679B}
J.~A. {Baldwin}, {Luminosity Indicators in the Spectra of Quasi-Stellar
  Objects}, \apj 214 (1977) 679--684.
\newblock \href {http://dx.doi.org/10.1086/155294} {\path{doi:10.1086/155294}}.

\bibitem{1999MNRAS.302L..24C}
S.~{Collier}, K.~{Horne}, I.~{Wanders}, B.~M. {Peterson}, {A new direct method
  for measuring the Hubble constant from reverberating accretion discs in
  active galaxies}, \mnras 302 (1999) L24--L28.
\newblock \href {http://arxiv.org/abs/arXiv:astro-ph/9811278}
  {\path{arXiv:arXiv:astro-ph/9811278}}, \href
  {http://dx.doi.org/10.1046/j.1365-8711.1999.02250.x}
  {\path{doi:10.1046/j.1365-8711.1999.02250.x}}.

\bibitem{2002ApJ...581L..67E}
M.~{Elvis}, M.~{Karovska}, {Quasar Parallax: A Method for Determining Direct
  Geometrical Distances to Quasars}, \apjl 581 (2002) L67--L70.
\newblock \href {http://arxiv.org/abs/arXiv:astro-ph/0211385}
  {\path{arXiv:arXiv:astro-ph/0211385}}, \href
  {http://dx.doi.org/10.1086/346015} {\path{doi:10.1086/346015}}.

\bibitem{wang13}
J.~M. {Wang}, et~al., PRL accepted.

\bibitem{2011ApJ...740L..49W}
D.~{Watson}, K.~D. {Denney}, M.~{Vestergaard}, T.~M. {Davis}, {A New
  Cosmological Distance Measure Using Active Galactic Nuclei}, \apjl 740 (2011)
  L49.
\newblock \href {http://arxiv.org/abs/1109.4632} {\path{arXiv:1109.4632}},
  \href {http://dx.doi.org/10.1088/2041-8205/740/2/L49}
  {\path{doi:10.1088/2041-8205/740/2/L49}}.

\bibitem{2000ApJ...533..631K}
S.~{Kaspi}, P.~S. {Smith}, H.~{Netzer}, D.~{Maoz}, B.~T. {Jannuzi},
  U.~{Giveon}, {Reverberation Measurements for 17 Quasars and the
  Size-Mass-Luminosity Relations in Active Galactic Nuclei}, \apj 533 (2000)
  631--649.
\newblock \href {http://arxiv.org/abs/arXiv:astro-ph/9911476}
  {\path{arXiv:arXiv:astro-ph/9911476}}, \href
  {http://dx.doi.org/10.1086/308704} {\path{doi:10.1086/308704}}.

\bibitem{2013ApJ...767..149B}
M.~C. {Bentz}, K.~D. {Denney}, C.~J. {Grier}, A.~J. {Barth}, B.~M. {Peterson},
  M.~{Vestergaard}, V.~N. {Bennert}, et~al., {The Low-luminosity End of the
  Radius-Luminosity Relationship for Active Galactic Nuclei}, \apj 767 (2013)
  149.
\newblock \href {http://arxiv.org/abs/1303.1742} {\path{arXiv:1303.1742}},
  \href {http://dx.doi.org/10.1088/0004-637X/767/2/149}
  {\path{doi:10.1088/0004-637X/767/2/149}}.

\bibitem{1982ApJ...255..419B}
R.~D. {Blandford}, C.~F. {McKee}, {Reverberation mapping of the emission line
  regions of Seyfert galaxies and quasars}, \apj 255 (1982) 419--439.
\newblock \href {http://dx.doi.org/10.1086/159843} {\path{doi:10.1086/159843}}.

\bibitem{1993PASP..105..247P}
B.~M. {Peterson}, {Reverberation mapping of active galactic nuclei}, PASP 105
  (1993) 247--268.
\newblock \href {http://dx.doi.org/10.1086/133140} {\path{doi:10.1086/133140}}.

\bibitem{2013ApJ...773...90G}
C.~J. {Grier}, P.~{Martini}, L.~C. {Watson}, B.~M. {Peterson}, M.~C. {Bentz},
  K.~M. {Dasyra}, M.~{Dietrich}, et~al., {Stellar Velocity Dispersion
  Measurements in High-luminosity Quasar Hosts and Implications for the AGN
  Black Hole Mass Scale}, \apj 773 (2013) 90.
\newblock \href {http://arxiv.org/abs/1305.2447} {\path{arXiv:1305.2447}},
  \href {http://dx.doi.org/10.1088/0004-637X/773/2/90}
  {\path{doi:10.1088/0004-637X/773/2/90}}.

\bibitem{2010IAUS..267..151P}
B.~M. {Peterson}, {Toward Precision Measurement of Central Black Hole Masses},
  in: B.~M. {Peterson}, R.~S. {Somerville}, T.~{Storchi-Bergmann} (Eds.), IAU
  Symposium, Vol. 267 of IAU Symposium, 2010, pp. 151--160.
\newblock \href {http://arxiv.org/abs/1001.3675} {\path{arXiv:1001.3675}},
  \href {http://dx.doi.org/10.1017/S1743921310006095}
  {\path{doi:10.1017/S1743921310006095}}.

\bibitem{2010ApJ...721..715D}
K.~D. {Denney}, B.~M. {Peterson}, R.~W. {Pogge}, A.~{Adair}, D.~W. {Atlee},
  K.~{Au-Yong}, M.~C. {Bentz}, et~al., {Reverberation Mapping Measurements of
  Black Hole Masses in Six Local Seyfert Galaxies}, \apj 721 (2010) 715--737.
\newblock \href {http://arxiv.org/abs/1006.4160} {\path{arXiv:1006.4160}},
  \href {http://dx.doi.org/10.1088/0004-637X/721/1/715}
  {\path{doi:10.1088/0004-637X/721/1/715}}.

\bibitem{2011ApJ...743L...4B}
A.~J. {Barth}, A.~{Pancoast}, S.~J. {Thorman}, V.~N. {Bennert}, D.~J. {Sand},
  W.~{Li}, G.~{Canalizo}, et~al., {The Lick AGN Monitoring Project 2011:
  Reverberation Mapping of Markarian 50}, \apjl 743 (2011) L4.
\newblock \href {http://arxiv.org/abs/1111.0061} {\path{arXiv:1111.0061}},
  \href {http://dx.doi.org/10.1088/2041-8205/743/1/L4}
  {\path{doi:10.1088/2041-8205/743/1/L4}}.

\bibitem{2006ApJ...647..901M}
K.~G. {Metzroth}, C.~A. {Onken}, B.~M. {Peterson}, {The Mass of the Central
  Black Hole in the Seyfert Galaxy NGC 4151}, \apj 647 (2006) 901--909.
\newblock \href {http://arxiv.org/abs/arXiv:astro-ph/0605038}
  {\path{arXiv:arXiv:astro-ph/0605038}}, \href
  {http://dx.doi.org/10.1086/505525} {\path{doi:10.1086/505525}}.

\bibitem{2007ApJ...659..997K}
S.~{Kaspi}, W.~N. {Brandt}, D.~{Maoz}, H.~{Netzer}, D.~P. {Schneider},
  O.~{Shemmer}, {Reverberation Mapping of High-Luminosity Quasars: First
  Results}, \apj 659 (2007) 997--1007.
\newblock \href {http://arxiv.org/abs/arXiv:astro-ph/0612722}
  {\path{arXiv:arXiv:astro-ph/0612722}}, \href
  {http://dx.doi.org/10.1086/512094} {\path{doi:10.1086/512094}}.

\bibitem{2013ApJ...764..160G}
E.~{Guerras}, E.~{Mediavilla}, J.~{Jimenez-Vicente}, C.~S. {Kochanek}, J.~A.
  {Mu{\~n}oz}, E.~{Falco}, V.~{Motta}, {Microlensing of Quasar Broad Emission
  Lines: Constraints on Broad Line Region Size}, \apj 764 (2013) 160.
\newblock \href {http://arxiv.org/abs/1207.2042} {\path{arXiv:1207.2042}},
  \href {http://dx.doi.org/10.1088/0004-637X/764/2/160}
  {\path{doi:10.1088/0004-637X/764/2/160}}.

\bibitem{2013arXiv1310.6774Z}
Y.~{Zu}, C.~S. {Kochanek}, S.~{Koz{\l}owski}, B.~M. {Peterson}, {Reverberation
  Mapping with Photometry}, ArXiv e-prints\href
  {http://arxiv.org/abs/1310.6774} {\path{arXiv:1310.6774}}.

\end{thebibliography}

\end{document}